\documentclass[journal]{IEEEtran}
\ifCLASSINFOpdf
\else
\fi
\ifCLASSOPTIONcompsoc
 \usepackage[caption=false,font=normalsize,labelfont=sf,textfont=sf]{subfig}
\else
 \usepackage[caption=false,font=footnotesize]{subfig}
\fi
\begin{document}
 
% --------------------------------------------------------------
%                         Start here
% --------------------------------------------------------------
 
\title{Variational Bayesian Multiuser Tracking for Reconfigurable Intelligent Surface Aided MIMO-OFDM Systems}
\author{Boyu~Teng,~\IEEEmembership{Student~Member,~IEEE},~Xiaojun~Yuan,~\IEEEmembership{Senior~Member,~IEEE},\\~and~Rui~Wang,~\IEEEmembership{Senior~Member,~IEEE}
% ,~and~Erik~G.~Larsson,~\IEEEmembership{Fellow,~IEEE}
	\thanks{B. Teng and X. Yuan are with the National Key Laboratory of Science and Technology on Communications, University of Electronic Science and Technology of China, Chengdu 610000, China (e-mail: byteng@std.uestc.edu.cn; xjyuan@uestc.edu.cn).}% <-this % stops a space 
	\thanks{R. Wang is with the College of Electronics and Information Engineering, Tongji University, Shanghai 201804, China. R. Wang is also with the Shanghai Institute of Intelligent Science and Technology, Tongji University, Shanghai 201804, China (e-mail: ruiwang@tongji.edu.cn).
}
% \thanks{E. G. Larsson is with the Department of Electrical Engineering (ISY), Link\"oping University, Link\"oping, Sweden (email: erik.g.larsson@liu.se).}
}
\maketitle
\begin{abstract}
Reconfigurable intelligent surface (RIS) has attracted enormous interest for its potential advantages in assisting both wireless communication and environmental sensing. In this paper, we study a challenging multiuser tracking problem in the multiple-input multiple-output (MIMO) orthogonal frequency division multiplexing (OFDM) system aided by multiple RISs. In particular, we assume that a multi-antenna base station (BS) receives the OFDM symbols from single-antenna users reflected by multiple RISs and tracks the positions of these users. Considering the users' mobility and the blockage of light-of-sight (LoS) paths, we establish a probability transition model to characterize the tracking process, where the geometric constraints between channel parameters and multiuser positions are utilized. We further develop an online message passing algorithm, termed the Bayesian multiuser tracking (BMT) algorithm, to estimate the multiuser positions, the angles-of-arrivals (AoAs) at multiple RISs, and the time delay and the blockage of the LoS path. The Bayesian Cram\'er Rao bound (BCRB) is derived as the fundamental performance limit of the considered tracking problem. Based on the BCRB, we optimize the passive beamforming (PBF) of the multiple RISs to improve the tracking performance. Simulation results show that the proposed PBF design significantly outperforms the counterpart schemes, and our BMT algorithm can achieve up to centimeter-level tracking accuracy.
\end{abstract}
\section{Introduction}
The sixth-generation (6G) mobile communication is expected to provide integrated sensing and communications (ISAC) service to enable emerging applications, such as digital twin, artificial intelligence (AI), virtual reality (VR), internet of vehicles (IoV), etc\cite{wymeersch2021integration,liu2022integrated,yang2021integrated,liu2022survey}.
As a critical issue in sensing, the acquisition of high-precision location information enables the interaction between the digital and physical worlds and is indispensable in the implementation of emerging applications \cite{wymeersch2019radio}.
With the carrier frequencies moving towards millimeter-wave (mmWave) and higher sub-THz spectrum, broad bandwidth signals and large antenna arrays will be widely used in 6G, leading to high sensing resolution in the angular and delay domains\cite{bourdoux20206g}. 
Without introducing additional hardware equipment, high-accuracy localization becomes possible based on radio signals.\par
Common radio localization methods are driven by the measurements of the angle of arrival (AoA), angle of departure (AoD), time of arrival (ToA), and received signal strength (RSS) from radio signals with several base stations (BSs) regarded as reference anchors \cite{wymeersch2019radio,bourdoux20206g}. 
In communication, the multipath effect provides spatial multiplexing which increases the communication capability of the channel. However, only the measurements corresponding to the line-of-sight (LoS) paths contribute to localization, and the non-LoS components reflected by scatters are regarded as interference since environment map information is unavailable \cite{liu2020two}.
% due to the unavailability of  . 
Unfortunately, in mobile communication scenarios, due to highly complicated propagation environments, the LoS path may be blocked by obstacles, such that the localization methods fail to work.\par
To cope with the above issue in localization, reconfigurable intelligent surface (RIS) is deployed in the electromagnetic environment as an energy-efficient and cost-effective device \cite{wu2021intelligent,di2020smart,yuan2021reconfigurable,huang2020holographic}. A RIS is generally an electromagnetic metasurface composed of a large number of passive reflecting elements that independently modify the incident radio waves by inducing controllable amplitude and additional phase changes \cite{xing2022passive}.
It provides an extra configurable reflecting path to assist both communication and localization. With efficient passive beamforming (PBF) design, the wireless channel fading impairment and interference issue can be mitigated leading to significant communication quality improvement \cite{abeywickrama2020intelligent,huang2020reconfigurable}. From the localization point of view, the RIS acts as an extra reference anchor and provides additional measurements to enable localization via a single BS and avoids the reliance on the LoS path \cite{wymeersch2019radio}.
\par
While RIS-aided communication has been extensively investigated, the research on the use of RIS for localization is still in its infancy. Early works are summarized as follows. Based on the RSS measurements, the authors in \cite{zhang2021metalocalization} propose a RIS-aided indoor user positioning scheme where the PBF of RIS is optimized to obtain a high-resolution signal strength map. However, the RSS measurements-based localization schemes rely on an idealized path loss model which is not accurate in practice. 
% In \cite{yu2022location} and \cite{alexandropoulos2022localization}, the semi-passive elements, equipped with radio-frequency (RF) transceiver hardware, are introduced to estimate the AoAs at RIS which are further utilized to obtain accurate position estimation.
In \cite{elzanaty2021reconfigurable} and \cite{wang2021joint}, a 3-dimension (3D) localization system is considered where the channel angle parameters and the user position are estimated based on the maximum likelihood principle. For the multiple-input multiple-output (MIMO) orthogonal frequency division multiplexing (OFDM) systems, the work \cite{lin2021channel} realizes environment mapping and user localization by using the so-called twin-RIS structure and exploiting array signal processing to acquire the channel parameters. 
The work \cite{lin2022conformal} proposes a new conformal structure for RIS which enables the estimations of AoDs and AoAs of a RIS. The scatterer and user localization problem is then tackled by combining the delay and angle estimations. The joint localization and synchronization problem is studied in \cite{fascista2022ris}, where the BS precoding and the RIS PBF design are optimized based on a novel beamforming codebook.
% In the above studies, localization is realized in each frame independently and the estimation algorithms are not designed in a Bayesian manner.
In our prior work \cite{teng2022bayesian}, we consider a user tracking problem and propose an online Bayesian tracking algorithm based on the AoAs estimation at the user antenna array exploiting the temporal correlation of the user location. \par
% However, \cite{teng2022bayesian} assumes the user is equipped with a large-scale antenna array and the LoS paths between the RISs and the user constantly exist, which makes the algorithm less practicable.
All the above works are limited to the category of single-user localization/tracking. Recently, \cite{yu2022location} studies a multiuser localization problem, where two semi-passive RISs, with each element equipped with radio-frequency transceiver hardware, are introduced to estimate the AoAs at RISs and further achieve multiuser localization. However, the deployment of semi-passive RISs requires an expensive hardware overhead, which compromises the cost-effectiveness of using RIS. Moreover, the existing literature adopts the assumption that the LoS paths between RISs and users always exist, which makes the proposed algorithm less practicable since a LoS path is possibly blocked in mobile scenarios.
% The cascaded channel parameter estimation is also simplified due to the existence of an RF chain in RISs.
\par
% To the best of the authors’ knowledge, no studies have tackled the problem of estimating multi-user positioning in a single BS.\par
Motivated by the limitations of the existing work, we study the tracking problem for multiple single-antenna users in a MIMO-OFDM system aided by multiple passive RISs, where the blockage of LoS paths between RISs and users is considered in the tracking process. 
% Both the angle estimations and the time delay estimations are utilized in tracking.
The main contributions of this paper are as follows:
% we study the multiple-RISs aided multiuser tracking problem following the Bayesian principle and design the PBF to enhance tracking performance.
\begin{itemize}
    \item 
    % We study the tracking problem for multiple single-antenna users in uplink MIMO-OFDM systems, where multiple RISs are deployed to assist user localization.
    For the considered tracking problem, we build a probability transition model to characterize the mobility of users and the live-and-death of the LoS paths between RISs and users caused by blockages.
    \item Given the factor graph representation of the probability model,
    we develop an online message passing algorithm, termed the Bayesian multiuser tracking (BMT) algorithm, to jointly estimate the multiuser positions, the AoAs at the multiple RISs, the time delay of each LoS path, and the existence of LoS paths in every frame. We exploit variational inference and Gaussian approximation to calculate the messages.
    \item We calculate the Bayesian information matrix (BIM) and derive the Bayesian Cram\`er Rao bound (BCRB) for the considered multiuser tracking problem. The BCRB provides a theoretic mean square error (MSE) lower bound for the tracking problem.
    \item Based on the derived BCRB, we optimize the passive beamforming design of multiple RISs to improve the tracking performance. The unknown multiuser positions in the optimization problem are substituted by the estimation in the previous frame. The numerical results show that the proposed PBF design outperforms benchmark schemes in sense of both the BCRB and the root mean square error (RMSE) of our multiuser tracking algorithm.
\end{itemize}
\par
The remainder of this paper is organized as follows. Section \ref{S2} introduces the multi-RIS aided MIMO-OFDM systems and formulates the multiuser tracking problem. In Section \ref{S3}, we derive the message calculations and develop the Bayesian multiuser tracking algorithm. In Section \ref{S4}, we analyze the BCRB for the multiuser tracking problem. In Section \ref{S5}, we present the BCRB-based passive beamforming design for the multiple RISs. Numerical results are presented in Section \ref{S6}, and the paper concludes in Section \ref{S7}.\par
\textit{Notations:} Throughout, vectors and matrices are denoted by bold lowercase letters and bold capital letters, respectively. We use $(\cdot)^{\mathrm{T}}$, $(\cdot)^{\mathrm{H}}$, and $(\cdot)^{*}$ to denote the operations of transpose, conjugate transpose, and conjugate, respectively. We use Tr($\mathbf{X}$) to denote the trace of $\mathbf{X}$, $\Re\{\mathbf{X}\}$ to denote the real part of $\mathbf{X}$, diag($\mathbf{x}$) to denote the diagonal matrix with its diagonal entries given by $\mathbf{x}$, $[\mathbf{X}]_{\mathcal{I},\mathcal{J}}$ to denote the submatrix of $\mathbf{X}$ with the rows indexed by the set $\mathcal{I}$ and columns indexed by the set $\mathcal{J}$, $a:b$ to denote the set $\{i|a\leq i\leq b,i\in\mathbb{Z}\}$ and $\mathbf{I}$ to denote the identity matrix. We use $\mathcal{N}\left(\boldsymbol{x};\boldsymbol{\mu},\boldsymbol{\Sigma}\right)$ and $\mathcal{CN}\left(\boldsymbol{x};\boldsymbol{\mu},\boldsymbol{\Sigma}\right)$ to denote the real Gaussian distribution and the circularly-symmetric Gaussian distribution with mean vector $\boldsymbol{\mu}$ and covariance matrix $\boldsymbol{\Sigma}$. We use $\mathcal{M}({x};{\mu},{\kappa})$ to denote the Von Mises (VM) distribution with mean direction ${\mu}$ and concentration parameters ${\kappa}$. We use $\mathbb{E}[\cdot]$ to denote the expectation operator, $\odot$ to denote the Hadamard product, $\circ$ to denote the cross product, {$\otimes$ to denote the Kronecker product,} $\|\cdot\|_p$ to denote the $\ell_{p}$ norm, and $\jmath$ to denote the imaginary unit.
\section{System model}\label{S2}
\subsection{System Description}
\begin{figure}[t]
    \centering
    \resizebox{9cm}{!}{\includegraphics{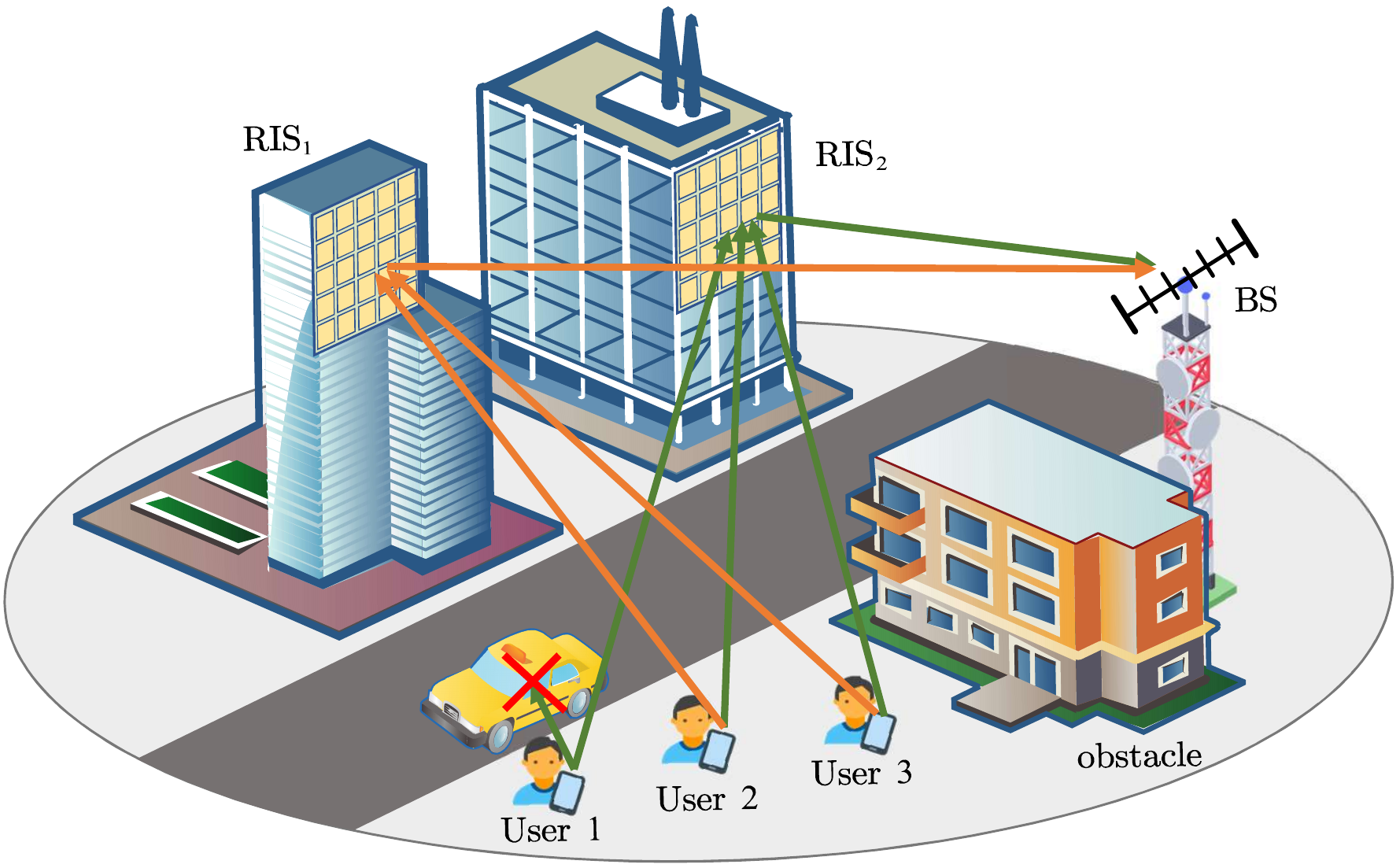}}
    \caption{System model.}
    \label{system}
\end{figure}
We consider a multi-RIS aided multiuser MIMO-OFDM system as illustrated in Fig. \ref{system}. In uplink transmission, a BS simultaneously receives the OFDM pilot signals transmitted by $K$ single antenna users to achieve multiuser tracking. The LoS paths between BS and users are assumed to be obstructed by following \cite{wei2021channel,he2021channel,wang2021joint}, and the user signals are reflected by $M$ RISs. We assume that the BS consists of $N_{\mathrm{B}}$ antennas arranged as a uniform linear array (ULA) and each RIS consists of $N_{\mathrm{R}}=N_\mathrm{x}\times N_\mathrm{y}$ reflecting elements arranged as uniform planar array (UPA). The antenna/element interval is set to half of the carrier frequency wavelength $\lambda$. The BS simultaneously tracks the $K$ users by estimating their positions in every pilot frame. Each pilot frame contains $G$ OFDM symbols and each OFDM symbol occupied $L$ subcarriers with bandwidth $f_s$. We assume a quasi-static flat fading channel model, where the channel coefficients and the user positions remain unchanged within a pilot frame. \par
Assume that a 3D Cartesian coordinate system has been set up appropriately. The positions of the BS, the $m$-th RIS and the $k$-th user in the $t$-th frame transmission are denoted by $\mathbf{p}_\mathrm{B}$, $\mathbf{p}_{\mathrm{R},m}$ and $\mathbf{p}_{k}^{(t)}$ respectively. The positions of the BS and the RISs can be acquired by the user and the BS accurately in advance since the BS and the RISs are deployed stationarily.
\subsection{Signal Model}
We adopt the geometric channel model used in \cite{he2021channel,wang2021joint}. In the $t$-th frame, the channel from the $k$-th user to the $m$-th RIS and from the $m$-th RIS to the BS over the $l$-th subcarrier are given by
\begin{align}
    \mathbf{h}_{m,k}^{(t,l)} &= \sqrt{\frac{\alpha}{1+\alpha}}\bar{\mathbf{h}}_{m,k}^{(t,l)} +\sqrt{\frac{1}{1+\alpha}}\tilde{\mathbf{h}}_{m,k}^{(t,l)},\\
	\mathbf{H}_{m}^{(t,l)} &= \sqrt{\frac{\alpha}{1+\alpha}}\bar{\mathbf{H}}_{m}^{(l)} +\sqrt{\frac{1}{1+\alpha}}\tilde{\mathbf{H}}_{m}^{(t,l)},
\end{align}
where $\tilde{\mathbf{h}}_{m,k}^{(t,l)}$ and $\tilde{\mathbf{H}}_{m}^{(t,l)}$ are the corresponding NLoS multi-path components; $\alpha$ is the Rician factor; $\bar{\mathbf{h}}_{m,k}^{(t,l)}$ and $\bar{\mathbf{H}}_{m}^{(l)}$ are the corresponding LoS path components given by
\begin{align}
    \bar{\mathbf{h}}_{m,k}^{(t,l)} &= \zeta_{m,k}^{(t)}\rho_{m,k}^{(t)} \mathrm{e}^{-\jmath\frac{2\pi f_s l}{L}\tau_{m,k}^{(t)}}\mathbf{a}_{\mathrm{R}}(\theta_{m,k,x}^{(t)},\theta_{m,k,y}^{(t)}),\\
    \bar{\mathbf{H}}_{m}^{(l)} &= \rho_{m}\mathrm{e}^{-\jmath\frac{2\pi f_s l}{L}\tau_{m}}\mathbf{a}_{\mathrm{B}}(\theta_{m})\mathbf{a}_{\mathrm{R}}^{\mathrm{H}}(\phi_{m,x},\phi_{m,y}),
\end{align}
where binary variable $\zeta_{m,k}^{(t)}\in\{0,1\}$ indicates the existence of the LoS path between the $m$-th
RIS and the $k$-th user; 
$\rho_{m,k}^{(t)}$ and $\tau_{m,k}^{(t)}$ denote the path gain and time delay of the LoS path between the $k$-th user and the $m$-th RIS in the $t$-th frame; $\rho_{m}$ and $\tau_{m}$ denote the path gain and time delay of the LoS path between the $m$-th RIS and the BS; $\theta_{m,k,x}^{(t)}$ and $\theta_{m,k,y}^{(t)}$ denote the cosine of the AoA of the LoS path along the $x$ direction and $y$ direction of the $m$-th RIS w.r.t. the $k$-th user; $\phi_{m,x}$ and $\phi_{m,y}$ are the cosine of the AoD at the $m$-th RIS; $\theta_{m}$ is the cosine of AoA of the LoS path from the $m$-th RIS to BS; $\mathbf{a}_{\mathrm{R}}(\cdot)$ and $\mathbf{a}_{\mathrm{B}}(\cdot)$ are the steering vector of BS antennas and RIS elements, respectively, given by $\mathbf{a}_{\mathrm{R}}(\theta_1,\theta_2)=\mathbf{a}_{\mathrm{x}}(\theta_1)\otimes\mathbf{a}_{\mathrm{y}}(\theta_2)$ and $\mathbf{a}_{{x}}(\theta) = [1,\mathrm{e}^{\jmath\pi {\theta}},\dots,\mathrm{e}^{\jmath\pi  {\theta}(N_{x}-1)}]^{\mathrm{T}}$ for $x\in\{\mathrm{B},{\mathrm{x},\mathrm{y}}\}$. We assume that the RIS PBF configuration remains the same for different subcarriers and varies over different OFDM symbols. Let $\boldsymbol{\omega}_{m}^{(t,g)}\in\mathbb{C}^{N_{\mathrm{R}}}$ denotes the PBF vector of the $m$-th RIS in the $g$-th OFDM symbol. The received signal over the $l$-th subcarrier of the $g$-th OFDM symbol is given by
\begin{equation}
\label{5}
    \boldsymbol{y}^{(t,g,l)}=\sum_{m=1}^{M}{\sum_{k=1}^{K}{\mathbf{H}_{m}^{(t,l)}\mathrm{diag}(\boldsymbol{\omega}_{m}^{(t,g)})\mathbf{h}_{m,k}^{(t,l)}x_{k}^{(t,l)}}}+\boldsymbol{w}^{(t,g,l)},
\end{equation}
where $x_{k}^{(t,l)}$ is the pilot signal of the $k$-th user and $\boldsymbol{w}^{(t,g,l)}$ is the additive white Gaussian noise. In user localization and tracking tasks, the information of user position $\mathbf{p}_{k}^{(t)}$ is contained in the channel parameters of the LoS path, i.e., $\theta_{m,k,x}^{(t)}$, $\theta_{m,k,y}^{(t)}$ and $\tau_{m,k}^{(t)}$. Therefore, we focus on the LoS path component and separate it from the NLoS multi-path components. With some manipulations, received signal \eqref{5} is simplified into a superposition of $M$ steering vectors expressed as
\begin{align}
\boldsymbol{y}^{(t,g,l)}
% \sum_{m=1}^{M}{\sum_{k=1}^{K}{\bar{\mathbf{H}}_{m}^{(t,l)}\mathrm{diag}(\boldsymbol{\omega}_{m}^{(t,g)})\bar{\mathbf{h}}_{m,k}^{(t,l)}x_{k}^{(t,l)}}}+\boldsymbol{w}^{(t,g,l)}\notag\\
    =\sum_{m=1}^{M}{[\boldsymbol{R}_{m}^{(t)}]_{g,l}\mathbf{a}_{\mathrm{B}}(\theta_{m})}+\boldsymbol{n}^{(t,g,l)}.\label{6}
\end{align}
Let $\boldsymbol{r}_{g,l}^{(t)}=\left[[\boldsymbol{R}_{1}^{(t)}]_{g,l},...,[\boldsymbol{R}_{M}^{(t)}]_{g,l}\right]^{\mathrm{T}}$ and $\mathbf{A}_{\mathrm{B}}(\boldsymbol{\theta})=[\mathbf{a}_{\mathrm{B}}(\theta_{1}),...,\mathbf{a}_{\mathrm{B}}(\theta_{M})]$, a compact form of \eqref{6} is
\begin{align}\label{10}
    \boldsymbol{y}^{(t,g,l)}=\mathbf{A}_{\mathrm{B}}(\boldsymbol{\theta})\boldsymbol{r}_{g,l}^{(t)}+\boldsymbol{n}^{(t,g,l)},
\end{align}
where $\boldsymbol{n}^{(t,g,l)}\in\mathbb{C}^{N_\mathrm{B}}$ is the interference-plus-noise term given by
\begin{align}
    \boldsymbol{n}^{(t,g,l)}\!=&\sum_{m=1}^{M}{\sum_{k=1}^{K}{\left(\tilde{\mathbf{H}}_{m}^{(t,l)}\mathrm{diag}(\boldsymbol{\omega}_{m}^{(t,g)})\tilde{\mathbf{h}}_{m,k}^{(t,l)}+\!
    \tilde{\mathbf{H}}_{m}^{(t,l)}\mathrm{diag}(\boldsymbol{\omega}_{m}^{(t,g)})\bar{\mathbf{h}}_{m,k}^{(t,l)}+\!
    \bar{\mathbf{H}}_{m}^{(l)}\mathrm{diag}(\boldsymbol{\omega}_{m}^{(t,g)})\tilde{\mathbf{h}}_{m,k}^{(t,l)}\right)x_{k}^{(t,l)}}}\notag\\
    &+\boldsymbol{w}^{(t,g,l)}.
\end{align}
Due to the severe large-scale fading and reflection loss, the power of interference is generally 20 dB weaker than the LoS path \cite{akdeniz2014millimeter}. We assume that the interference-plus-noise term follows complex Gaussian distribution \cite{wang2021joint}, i.e., $\mathcal{CN}(\boldsymbol{n}^{(t,g,l)};\mathbf{0},\nu\mathbf{I})$. The complex amplitude $[\boldsymbol{R}_{m}^{(t)}]_{g,l}$ corresponding to the $m$-th RIS in \eqref{6} is given by
\begin{align}
\label{8}
    [\boldsymbol{R}_{m}^{(t)}]_{g,l}=\sum_{k=1}^{K}{x_{k}^{(t,l)}\zeta_{m,k}^{(t)}\varrho_{m,k}^{(t)} \mathrm{e}^{-\jmath\frac{2\pi f_s l}{L}\varsigma_{m,k}^{(t)}}(\boldsymbol{\omega}_{m}^{(t,g)})^{\mathrm{T}} \mathbf{a}_{\mathrm{R}}(\vartheta_{m,k,x}^{(t)},\vartheta_{m,k,y}^{(t)})},
\end{align}
with $\varrho_{m,k}^{(t)}=\rho_{m,k}^{(t)}\rho_{m}$, $\vartheta_{m,k,x}^{(t)}=\theta_{m,k,x}^{(t)}-\phi_{m,x}$, $\vartheta_{m,k,y}^{(t)}=\theta_{m,k,y}^{(t)}-\phi_{m,y}$, and $\varsigma_{m,k}^{(t)}=\tau_{m}+\tau_{m,k}^{(t)}$. We define $\mathbf{a}_{L}(\varsigma)=[1,\mathrm{e}^{-\jmath\frac{2\pi f_s}{L}\varsigma},...,\mathrm{e}^{-\jmath\frac{2\pi f_s}{L}\varsigma(L-1)}]$.
% Let $\boldsymbol{W}_{k}^{(t)}=[\boldsymbol{\omega}_{k}^{(t,1)},...,\boldsymbol{\omega}_{m}^{(t,g)}]$ and $[\boldsymbol{X}^{(t)}]_{l,m}=x_{k}^{(t,l)}$. Let $\boldsymbol{\varrho}_{m}^{(t)}=[\varrho_{k,1}^{(t)},...,\varrho_{m,k}^{(t)}]^{\mathrm{T}}$ and apply similar notation to $\vartheta_{m,k,x}^{(t)}$, $\vartheta_{m,k,y}^{(t)}$, $\varsigma_{m,k}^{(t)}$ and $\zeta_{m,k}^{(t)}$.
Let $\boldsymbol{W}_{m}^{(t)}=[\boldsymbol{\omega}_{m}^{(t,1)},...,\boldsymbol{\omega}_{m}^{(t,G)}]$ and $\boldsymbol{x}_{k}^{(t)}=[x_{k}^{(t,1)},...,x_{k}^{(t,L)}]^{\mathrm{T}}$.
The matrix $\boldsymbol{R}_{m}^{(t)}\in\mathbb{C}^{G\times L}$ is expressed by 
\begin{align}
\label{9}
    % \boldsymbol{R}_{m}^{(t)}=(\boldsymbol{W}_{k}^{(t)})^{\mathrm{T}} \mathbf{A}_{\mathrm{R}}(\boldsymbol{\vartheta}_{m,x}^{(t)},\boldsymbol{\vartheta}_{m,y}^{(t)})\mathrm{diag}(\boldsymbol{\varrho}_{m}^{(t)})(\boldsymbol{X}^{(t)}\odot\mathbf{A}_{L}(\boldsymbol{\varsigma}_{k}^{(t)}))^{\mathrm{T}}.\\
    \boldsymbol{R}_{m}^{(t)}=(\boldsymbol{W}_{m}^{(t)})^{\mathrm{T}}\sum_{k=1}^{K} \zeta_{m,k}^{(t)}\varrho_{m,k}^{(t)}\mathbf{a}_{\mathrm{R}}(\vartheta_{m,k,x}^{(t)},\vartheta_{m,k,y}^{(t)})(\boldsymbol{x}_{k}^{(t)}\odot\mathbf{a}_{L}(\varsigma_{m,k}^{(t)}))^{\mathrm{T}}.
\end{align}
% where the matrices $\mathbf{A}_{\mathrm{R}}(\boldsymbol{\vartheta}_{m,x}^{(t)},\boldsymbol{\vartheta}_{m,y}^{(t)})=[\mathbf{a}_{\mathrm{R}}(\vartheta_{k,1,x}^{(t)},\vartheta_{k,1,y}^{(t)}),...,\mathbf{a}_{\mathrm{R}}(\vartheta_{m,k,x}^{(t)},\vartheta_{m,k,y}^{(t)})]$ and $\mathbf{A}_{L}(\boldsymbol{\varsigma}^{(t)})=[\mathbf{a}_{L}(\varsigma_{k,1}^{(t)}),...,\mathbf{a}_{L}(\varsigma_{m,k}^{(t)})]$ are the collection of steering vector .
% The complex amplitude $\boldsymbol{r}_{g,l}^{(t)}$ and $\boldsymbol{R}_{m}^{(t)}$ are organized as shown in Fig. .
\par
From \eqref{9}, we note that the channel parameters $\theta_{m,k,x}^{(t)}$, $\theta_{m,k,y}^{(t)}$ and $\tau_{m,k}^{(t)}$ are wrapped into the complex amplitude $\boldsymbol{R}_{m}^{(t)}$. The estimation of these channel parameters from $\boldsymbol{R}_{m}^{(t)}$ falls into the category of multidimensional line spectrum inference. The user position is further estimated based on the geometric constraints between $\varsigma_{m,k}^{(t)}$, $\vartheta_{m,k,x}^{(t)}$ and $\vartheta_{m,k,y}^{(t)}$ and $\mathbf{p}_{k}^{(t)}$ given by
%  By collecting $\alpha_k^{(t,g,l)}$ in different subcarriers and different OFDM symbols, we can well est
% The user positions are related to the channel parameters among RISs and users as
\begin{subequations}
\label{11}
\begin{align}
    \label{11a}\vartheta_{m,k,x}^{(t)} &= \frac{(\mathbf{p}_{k}^{(t)}-\mathbf{p}_{\mathrm{R},m})^{\mathrm{T}}\mathbf{e}_{m,x}}{\left\|\mathbf{p}_{k}^{(t)}-\mathbf{p}_{\mathrm{R},m}\right\|_{2}}-\phi_{m,x},\\
    \label{11b}\vartheta_{m,k,y}^{(t)} &= \frac{(\mathbf{p}_{k}^{(t)}-\mathbf{p}_{\mathrm{R},m})^{\mathrm{T}}\mathbf{e}_{m,y}}{\left\|\mathbf{p}_{k}^{(t)}-\mathbf{p}_{\mathrm{R},m}\right\|_{2}}-\phi_{m,y},\\
    \label{11c}\varsigma_{m,k}^{(t)} &=\tau_{m}+ \frac{\left\|\mathbf{p}_{k}^{(t)}-\mathbf{p}_{\mathrm{R},m}\right\|_{2}}{\mathrm{c}_{0}},
\end{align}
\end{subequations}
where $\mathbf{e}_{m,x}$ and $\mathbf{e}_{m,y}$ are the x-axis and y-axis unit vectors of the $m$-th RIS, respectively, and $\mathrm{c}_{0}$ denotes the speed of light.
\subsection{Probabilistic Problem Formulation}
We build a probability model for $\mathbf{p}_{k}^{(t)}$, $\vartheta_{m,k,x}^{(t)}$, $\vartheta_{m,k,y}^{(t)}$, $\varsigma_{m,k}^{(t)}$, $\zeta_{m,k}^{(t)}$, $\varrho_{m,k}^{(t)}$ and $\boldsymbol{y}^{(t,g,l)}$ in the whole tracking duration. Given the geometric constraint in \eqref{9}, the conditional probability density functions (pdf) $p(\vartheta_{m,k,x}^{(t)}|\mathbf{p}_{k}^{(t)})$, $p(\vartheta_{m,k,y}^{(t)}|\mathbf{p}_{k}^{(t)})$ and $p(\varsigma_{m,k}^{(t)}|\mathbf{p}_{k}^{(t)})$ are given by
\begin{subequations}\label{12}
\begin{align}
    p(\vartheta_{m,k,x}^{(t)}|\mathbf{p}_{k}^{(t)}) &= \delta\left(\vartheta_{m,k,x}^{(t)}-\frac{(\mathbf{p}_{k}^{(t)}-\mathbf{p}_{\mathrm{R},m})^{\mathrm{T}}\mathbf{e}_{m,x}}{\left\|\mathbf{p}_{k}^{(t)}-\mathbf{p}_{\mathrm{R},m}\right\|_{2}}+\phi_{m,x}\right),\\
    p(\vartheta_{m,k,y}^{(t)}|\mathbf{p}_{k}^{(t)}) &= \delta\left(\vartheta_{m,k,y}^{(t)}-\frac{(\mathbf{p}_{k}^{(t)}-\mathbf{p}_{\mathrm{R},m})^{\mathrm{T}}\mathbf{e}_{m,y}}{\left\|\mathbf{p}_{k}^{(t)}-\mathbf{p}_{\mathrm{R},m}\right\|_{2}}+\phi_{m,y}\right),\\
    p(\varsigma_{m,k}^{(t)}|\mathbf{p}_{k}^{(t)}) &= \delta\left(\varsigma_{m,k}^{(t)}-\frac{\left\|\mathbf{p}_{k}^{(t)}-\mathbf{p}_{\mathrm{R},m}\right\|_{2}}{\mathrm{c}_{0}}-\tau_{m}\right),
\end{align}
\end{subequations}
where $\delta(\cdot)$ is a Dirac delta function. The existence of the LoS paths between RISs and users may change over time due to the blockage of obstacles. The prior distribution of the cascaded channel path gain $\varrho_{m,k}^{(t)}$ is modeled by a complex Gaussian distribution, i.e., $p(\varrho_{m,k}^{(t)})=\mathcal{CN}(\varrho_{m,k}^{(t)};0,\sigma_{m}^{(t)})$.
% where $\zeta_{m,k}^{(l)}$ is the Bernoulli variable governed by $p(\zeta_{m,k}^{(l)})=\lambda^{\zeta_{m,k}^{(l)}}(1-\lambda)^{(1-\zeta_{m,k}^{(l)})}$.
$\varrho_{m,k}^{(t)}$ is independent with $\varrho_{m',k'}^{(t')}$ for any $m\neq m'$, $k\neq k'$, $t\neq t'$. Given the observation model in \eqref{6}, the likelihood function is represented by
\begin{align}
    p(\boldsymbol{y}^{(t,g,l)}|\{\boldsymbol{\vartheta}_{m,x}^{(t)},\boldsymbol{\vartheta}_{m,y}^{(t)},\boldsymbol{\varsigma}_{m}^{(t)},\boldsymbol{\zeta}_{m}^{(t)},\boldsymbol{\varrho}_{m}^{(t)}\}_{m=1}^{M})=p(\boldsymbol{y}^{(t,g,l)}|\boldsymbol{r}_{g,l}^{(t)})\prod_{m=1}^{M}p([\boldsymbol{r}_{g,l}^{(t)}]_{m}|\boldsymbol{\vartheta}_{m,x}^{(t)},\boldsymbol{\vartheta}_{m,y}^{(t)},\boldsymbol{\varsigma}_{m}^{(t)},\boldsymbol{\zeta}_{m}^{(t)},\boldsymbol{\varrho}_{m}^{(t)}),
\end{align}
where $\boldsymbol{\varrho}_{m}^{(t)}=[\varrho_{m,1}^{(t)},...,\varrho_{m,k}^{(t)}]^{\mathrm{T}}$ and similar notations are applied to $\vartheta_{m,k,x}^{(t)}$, $\vartheta_{m,k,y}^{(t)}$, $\varsigma_{m,k}^{(t)}$ and $\zeta_{m,k}^{(t)}$,
$p([\boldsymbol{r}_{g,l}^{(t)}]_{m}|\boldsymbol{\vartheta}_{m,x}^{(t)},\boldsymbol{\vartheta}_{m,y}^{(t)},\boldsymbol{\varsigma}_{m}^{(t)},\boldsymbol{\zeta}_{m}^{(t)},\boldsymbol{\varrho}_{m}^{(t)})$ is a Dirac delta function derived from \eqref{8} and $p(\boldsymbol{y}^{(t,g,l)}|\boldsymbol{r}_{g,l}^{(t)})=\mathcal{CN}(\boldsymbol{y}^{(t,g,l)};\mathbf{A}_{\mathrm{B}}(\boldsymbol{\theta}_k)\boldsymbol{r}_{g,l}^{(t)},\nu\mathbf{I})$.\par
Considering the temporal correlation of user positions, we build a Markov transition probability model to characterize user movements. The difference of a user position between any two adjacent frames is modeled as an independent Gaussian noise \cite{teng2022bayesian}. The conditional probability of $\mathbf{p}_{k}^{(t)}$ given $\mathbf{p}_{k}^{(t-1)}$ is given by
\begin{align}\label{15}
    p(\mathbf{p}_{k}^{(t)}|\mathbf{p}_{k}^{(t-1)})=\mathcal{N}(\mathbf{p}_{k}^{(t)};\mathbf{p}_{k}^{(t-1)},\mathbf{C}_{k})
\end{align}
with $\mathbf{C}_{k}=\mathrm{diag}([\sigma_{k,x}^2,\sigma_{k,y}^2,\sigma_{k,z}^2]^{\mathrm{T}})$.\par 
In the tracking duration, the binary variable $\zeta_{m,k}^{(t)}$ does not change frequently among different frames since the user continuously moves and the time gap of adjacent frames is small. The existence of the LoS path is correlated over time. Therefore, we use a birth and death process to model the state transition of $\zeta_{m,k}^{(t)}$, i.e.,
\begin{align}\label{deth}
   p(\zeta_{m,k}^{(t)} \mid \zeta_{m,k}^{(t-1)})= \begin{cases}\left(1-p_{l,m,k}^{(t)}\right)^{1-\zeta_{m,k}^{(t)}}\left(p_{l,m,k}^{(t)}\right)^{\zeta_{m,k}^{(t)}}, & \zeta_{m,k}^{(t-1)}=0; \\ \left(p_{d,m,k}^{(t)}\right)^{1-\zeta_{m,k}^{(t)}}\left(1-p_{d,m,k}^{(t)}\right)^{\zeta_{m,k}^{(t)}}, & \zeta_{m,k}^{(t-1)}=1,\end{cases}
\end{align}
where $p_{l,m,k}^{(t)}=p(\zeta_{m,k}^{(t)}=1|\zeta_{m,k}^{(t-1)}=0)$ and $p_{d,m,k}^{(t)}=p(\zeta_{m,k}^{(t)}=0|\zeta_{m,k}^{(t-1)}=1)$. 
Then, the joint pdf of the random variables in each frame is given by
\begin{align}\label{jointpdf}
    p(\{\mathbf{p}_\mathrm{k}^{(t)}\},&\{\zeta_{m,k}^{(t)}\},\{\vartheta_{m,k,x}^{(t)}\},\{\vartheta_{m,k,y}^{(t)}\},\{\varsigma_{m,k}^{(t)}\},\{\varrho_{m,k}^{(t)}\},\{\boldsymbol{y}^{(t,g,l)}\})\notag\\
    =&{f_m\left(\{\mathbf{p}_\mathrm{k}^{(t)}\},\{\zeta_{m,k}^{(t)}\}\right)}{f_g\left(\{\vartheta_{m,k,x}^{(t)}\},\{\vartheta_{m,k,y}^{(t)}\},\{\varsigma_{m,k}^{(t)}\},\{\varrho_{m,k}^{(t)}\}|\{\mathbf{p}_\mathrm{k}^{(t)}\}\right)}\notag\\
    &\times{f_l\left(\{\boldsymbol{y}^{(t,g,l)}\}|\{\vartheta_{m,k,x}^{(t)}\},\{\vartheta_{m,k,y}^{(t)}\},\{\varsigma_{m,k}^{(t)}\},\{\zeta_{m,k}^{(t)}\},\{\varrho_{m,k}^{(t)}\}\right)},
\end{align}
where $\{x\}$ represents the collection of variable $x$ over all considered indexes. Factors $f_m$, $f_g$ and $f_l$ are respectively given by
\begin{subequations}
\begin{align}
    % f_m(\cdot)&=\prod_{k=1}^{K}{\left(p(\mathbf{p}_{k}^{(0)})\prod_{m=1}^{M}{p(\zeta_{m,k}^{(0)})}\prod_{j=1}^{t}\left({p(\mathbf{p}_{k}^{(j)}|\mathbf{p}_{k}^{(j-1)})}\prod_{m=1}^{M}{p(\zeta_{m,k}^{(j)}|\zeta_{m,k}^{(j-1)})}\right)\right)},\\
    f_m(\cdot)&=\prod_{k=1}^{K}{\left(p(\mathbf{p}_{k}^{(0)})\prod_{m=1}^{M}{p(\zeta_{m,k}^{(0)})}\prod_{j=1}^{t}\left({p(\mathbf{p}_{k}^{(j)}|\mathbf{p}_{k}^{(j-1)})}{p(\zeta_{m,k}^{(j)}|\zeta_{m,k}^{(j-1)})}\right)\right)},\\
    f_g(\cdot)&=\prod_{j=1}^{t}\prod_{m=1}^{M}{\prod_{k=1}^{K}{p(\vartheta_{m,k,x}^{(j)}|\mathbf{p}_{k}^{(j)})p(\vartheta_{m,k,y}^{(j)}|\mathbf{p}_{k}^{(j)})p(\varsigma_{m,k}^{(j)}|\mathbf{p}_{k}^{(j)})p(\varrho_{m,k}^{(j)})}},\\
    f_l(\cdot)&=\prod_{j=1}^{t}{\prod_{g=1}^{G}{\prod_{l=1}^{L}{p(\boldsymbol{y}^{(t,g,l)}|\{\boldsymbol{\vartheta}_{m,x}^{(t)},\boldsymbol{\vartheta}_{m,y}^{(t)},\boldsymbol{\varsigma}_{m}^{(t)},\boldsymbol{\zeta}_{m}^{(t)},\boldsymbol{\varrho}_{m}^{(t)}\}_{m=1}^{M})}}},
\end{align}
\end{subequations}
where $p(\mathbf{p}_{k}^{(0)})$ and $p(\zeta_{m,k}^{(0)})$ are respectively the distribution of $\mathbf{p}_{k}^{(t)}$ and $\zeta_{m,k}^{(t)}$ at the initial frame. In the multiuser tracking task, we aim to develop an online algorithm that in each frame the BS can estimate the position of each user $\mathbf{p}_{k}^{(t)}$ and the binary variable $\zeta_{m,k}^{(0)}$ given the historically received signal and the initial distribution. Despite the traditional minimum mean-square error (MMSE) and maximum \textit{a posteriori} (MAP) estimators being desirable, the high computational complexity of the high dimension integral or the high dimensional search makes them intractable. In this paper, we provide a low-complexity approximate solution by following the message passing principle and the variational Bayesian principle.
% For each user, the positions $\{\mathbf{p}_{k}^{(t)}\}$ form a Markov chain satisfying
% \begin{align}
% p(\mathbf{p}_\mathrm{m}^{(t)}|\mathbf{p}_\mathrm{m}^{(0:t-1)},\boldsymbol{y}^{(1:t-1)})=p(\mathbf{p}_\mathrm{m}^{(t)}|\mathbf{p}_\mathrm{m}^{(t-1)}),
% \end{align}
\begin{figure}[t]
    \centering
    \resizebox{14cm}{!}{\includegraphics{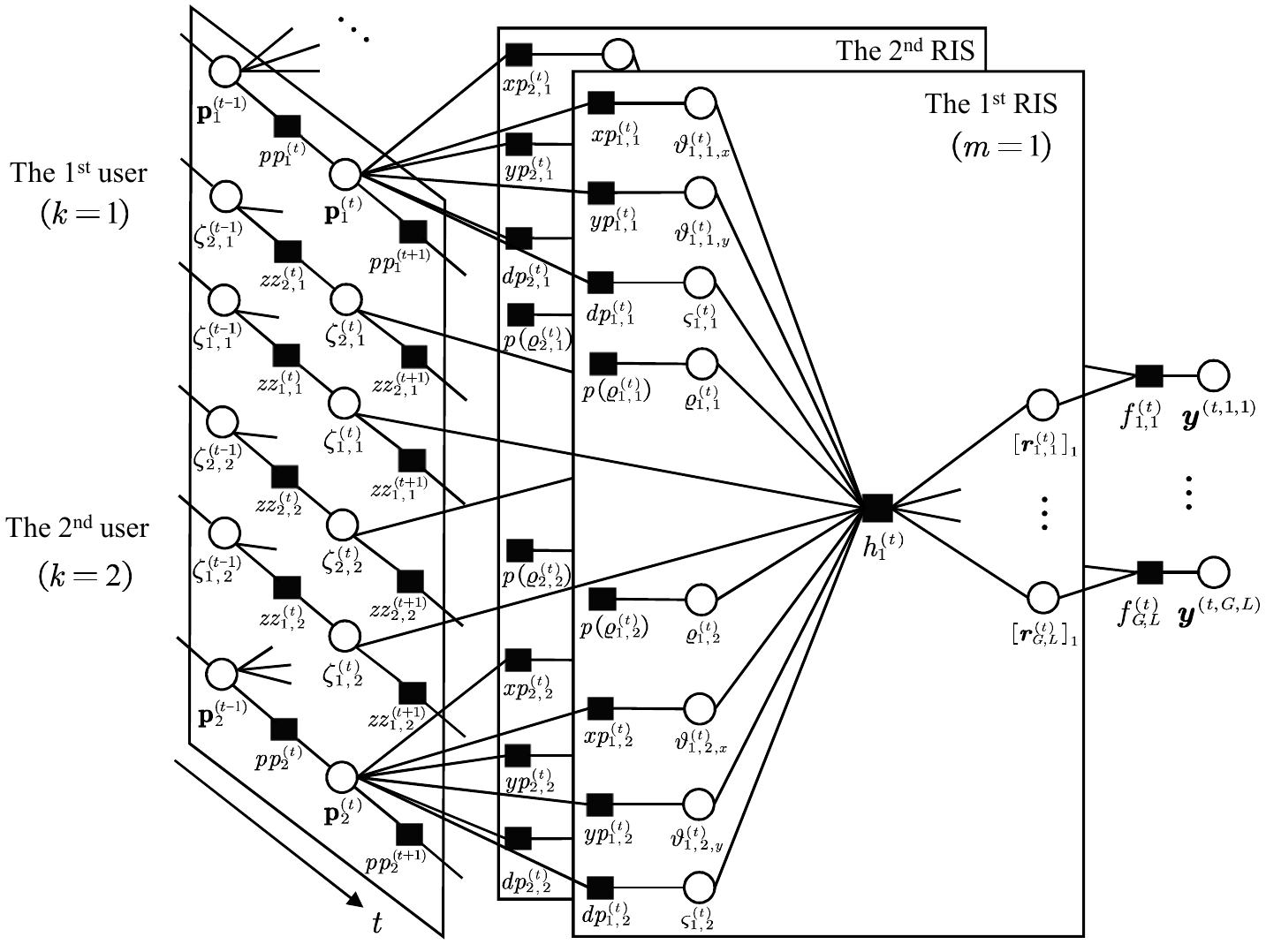}}
    \caption{Factor graph representation for $M=2$ and $K=2$, where we only depict the detail at the $t$-th frame for brevity.}
    \label{factor_graph}
\end{figure}
\section{Bayesian Multiuser Tracking Algorithm}\label{S3}
The factor graph representation of the joint pdf \eqref{jointpdf} is shown in Fig. \ref{factor_graph}. We use blank circles to represent the variable nodes and black rectangles to represent factor nodes. The factor function of each factor node is listed in Table \ref{table1}.  Denote by $\Delta_{a \rightarrow b}(\cdot)$ the message from node $a$ to $b$, and by $\Delta_{a}(\cdot)$ the marginal message at variable node $a$. The mean vector and the covariance matrix of message $\Delta_{a \rightarrow b}(\cdot)$ are represented by $\mathbf{m}_{{a \rightarrow b}}$ and $\mathbf{C}_{{a \rightarrow b}}$ . We perform sum-product message passing on the factor graph to estimate user positions in an online fashion. 
% We note that there are some loops in Fig. \ref{factor_graph}, which increase the computational cost of exact inference. To tackle this, message $\Delta_{[\boldsymbol{r}_{g,l}^{(t)}]_{m}\rightarrow h_{m,g,l}^{(t)}}([\boldsymbol{r}_{g,l}^{(t)}]_{m})$ is approximated calculated based on the linear model in \eqref{10} and messages $\Delta_{\vartheta_{m,k,x}^{(t)} \rightarrow xp_{m,k}^{(t)}}(\vartheta_{m,k,x}^{(t)})$, $\Delta_{\vartheta_{m,k,y}^{(t)} \rightarrow yp_{m,k}^{(t)}}(\vartheta_{m,k,y}^{(t)})$ and $\Delta_{\varsigma_{m,k}^{(t)} \rightarrow rz_{m,k}^{(t)}}(\varsigma_{m,k}^{(t)})$ are approximately calculated by variational inference approach.
\begin{table}[t]
	\centering
	\small
	\caption{Factor nodes in Fig. \ref{factor_graph}.}
	\begin{tabular}{cc}
		\hline
		\textbf{Factor Node}&\textbf{Factor Function}\\
		\hline
		 $pp_{k}^{(t)}$          & $p(\mathbf{p}_{k}^{(t)}|\mathbf{p}_{k}^{(t-1)})=\mathcal{N}(\mathbf{p}_{k}^{(t)};\mathbf{p}_{k}^{(t-1)},\mathbf{C}_{k})$\\
		 $xp_{m,k}^{(t)}$ & $p(\vartheta_{m,k,x}^{(t)}|\mathbf{p}_{k}^{(t)}) = \delta(\vartheta_{m,k,x}^{(t)}-\frac{(\mathbf{p}_{k}^{(t)}-\mathbf{p}_{\mathrm{R},m})^{\mathrm{T}}\mathbf{e}_{m,x}}{\left\|\mathbf{p}_{k}^{(t)}-\mathbf{p}_{\mathrm{R},m}\right\|_{2}}+\phi_{m,x})$\\
		 $yp_{m,k}^{(t)}$ & $p(\vartheta_{m,k,y}^{(t)}|\mathbf{p}_{k}^{(t)}) = \delta(\vartheta_{m,k,y}^{(t)}-\frac{(\mathbf{p}_{k}^{(t)}-\mathbf{p}_{\mathrm{R},m})^{\mathrm{T}}\mathbf{e}_{m,y}}{\left\|\mathbf{p}_{k}^{(t)}-\mathbf{p}_{\mathrm{R},m}\right\|_{2}}+\phi_{m,y})$\\
		 $dp_{m,k}^{(t)}$ & $p(\varsigma_{m,k}^{(t)}|\mathbf{p}_{k}^{(t)}) = \delta(\varsigma_{m,k}^{(t)}-\frac{\left\|\mathbf{p}_{k}^{(t)}-\mathbf{p}_{\mathrm{R},m}\right\|_{2}}{\mathrm{c}_0}-\tau_{m})$\\
		 $zz_{m,k}^{(t)}$   & $   p(\zeta_{m,k}^{(t)} \mid \zeta_{m,k}^{(t-1)})= \begin{cases}\left(1-p_{l,m,k}^{(t)}\right)^{1-\zeta_{m,k}^{(t)}}\left(p_{l,m,k}^{(t)}\right)^{\zeta_{m,k}^{(t)}}, & \zeta_{m,k}^{(t-1)}=0 \\ \left(p_{d,m,k}^{(t)}\right)^{1-\zeta_{m,k}^{(t)}}\left(1-p_{d,m,k}^{(t)}\right)^{\zeta_{m,k}^{(t)}}, & \zeta_{m,k}^{(t-1)}=1\end{cases}$\\
		 $p(\varrho_{m,k}^{(t)})$ & 
   $\mathcal{CN}(\varrho_{m,k}^{(t)};0,\sigma_{m}^{(t)})$\\
		 $h_{m}^{(t)}$  & $\prod_{g=1}^{G}\prod_{l=1}^{L}p([\boldsymbol{r}_{g,l}^{(t)}]_{m}|\boldsymbol{\vartheta}_{m,x}^{(t)},\boldsymbol{\vartheta}_{m,y}^{(t)},\boldsymbol{\varsigma}_{m}^{(t)},\boldsymbol{\zeta}_{m}^{(t)},\boldsymbol{\varrho}_{m}^{(t)})=\prod_{g=1}^{G}\prod_{l=1}^{L}\delta([\boldsymbol{r}_{g,l}^{(t)}]_{m}-[\boldsymbol{R}_{m}^{(t)}]_{g,l})$
%    $\begin{array}{c}\prod_{g=1}^{G}\prod_{l=1}^{L}p([\boldsymbol{r}_{g,l}^{(t)}]_{m}|\boldsymbol{\vartheta}_{m,x}^{(t)},\boldsymbol{\vartheta}_{m,y}^{(t)},\boldsymbol{\varsigma}_{m}^{(t)},\boldsymbol{\zeta}_{m}^{(t)},\boldsymbol{\varrho}_{m}^{(t)})\\=\prod_{g=1}^{G}\prod_{l=1}^{L}\delta([\boldsymbol{r}_{g,l}^{(t)}]_{m}-[\boldsymbol{R}_{m}^{(t)}]_{g,l})\end{array}$
\\
% 		 \sum_{m=1}^{M}{x_{k}^{(t,l)}\varrho_{m,k}^{(t)} [\mathrm{a}_{L}(\varsigma_{m,k}^{(t)})]_{l}(\boldsymbol{\omega}_{m}^{(t,g)})^{\mathrm{T}} \mathbf{a}_{\mathrm{R}}(\vartheta_{m,k,x}^{(t)},\vartheta_{m,k,y}^{(t)})}
		 $f_{g,l}^{(t)}$  & $p(\boldsymbol{y}^{(t,g,l)}|\boldsymbol{r}_{g,l}^{(t)})=\mathcal{CN}(\boldsymbol{y}^{(t,g,l)};\mathbf{A}_{\mathrm{B}}(\boldsymbol{\theta}_k)\boldsymbol{r}_{g,l}^{(t)},\nu\mathbf{I})$ \\
		\hline
	\end{tabular}
	\label{table1}
\end{table}
\subsection{Calculation of Messages $\Delta_{[\boldsymbol{r}_{g,l}^{(t)}]_{m} \rightarrow h_{m}^{(t)}}([\boldsymbol{r}_{g,l}^{(t)}]_{m})$}
The message passing from variable node $[\boldsymbol{r}_{g,l}^{(t)}]_{m}$ to factor node $h_{m}^{(t)}$ is calculated by
\begin{align}\label{19}
\Delta_{[\boldsymbol{r}_{g,l}^{(t)}]_{m} \rightarrow h_{m}^{(t)}}([\boldsymbol{r}_{g,l}^{(t)}]_{m})\propto
\frac{\int_{\backslash[\boldsymbol{r}_{g,l}^{(t)}]_{m}}{\prod_{i=1}^{M}{\Delta_{ h_{i}^{(t)} \rightarrow [\boldsymbol{r}_{g,l}^{(t)}]_{i}}([\boldsymbol{r}_{g,l}^{(t)}]_{i})}p(\boldsymbol{y}^{(t,g,l)}|\boldsymbol{r}_{g,l}^{(t)})}}{\Delta_{ h_{m}^{(t)} \rightarrow [\boldsymbol{r}_{g,l}^{(t)}]_m }([\boldsymbol{r}_{g,l}^{(t)}]_m)},
\end{align}
where $\int_{\backslash[\boldsymbol{r}_{g,l}^{(t)}]_{m}}$ denotes the integral over all involved variables except $[\boldsymbol{r}_{g,l}^{(t)}]_{m}$. 
% We treat the message $\Delta_{ h_{m,g,l}^{(t)} \rightarrow [\boldsymbol{r}_{g,l}^{(t)}]_{m}}([\boldsymbol{r}_{g,l}^{(t)}]_{m})$ as the prior distribution of $[\boldsymbol{r}_{g,l}^{(t)}]_{m}$ and equation \eqref{19} is expressed as 
% \begin{align}
%     \Delta_{[\boldsymbol{r}_{g,l}^{(t)}]_{m} \rightarrow h_{m,g,l}^{(t)}}([\boldsymbol{r}_{g,l}^{(t)}]_{m})\propto\frac{p([\boldsymbol{r}_{g,l}^{(t)}]_{m}|\boldsymbol{y}^{(t,g,l)})}{\Delta_{ h_{m,g,l}^{(t)} \rightarrow [\boldsymbol{r}_{g,l}^{(t)}]_i }([\boldsymbol{r}_{g,l}^{(t)}]_i)}.
% \end{align}
In practice, the BS is generally equipped with large-scale antennas, and the RISs are separately deployed with different $\theta_{m}$. Therefore, the columns of $\mathbf{A}_{\mathrm{B}}(\boldsymbol{\theta})$ are approximately orthogonal. Based on the central limit theorem \cite{donoho2009message}, the message $\Delta_{ h_{m}^{(t)} \rightarrow [\boldsymbol{r}_{g,l}^{(t)}]_{m}}([\boldsymbol{r}_{g,l}^{(t)}]_{m})$ is approximated as a Gaussian message. 
Given the linear model in \eqref{10}, the message $\Delta_{[\boldsymbol{r}_{g,l}^{(t)}]_{m} \rightarrow h_{m}^{(t)}}([\boldsymbol{r}_{g,l}^{(t)}]_{m})$ is approximately calculated by
\begin{align}\label{21}
     \Delta_{[\boldsymbol{r}_{g,l}^{(t)}]_{m} \rightarrow h_{m}^{(t)}}([\boldsymbol{r}_{g,l}^{(t)}]_{m})=\mathcal{CN}\left([\boldsymbol{r}_{g,l}^{(t)}]_{m};\frac{\mathbf{a}_{\mathrm{B}}^{\mathrm{H}}(\theta_{i})\boldsymbol{y}^{(t,g,l)}}{N_{\mathrm{B}}},\frac{\nu}{N_{\mathrm{B}}}\right).
\end{align}
% we obtain the proposition \ref{prop1} to indicate the Gaussian approximation of message .
% \begin{proposition}\label{prop1}
% Suppose $\mathbf{A}_{\mathrm{B}}(\boldsymbol{\theta}_k)$ is a matrix with orthogonal column and $\Delta_{ h_{m,g,l}^{(t)} \rightarrow [\boldsymbol{r}_{g,l}^{(t)}]_{m}}([\boldsymbol{r}_{g,l}^{(t)}]_{m})$ is a Gaussian message, the message $ \Delta_{[\boldsymbol{r}_{g,l}^{(t)}]_{m} \rightarrow h_{m,g,l}^{(t)}}([\boldsymbol{r}_{g,l}^{(t)}]_{m})$ is calculated by
% \begin{align}\label{21}
%      \Delta_{[\boldsymbol{r}_{g,l}^{(t)}]_{m} \rightarrow h_{m,g,l}^{(t)}}([\boldsymbol{r}_{g,l}^{(t)}]_{m})=\mathcal{CN}([\boldsymbol{r}_{g,l}^{(t)}]_{m};\frac{\mathbf{a}_{\mathrm{B}}^{\mathrm{H}}(\theta_{i})\boldsymbol{y}^{(t,g,l)}}{N_{\mathrm{B}}},\frac{\nu}{N_{\mathrm{B}}}).
% \end{align}
% \end{proposition}
% \noindent\textit{Proof.} Please see Appendix \ref{append1}.\par
% We obtain the approximation of message $ \Delta_{[\boldsymbol{r}_{g,l}^{(t)}]_{m} \rightarrow h_{m,g,l}^{(t)}}([\boldsymbol{r}_{g,l}^{(t)}]_{m})$ according to proposition \ref{prop1}.
\subsection{Messages Approximated by Variational Inference}\label{varia}
% \subsubsection{Calculation of Messages $\Delta_{xp_{k,j}^{(t)}\rightarrow \vartheta_{k,j,x}^{(t)} }(\vartheta_{k,j,x}^{(t)})$, $\Delta_{yp_{k,j}^{(t)}\rightarrow \vartheta_{k,j,y}^{(t)} }(\vartheta_{k,j,y}^{(t)})$, and $\Delta_{dp_{k,j}^{(t)}\rightarrow \varsigma_{k,j}^{(t)} }(\varsigma_{k,j}^{(t)})$} For
% where 
For $\forall t$, $1\leq k \leq K$, $1\leq m \leq M$, the messages from variable node $\vartheta_{m,k,x}^{(t)}$ to factor node $xp_{m,k}^{(t)}$ are calculated by
\begin{align}
    \Delta&_{\vartheta_{m,k,x}^{(t)}\rightarrow xp_{m,k}^{(t)}}(\vartheta_{m,k,x}^{(t)})\propto\frac{\Delta_{\vartheta_{m,k,x}^{(t)} }(\vartheta_{m,k,x}^{(t)})}{\Delta_{xp_{m,k}^{(t)}\rightarrow \vartheta_{m,k,x}^{(t)} }(\vartheta_{m,k,x}^{(t)})},
    \label{21_n}
\end{align}
where $\Delta_{xp_{m,k}^{(t)}\rightarrow \vartheta_{m,k,x}^{(t)} }(\vartheta_{m,k,x}^{(t)})$ is calculated by
\begin{align}\label{22nnn}
    \Delta_{xp_{m,k}^{(t)}\rightarrow \vartheta_{m,k,x}^{(t)} }(\vartheta_{m,k,x}^{(t)})\propto\int_{\mathbf{p}_{k}^{(t)}}\Delta_{pp_{k}^{(t)}\rightarrow\mathbf{p}_{k}^{(t)}}(\mathbf{p}_{k}^{(t)})p(\vartheta_{m,k,x}^{(t)}|\mathbf{p}_{k}^{(t)})
\end{align}
The messages passing from factor nodes $xp_{m,k}^{(t)}$, $xp_{m,k}^{(t)}$, and $xp_{m,k}^{(t)}$ to variable node $\mathbf{p}_{k}^{(t)}$ are omitted in \eqref{22nnn}. Given the Gaussian message $\Delta_{pp_{k}^{(t)}\rightarrow\mathbf{p}_{k}^{(t)}}(\mathbf{p}_{k}^{(t)})$ derived in Section \ref{III_C},
$\Delta_{xp_{m,k}^{(t)}\rightarrow \vartheta_{m,k,x}^{(t)} }(\vartheta_{m,k,x}^{(t)})$ is well approximated by a VM distribution $\mathcal{M}\left(\vartheta_{m,k,x}^{(t)};\mu_{xp_{m,k}^{(t)}\rightarrow\vartheta_{m,k,x}^{(t)}} ,\kappa_{xp_{m,k}^{(t)}\rightarrow\vartheta_{m,k,x}^{(t)}}\right)$ where \cite{teng2022bayesian}
\begin{subequations}
\begin{align}
    \mu_{xp_{m,k}^{(t)}\rightarrow\vartheta_{m,k,x}^{(t)}}&=\frac{(\mathbf{m}_{pp_{k}^{(t)}\rightarrow\mathbf{p}_{k}^{(t)}}-\mathbf{p}_{\mathrm{R},m})^{\mathrm{T}}\mathbf{e}_{m,x}}{\big\|\mathbf{m}_{pp_{k}^{(t)}\rightarrow\mathbf{p}_{k}^{(t)}}-\mathbf{p}_{\mathrm{R},m}\big\|_{2}}-\phi_{m,x},\\
    \kappa_{xp_{m,k}^{(t)}\rightarrow\vartheta_{m,k,x}^{(t)}}&=\frac{\big\|\mathbf{m}_{pp_{k}^{(t)}\rightarrow\mathbf{p}_{k}^{(t)}}-\mathbf{p}_{\mathrm{R},m}\big\|_{2}^{6}}{\big(1-\big(\mu_{xp_{m,k}^{(t)}\rightarrow\vartheta_{m,k,x}^{(t)}}+\phi_{m,x}\big)^2\big)\boldsymbol{q}^{(t)}_{m,k}{}^{\mathrm{T}}\mathbf{C}_{pp_{k}^{(t)}\rightarrow\mathbf{p}_{k}^{(t)}}\boldsymbol{q}^{(t)}_{m,k}},
\end{align}
\end{subequations}
with $\boldsymbol{q}^{(t)}_{m,k}=\big(\big(\mathbf{m}_{pp_{k}^{(t)}\rightarrow\mathbf{p}_{k}^{(t)}}-\mathbf{p}_{\mathrm{R},m}\big)\circ\mathbf{e}_{m,x}\big)\circ\big(\mathbf{m}_{pp_{k}^{(t)}\rightarrow\mathbf{p}_{k}^{(t)}}-\mathbf{p}_{\mathrm{R},m}\big)$.
Messages $\Delta_{\vartheta_{m,k,y}^{(t)}\rightarrow yp_{m,k}^{(t)} }(\vartheta_{m,k,y}^{(t)})$ and $\Delta_{\varsigma_{m,k}^{(t)}\rightarrow dp_{m,k}^{(t)}  }(\varsigma_{m,k}^{(t)})$ are calculated similarly by
$\Delta_{\vartheta_{m,k,y}^{(t)}\rightarrow yp_{m,k}^{(t)}}(\vartheta_{m,k,y}^{(t)})\propto\frac{\Delta_{\vartheta_{m,k,y}^{(t)} }(\vartheta_{m,k,y}^{(t)})}{\Delta_{yp_{m,k}^{(t)}\rightarrow \vartheta_{m,k,y}^{(t)} }(\vartheta_{m,k,y}^{(t)})}$ and $\Delta_{\varsigma_{m,k}^{(t)}\rightarrow dp_{m,k}^{(t)}}(\varsigma_{m,k}^{(t)})\propto\frac{\Delta_{\varsigma_{m,k}^{(t)} }(\varsigma_{m,k}^{(t)})}{\Delta_{dp_{m,k}^{(t)}\rightarrow \varsigma_{m,k}^{(t)} }(\varsigma_{m,k}^{(t)})}$ with $\Delta_{yp_{m,k}^{(t)}\rightarrow \vartheta_{m,k,y}^{(t)} }(\vartheta_{m,k,y}^{(t)})$ and $\Delta_{dp_{m,k}^{(t)}\rightarrow \varsigma_{m,k}^{(t)} }(\varsigma_{m,k}^{(t)})$ are respectively approximated by VM distributions as $\mathcal{M}(\vartheta_{m,k,y}^{(t)};\mu_{yp_{m,k}^{(t)}\rightarrow \vartheta_{m,k,y}^{(t)} } ,\kappa_{yp_{m,k}^{(t)}\rightarrow \vartheta_{m,k,y}^{(t)} })$ and $\mathcal{M}(\varsigma_{m,k}^{(t)};\mu_{dp_{m,k}^{(t)}\rightarrow \varsigma_{m,k}^{(t)} } ,\kappa_{dp_{m,k}^{(t)}\rightarrow \varsigma_{m,k}^{(t)} })$. Due to the complex form of factor nodes $\{h_{m}^{(t)}\}$, it is difficult to perform the sum-product rule on these nodes. Inspired by the variational inference of line spectra \cite{badiu2017variational,zhu2019grid}, in the following, we adopt the variational inference approach to obtain the approximation of ${\Delta}_{\vartheta_{m,k,x}^{(t)}}(\vartheta_{m,k,x}^{(t)})$, ${\Delta}_{\vartheta_{m,k,y}^{(t)}}(\vartheta_{m,k,y}^{(t)})$, ${\Delta}_{\varsigma_{m,k}^{(t)}}(\varsigma_{m,k}^{(t)})$, ${\Delta}_{\varrho_{m,k}^{(t)}}(\varrho_{m,k}^{(t)})$, and ${\Delta}_{\zeta_{m,k}^{(t)}}(\zeta_{m,k}^{(t)})$. Note that the algorithms in \cite{badiu2017variational,zhu2019grid} are not applicable due to the existence of $\boldsymbol{W}_{m}^{(t)}$ in \eqref{9}. Our derivation extends the variational inference of line spectra to a more general form. The variational inference is parallelly performed over each RIS for each frame. For notation brevity, we omit indexes $t$ and $m$ in the derivation.\par
% Lines 7-10 in Algorithm \ref{tracking_algorithm} are inspired by the variational inference of line spectra \cite{badiu2017variational,zhu2019grid}. Since the PBF of RISs made up the compression matrix $\boldsymbol{W}_{m}^{(t)}$ as shown in \eqref{9}, the existing algorithms are not applicable. The derivation in lines 7-10 extends the variational inference of line spectra to a more general form. Therefore, the channel parameters $\theta_{m,k,x}^{(t)}$, $\theta_{m,k,y}^{(t)}$, and $\tau_{m,k}^{(t)}$ are well estimated from $\boldsymbol{R}_{m}^{(t)}$ to help with multiuser tracking.\par
Defining 
% $\Delta(\boldsymbol{\vartheta}_{x},\boldsymbol{\vartheta}_{y},\boldsymbol{\varrho},\boldsymbol{\zeta})=\prod_{k=1}^{K}{\Delta}_{\vartheta_{k,x}}(\vartheta_{k,x}) {\Delta}_{\vartheta_{k,y}}(\vartheta_{k,y}){\Delta}_{\varsigma_{k}}(\varsigma_{k}){\Delta}_{\varrho_{k}}(\varrho_{k}){\Delta}_{\zeta_{k}}(\zeta_{k})$,
\begin{align}\label{22nn}
    \Delta(\boldsymbol{\vartheta}_{x},\boldsymbol{\vartheta}_{y},\boldsymbol{\varsigma},\boldsymbol{\zeta},\boldsymbol{\varrho})=& \prod_{k=1}^{K}{\Delta_{xp_{k}\rightarrow \vartheta_{k,x} }(\vartheta_{k,x})\Delta_{yp_{k}\rightarrow \vartheta_{k,y} }(\vartheta_{k,y})\Delta_{dp_{k}\rightarrow \varsigma_{k} }(\varsigma_{k})\Delta_{zz_{k}\rightarrow \zeta_{k} }(\zeta_{k})}p(\varrho_{k})
    \notag\\
&\times\int_{\boldsymbol{R}}\prod_{g=1}^{G}\prod_{l=1}^{L} p({r}_{g,l}|\boldsymbol{\vartheta}_{x},\boldsymbol{\vartheta}_{y},\boldsymbol{\varsigma}_{k},\boldsymbol{\zeta}_{k},\boldsymbol{\varrho})\Delta_{{r}_{g,l} \rightarrow h}({r}_{g,l}),
\end{align}
the basic idea of variational inference is to find the surrogate pdf $q(\boldsymbol{\vartheta}_{x},\boldsymbol{\vartheta}_{y},\boldsymbol{\varsigma},\boldsymbol{\zeta},\boldsymbol{\varrho})$ that minimizes the Kullback-Leibler (KL) divergence between the surrogate pdf and the objective pdf $\Delta(\boldsymbol{\vartheta}_{x},\boldsymbol{\vartheta}_{y},\boldsymbol{\varsigma},\boldsymbol{\zeta},\boldsymbol{\varrho})$, which is equivalent to maximizing the evidence lower bound \cite{bishop2006pattern}
\begin{align}\label{26}
    \mathcal{L}(q(\boldsymbol{\vartheta}_{x},\boldsymbol{\vartheta}_{y},\boldsymbol{\varsigma},\boldsymbol{\zeta},\boldsymbol{\varrho}))=\mathbb{E}_{q(\boldsymbol{\vartheta}_{x},\boldsymbol{\vartheta}_{y},\boldsymbol{\varsigma},\boldsymbol{\zeta},\boldsymbol{\varrho})}\Big[\ln{\frac{\Delta(\boldsymbol{\vartheta}_{x},\boldsymbol{\vartheta}_{y},\boldsymbol{\varsigma},\boldsymbol{\zeta},\boldsymbol{\varrho})}{q(\boldsymbol{\vartheta}_{x},\boldsymbol{\vartheta}_{y},\boldsymbol{\varsigma},\boldsymbol{\zeta},\boldsymbol{\varrho})}}\Big].
\end{align}
% In \eqref{10}, the matrix $\boldsymbol{R}_{m}^{(t)}$ is a superposition of $M$ 3-dimension line spectra. The variational inference based line spectra estimation is first proposed in [] and is extended to multiple measurements in [] and multidimensional line spectra in [].
Assume that the surrogate pdf admits a factorization as
\begin{align}\label{27}
q(\boldsymbol{\vartheta}_{x},\boldsymbol{\vartheta}_{y},\boldsymbol{\varsigma},\boldsymbol{\zeta},\boldsymbol{\varrho})=q(\boldsymbol{\zeta})q(\boldsymbol{\varrho})\prod_{k=1}^{K}q({\vartheta}_{k,x})q({\vartheta}_{k,y})q({\varsigma}_{k}),
\end{align}
where $q(\boldsymbol{\zeta})$ is restricted to satisfy $q(\boldsymbol{\zeta})=\delta(\boldsymbol{\zeta}-\hat{\boldsymbol{\zeta}})$. Given the factor $q(\vartheta_{k,x})$, $q(\vartheta_{k,y})$ and $q(\varsigma_{k})$, we obtain the estimations of random variables $\hat{\vartheta}_{k,x}=\arg(\mathbb{E}_{q(\vartheta_{k,x})}[\mathrm{e}^{\jmath\vartheta_{k,x}}])$, $\hat{\vartheta}_{k,y}=\arg(\mathbb{E}_{q(\vartheta_{k,y})}[\mathrm{e}^{\jmath\vartheta_{k,y}}])$ and $\hat{\varsigma}_{k}=\arg(\mathbb{E}_{q(\varsigma_{k})}[\mathrm{e}^{\jmath\varsigma_{k}}])$, and the estimations of steering vectors $\hat{\mathbf{a}}_{\mathrm{x}}(\vartheta_{k,x})=\mathbb{E}_{q(\vartheta_{k,x})}[\mathbf{a}_{\mathrm{x}}(\vartheta_{k,x})]$, $\hat{\mathbf{a}}_{\mathrm{y}}(\vartheta_{k,y})=\mathbb{E}_{q(\vartheta_{k,y})}[\mathbf{a}_{\mathrm{y}}(\vartheta_{k,y})]$, and $\hat{\mathbf{a}}_{L}(\varsigma_{k})=\mathbb{E}_{q(\varsigma_{k})}[\mathbf{a}_{L}(\varsigma_{k})]$. Given the factor $q(\boldsymbol{\varrho})$, we obtain $\hat{\boldsymbol{\varrho}}=\mathbb{E}_{q(\boldsymbol{\varrho})}[\boldsymbol{\varrho}]$ and $\hat{\mathbf{C}}_{\boldsymbol{\varrho}}=\mathbb{E}_{q(\boldsymbol{\varrho})}[\boldsymbol{\varrho}\boldsymbol{\varrho}^{\mathrm{H}}]-\hat{\boldsymbol{\varrho}}\hat{\boldsymbol{\varrho}}^{\mathrm{H}}$. Given the factor $q(\boldsymbol{\zeta})$, the support set of the non-zero components in $\boldsymbol{\zeta}$ is estimated by $\hat{\mathcal{S}}=\{j|1\leq j \leq K,\hat{\zeta}_{j}=1\}$. As maximizing $\mathcal{L}$ over all the factors simultaneously is intractable, we adopt alternating optimization and maximize $\mathcal{L}$ over each factor independently with the other factors keeping invariant. The factors are iteratively updated until converge.
% The optimal form of $q(z)$ maximizing $\mathcal{L}$ is calculated by \cite{bishop2006pattern}
% \begin{align}\label{28}
%     \ln{q(z)}=\mathbb{E}_{\backslash z}[\ln{\Delta(\boldsymbol{\vartheta}_{x},\boldsymbol{\vartheta}_{y},\boldsymbol{\varsigma},\boldsymbol{\zeta},\boldsymbol{\varrho})}]+\text{const},\ z \in
% \end{align}
% where $\mathbb{E}_{\backslash z}[\cdot]$ denote the expectation over $\frac{q(\boldsymbol{\vartheta}_{x},\boldsymbol{\vartheta}_{y},\boldsymbol{\varsigma},\boldsymbol{\zeta},\boldsymbol{\varrho})}{q(z)}$.
We present the details of the inference in the following.\par
\subsubsection{Calculate $q({\vartheta}_{k,x})$ and $q({\vartheta}_{k,y})$} As derived in Appendix \ref{append2},
for $k\in\hat{\mathcal{S}}$, maximizing $\mathcal{L}$ w.r.t.  $q({\vartheta}_{k,x})$ gives
\begin{align}\label{29}
    \ln q(\vartheta_{k,x})\!=\! \ln\! \Delta_{xp_{k}\rightarrow \vartheta_{k,x} }(\vartheta_{k,x})\!+\!\Re\{\boldsymbol{\beta}_{k}^{\mathrm{H}}\mathbf{a}_{\mathrm{x}}(\vartheta_{k,x})\!\otimes\!\hat{\mathbf{a}}_{\mathrm{y}}(\vartheta_{k,y})\}\!-\!\chi_{k}\mathbb{E}_{q(\vartheta_{k,y})}\!\left[\big\|\boldsymbol{W}^{\mathrm{T}}\mathbf{a}_{\mathrm{R}}(\vartheta_{k,x},\vartheta_{k,y})\big\|_{2}^{2}\right].
\end{align}
The calculation of $\mathbb{E}_{q(\vartheta_{k,y})}\left[\big\|\boldsymbol{W}^{\mathrm{T}}\mathbf{a}_{\mathrm{R}}(\vartheta_{k,x},\vartheta_{k,y})\big\|_{2}^{2}\right]$ can be found in Appendix \ref{append3}; $\boldsymbol{\beta}_{k}$ and $\chi_{k}$ are given by
\begin{subequations}
\begin{align}
    \boldsymbol{\beta}_{k} \!&= \frac{2N_{\mathrm{B}}}{\nu}\boldsymbol{W}^{*}\Big(\hat{\varrho}_{k}^{*}\boldsymbol{R}-\!\boldsymbol{W}^{\mathrm{T}}\!\sum_{j\in\hat{\mathcal{S}}\backslash k}{([\hat{\mathbf{C}}_{\boldsymbol{\varrho}}]_{j,k}+\hat{\varrho}_{j}\hat{\varrho}_{k}^{*})\hat{\mathbf{a}}_{\mathrm{R}}(\vartheta_{j,x},\vartheta_{j,y})\boldsymbol{x}_{j}^{\mathrm{T}}\odot\hat{\mathbf{a}}_{L}^{\mathrm{T}}({\varsigma_{j}})}\Big)\boldsymbol{x}_{k}^{*}\odot\hat{\mathbf{a}}_{L}^{*}(\varsigma_{k})\\
    \chi_{k}&=\frac{N_{\mathrm{B}}}{\nu}([\hat{\mathbf{C}}_{\boldsymbol{\varrho}}]_{k,k}+\hat{\varrho}_{k}\hat{\varrho}_{k}^{*})\big\|\boldsymbol{x}_{k}\odot\hat{\mathbf{a}}_{L}(\varsigma_{k})\big\|_{2}^{2}.
\end{align}
\end{subequations}
For $k\notin\hat{\mathcal{S}}$, we have $\ln q(\vartheta_{k,x})= \ln \Delta_{xp_{k}\rightarrow \vartheta_{k,x} }(\vartheta_{k,x})$. The posterior pdf $q(\vartheta_{k,x})$ in \eqref{29} is well approximated by a VM distribution $\mathcal{M}(\vartheta_{k,x};\mu_{\vartheta_{k,x}},\kappa_{\vartheta_{k,x}})$,
% as
% \begin{align}
%     \mathcal{M}(\vartheta_{k,x};\hat{\vartheta}_{k,x},\hat{\kappa}_{k,x})=\frac{1}{2\pi I_{0}(\hat{\kappa}_{k,x})}\exp{(\hat{\kappa}_{k,x}\cos{(\vartheta_{k,x}-\hat{\vartheta}_{k,x})})},
% \end{align}
% where $I_{0}(\cdot)$ is the modified Bessel function of the first kind in order 0, $\hat{\vartheta}_{k,x}$ and $\hat{\kappa}_{k,x}$ are the mean direction and concentration parameters respectively. 
where $\mu_{\vartheta_{k,x}}$ and $\kappa_{\vartheta_{k,x}}$ are set to be the maximum of $\ln q(\vartheta_{k,x})$ and $A^{-1}(\exp(\frac{1}{2(\ln q(\mu_{\vartheta_{k,x}})^{''}}))$ respectively. $A^{-1}(\cdot)$ is the inverse of function $A(\cdot)=I_{1}(\cdot)/ I_{0}(\cdot)$ given in \cite{mardia2009directional}, where $I_{i}(\cdot)$ is the modified Bessel function of the first kind in order $i$.
We resort to the \textit{gradient descent method} (GDM) to find the local maximum of $\ln q(\vartheta_{k,x})$. The gradient descent search starts from the the mean of $\Delta_{xp_{k}\rightarrow \vartheta_{k,x} }(\vartheta_{k,x})$. Similarly, $q({\vartheta}_{k,y})$ can be approximated by $\mathcal{M}(\vartheta_{k,y};\mu_{\vartheta_{k,y}},\kappa_{\vartheta_{k,y}})$.
% \begin{align}
%     \mathcal{M}(\vartheta_{k,y};\hat{\vartheta}_{m,y},\hat{\kappa}_{m,y})=\frac{1}{2\pi I_{0}(\hat{\kappa}_{m,y})}\exp{(\hat{\kappa}_{m,y}\cos{(\vartheta_{k,y}-\hat{\vartheta}_{m,y})})}.
% \end{align}
% where $\hat{\vartheta}_{m,y}$ and $\hat{\kappa}_{m,y}$ are calculated from
% \begin{align}
%     \ln q(\vartheta_{k,y}|\boldsymbol{R})= \ln p(\vartheta_{k,y})+\Re\{\boldsymbol{\eta}_{m}^{\mathrm{H}}\hat{\mathbf{a}}_{\mathrm{x}}(\vartheta_{k,x})\otimes\mathbf{a}_{\mathrm{y}}(\vartheta_{k,y})\}-\gamma_{m}\big\|\boldsymbol{W}^{\mathrm{T}}\hat{\mathbf{a}}_{\mathrm{x}}(\vartheta_{k,x})\otimes\mathbf{a}_{\mathrm{y}}(\vartheta_{k,y})\big\|_{2}^{2}.
% \end{align}
\subsubsection{Calculate $q(\varsigma_{k})$} Similar to the calculation of $q(\vartheta_{k,x})$ and $q(\vartheta_{k,y})$, for $k\in\hat{\mathcal{S}}$, maximizing $\mathcal{L}$ w.r.t.  $q(\varsigma_{k})$ gives
\begin{align}
    \ln q(\varsigma_{k})= \ln \Delta_{dp_{k}\rightarrow \varsigma_{k} }(\varsigma_{k})+\Re\{\boldsymbol{\beta}_{\varsigma, k}^{\mathrm{H}}\mathbf{a}_{L}(\varsigma_{k})\}+\text{const},
\end{align}
where $\boldsymbol{\beta}_{\varsigma,k}$ is given by
\begin{align}
    \boldsymbol{\beta}_{\varsigma,k} \!=\! \frac{2N_{\mathrm{B}}}{\nu}\mathrm{diag}(\boldsymbol{x}_{k}^{*})\Big(\!\hat{\varrho}_{k}^{*}\boldsymbol{R}^{\mathrm{T}}\!-\!\sum_{j\in\hat{\mathcal{S}}\backslash k}\!([\hat{\mathbf{C}}_{\boldsymbol{\varrho}}]_{j,k}\!+\!\hat{\varrho}_{j}\hat{\varrho}_{k}^{*})\boldsymbol{x}_j\!\odot\!\mathbf{a}_{L}(\varsigma_{j})\mathbf{a}_{\mathrm{R}}^{\mathrm{T}}(\vartheta_{j,x},\vartheta_{j,y})\boldsymbol{W}\!\Big)\boldsymbol{W}^{\mathrm{H}}\hat{\mathbf{a}}_{\mathrm{R}}^{*}(\vartheta_{k,x},\vartheta_{k,y}).
\end{align}
For $k\notin\hat{\mathcal{S}}$, we have $\ln q(\varsigma_{k})= \ln \Delta_{dp_{k}\rightarrow \varsigma_{k} }(\varsigma_{k})$. The posterior pdf $q(\varsigma_{k})$ can be approximated by a VM distribution $\mathcal{M}(\varsigma_{k};
\mu_{\varsigma_{k}},\kappa_{\varsigma_{k}})$ using Algorithm 2 from \cite{badiu2017variational}.
\subsubsection{Calculate $q(\boldsymbol{\varrho})$} Maximizing $\mathcal{L}$ w.r.t.  $q(\boldsymbol{\varrho})$  gives
\begin{align}\label{35}
    q(\boldsymbol{\varrho})=\mathcal{CN}([\boldsymbol{\varrho}]_{\hat{\mathcal{S}}};[\hat{\boldsymbol{\varrho}}]_{\hat{\mathcal{S}}},[\hat{\mathbf{C}}_{\boldsymbol{\varrho}}]_{\hat{\mathcal{S}},\hat{\mathcal{S}}})\prod_{l\notin\hat{\mathcal{S}}}\delta(\varrho_{l}),
\end{align}
where $\hat{\boldsymbol{\varrho}}=\frac{N_{\mathrm{B}}}{\nu}\hat{\mathbf{C}}_{\boldsymbol{\varrho}}\boldsymbol{h}$ and $\hat{\mathbf{C}}_{\boldsymbol{\varrho}}=\Big(\frac{N_{\mathrm{B}}}{\nu}\boldsymbol{J}+\frac{1}{\sigma}\mathbf{I}\Big)^{-1}$ with the entries of
$\boldsymbol{h}$ and $\boldsymbol{J}$ given by
\begin{subequations}\label{36}
\begin{align}
[\boldsymbol{h}]_{j}&=\mathrm{Tr}\big(\boldsymbol{x}_{j}^{*}\odot\hat{\mathbf{a}}_{L}^{*}(\varsigma_{j})\hat{\mathbf{a}}_{\mathrm{R}}^{\mathrm{H}}(\vartheta_{j,x},\vartheta_{j,y})\boldsymbol{W}^{*}\boldsymbol{R}\big),\\
[\boldsymbol{J}]_{i,j}&=\begin{cases} \big\|\boldsymbol{x}_{i}\odot\hat{\mathbf{a}}_{L}(\varsigma_{i})\big\|_{2}^{2}\mathbb{E}_{q(\vartheta_{i,x})q(\vartheta_{i,y})}\big[\left\|\boldsymbol{W}^{\mathrm{T}}\mathbf{a}_{\mathrm{R}}(\vartheta_{i,x},\vartheta_{i,y})\right\|_{2}^{2}\big],&i=j;\\
\boldsymbol{x}_{j}^{\mathrm{T}}\odot\hat{\mathbf{a}}_{L}^{\mathrm{T}}(\varsigma_{j})\boldsymbol{x}_{i}^{*}\odot\hat{\mathbf{a}}_{L}^{*}(\varsigma_{i})\hat{\mathbf{a}}_{\mathrm{R}}^{\mathrm{H}}(\vartheta_{i,x},\vartheta_{i,y})\boldsymbol{W}^{*}\boldsymbol{W}^{\mathrm{T}}\hat{\mathbf{a}}_{\mathrm{R}}(\vartheta_{j,x},\vartheta_{j,y}),&i\neq j.\end{cases}
\end{align}
\end{subequations}
The calculation of $\mathbb{E}_{q(\vartheta_{i,x})q(\vartheta_{i,y})}\big[\left\|\boldsymbol{W}^{\mathrm{T}}\mathbf{a}_{\mathrm{R}}(\vartheta_{i,x},\vartheta_{i,y})\right\|_{2}^{2}\big]$ can be found in Appendix \ref{append3}. \par
\subsubsection{Calculate $q(\boldsymbol{\zeta})$}
We calculate $q(\boldsymbol{\zeta})$ by plugging \eqref{27} into \eqref{26} and maximizing $\mathcal{L}$ w.r.t. $q(\boldsymbol{\zeta})$. Since $q(\boldsymbol{\zeta})=\delta(\boldsymbol{\zeta}-\hat{\boldsymbol{\zeta}})$,
% Since all the factors are fixed except $q(\boldsymbol{\zeta})$, 
we obtain
\begin{align}\label{37}
    \mathcal{L}(\hat{\boldsymbol{\zeta}})=&\ln{\det{\Big(\boldsymbol{J}_{\hat{\mathcal{S}},\hat{\mathcal{S}}}+\frac{\nu}{N_{\mathrm{B}}\sigma}\mathbf{I}\Big)^{-1}}}
    +\frac{N_{\mathrm{B}}}{\nu}\boldsymbol{h}_{\hat{\mathcal{S}}}^{\mathrm{H}}\Big(\boldsymbol{J}_{\hat{\mathcal{S}},\hat{\mathcal{S}}}+\frac{\nu}{N_{\mathrm{B}}\sigma}\mathbf{I}\Big)^{-1}\boldsymbol{h}_{\hat{\mathcal{S}}}\notag\\
    &+\|\boldsymbol{\zeta}\|_{0}\ln\frac{\nu}{N_{\mathrm{B}}\sigma}
    + \prod_{k=1}^{K}\Delta_{zz_{k}\rightarrow \zeta_{k} }(\zeta_{k})
    +\text{const}.
\end{align}
We adopt the greedy iterative search provided in \cite{badiu2017variational} to find a local maximum of $\mathcal{L}(\hat{\boldsymbol{\zeta}})$. In the $i$-th iteration, $K$ test variables $\{\hat{\boldsymbol{\zeta}}_{k}^{i}\}_{k=1}^{K}$ are generated by flip the $k$-th element of $\hat{\boldsymbol{\zeta}}^{i-1}$ for $k=1,...,K$. $\hat{\boldsymbol{\zeta}}^{i}$ is obtained by calculating $\mathcal{L}(\hat{\boldsymbol{\zeta}}_{k}^{i})-\mathcal{L}(\hat{\boldsymbol{\zeta}}^{i-1})$ for each $\hat{\boldsymbol{\zeta}}_{k}^{i}$ and find the largest non-negative term.\par
% $\hat{\boldsymbol{\zeta}}^{i}={\arg\max}_{\boldsymbol{\zeta}\in\{\hat{\boldsymbol{\zeta}}_{k}^{i}\}_{k=1}^{K}} \mathcal{L}(\boldsymbol{\zeta})-\mathcal{L}(\hat{\boldsymbol{\zeta}}^{i-1})$.
% The posterior estimations of $\boldsymbol{\zeta}_{m}^{(t)}$ is derived from the nonzero term in $q(\boldsymbol{\zeta})$ denoted by $\hat{\boldsymbol{\zeta}}_{m}^{(t)}$.\par
% The posterior estimations of $\boldsymbol{\varrho}_{m}^{(t)}$ is given by $\hat{\boldsymbol{\varrho}}_{m}^{(t)}$.
For $\forall t$ and $1\leq m\leq M$, the factors $q({\vartheta}_{m,k,x}^{(t)})$, $q({\vartheta}_{m,k,y}^{(t)})$, and $q(\varsigma_{m,k}^{(t)})$ are first initialized by the message $\Delta_{xp_{m,k}^{(t)}\rightarrow \vartheta_{m,k,x}^{(t)} }(\vartheta_{m,k,x}^{(t)})$, $\Delta_{yp_{m,k}^{(t)}\rightarrow \vartheta_{m,k,y}^{(t)} }(\vartheta_{m,k,y}^{(t)})$, and
$\Delta_{dp_{m,k}^{(t)}\rightarrow \varsigma_{m,k}^{(t)} }(\varsigma_{m,k}^{(t)})$. Factor $q(\boldsymbol{\varrho}_{m}^{(t)})$ and $q(\boldsymbol{\zeta}_{m}^{(t)})$ are then initialized by \eqref{35} and maximizing $\mathcal{L}(\hat{\boldsymbol{\zeta}}_{m}^{(t)})$ respectively. Since the family of VM distributions is closed under multiplication up to scaling, we calculate \eqref{21_n} and obtain the approximated message $\Delta_{\vartheta_{m,k,x}^{(t)}\rightarrow xp_{m,k}^{(t)}}(\vartheta_{m,k,x}^{(t)})$ as
\begin{subequations}\label{366}
\begin{align}
    \Delta_{\vartheta_{m,k,x}^{(t)}\rightarrow xp_{m,k}^{(t)}}(\vartheta_{m,k,x}^{(t)})&=\mathcal{M}(\vartheta_{m,k,x}^{(t)};\mu_{\vartheta_{m,k,x}^{(t)}\rightarrow xp_{m,k}^{(t)}},\kappa_{\vartheta_{m,k,x}^{(t)}\rightarrow xp_{m,k}^{(t)}}),
\end{align}
\end{subequations}
where $\mu_{\vartheta_{m,k,x}^{(t)}\rightarrow xp_{m,k}^{(t)}}$ and $\kappa_{\vartheta_{m,k,x}^{(t)}\rightarrow xp_{m,k}^{(t)}}$ satisfy
\begin{align}
    \kappa_{\vartheta_{m,k,x}^{(t)}\rightarrow xp_{m,k}^{(t)}}\!\exp\big(\jmath\mu_{\vartheta_{m,k,x}^{(t)}\rightarrow xp_{m,k}^{(t)}}\big)=\kappa_{\vartheta_{m,k,x}^{(t)}}\!\exp\big(\jmath\mu_{\vartheta_{m,k,x}^{(t)}}\big)-\kappa_{xp_{m,k}^{(t)}\rightarrow\vartheta_{m,k,x}^{(t)} }\!\exp\big(\jmath\mu_{xp_{m,k}^{(t)}\rightarrow\vartheta_{m,k,x}^{(t)}}\big).\notag
\end{align}
Messages $\Delta_{\vartheta_{m,k,y}^{(t)}\rightarrow yp_{m,k}^{(t)} }(\vartheta_{m,k,y}^{(t)})$ and $\Delta_{\varsigma_{m,k}^{(t)}\rightarrow dp_{m,k}^{(t)}  }(\varsigma_{m,k}^{(t)})$ are calculated similarly.
% $\Delta_{\vartheta_{m,k,y}^{(t)}\rightarrow yp_{m,k}^{(t)} }(\vartheta_{m,k,y}^{(t)})$ and $\Delta_{\varsigma_{m,k}^{(t)}\rightarrow dp_{m,k}^{(t)}  }(\varsigma_{m,k}^{(t)})$ expressed by
% \begin{subequations}\label{366}
% \begin{align}
%     \Delta_{\vartheta_{m,k,x}^{(t)}\rightarrow xp_{m,k}^{(t)}}(\vartheta_{m,k,x}^{(t)})&=\mathcal{M}(\vartheta_{m,k,x}^{(t)};\mu_{\vartheta_{m,k,x}^{(t)}\rightarrow xp_{m,k}^{(t)}},\kappa_{\vartheta_{m,k,x}^{(t)}\rightarrow xp_{m,k}^{(t)}}),\\
%     \Delta_{\vartheta_{m,k,y}^{(t)}\rightarrow yp_{m,k}^{(t)} }(\vartheta_{m,k,y}^{(t)})&=\mathcal{M}(\vartheta_{m,k,y}^{(t)};\mu_{\vartheta_{m,k,y}^{(t)}\rightarrow yp_{m,k}^{(t)}},\kappa_{\vartheta_{m,k,y}^{(t)}\rightarrow yp_{m,k}^{(t)} }),\\
%     \Delta_{\varsigma_{m,k}^{(t)}\rightarrow dp_{m,k}^{(t)}}(\varsigma_{m,k}^{(t)})&=\mathcal{M}(\varsigma_{m,k}^{(t)};\mu_{\varsigma_{m,k}^{(t)}\rightarrow dp_{m,k}^{(t)}},\kappa_{\varsigma_{m,k}^{(t)}\rightarrow dp_{m,k}^{(t)}}).
% \end{align}
% \end{subequations}
\subsection{Calculation of Messages Passing Between Frames}\label{III_C}
\subsubsection{Messages from $\mathbf{p}_{k}^{(t)}$ to $pp_{k}^{(t+1)}$} For $1\leq k\leq K$,
\begin{align}
    \Delta_{\mathbf{p}_{k}^{(t)}\rightarrow pp_{k}^{(t+1)}}(\mathbf{p}_{k}^{(t)})\propto&\Delta_{pp_{k}^{(t)}\rightarrow\mathbf{p}_{k}^{(t)}}(\mathbf{p}_{k}^{(t)})\prod_{m=1}^{M}
    \Delta_{xp_{m,k}^{(t)}\rightarrow\mathbf{p}_{k}^{(t)}}(\mathbf{p}_{k}^{(t)})
    \Delta_{yp_{m,k}^{(t)}\rightarrow\mathbf{p}_{k}^{(t)}}(\mathbf{p}_{k}^{(t)})
    \Delta_{dp_{m,k}^{(t)}\rightarrow\mathbf{p}_{k}^{(t)}}(\mathbf{p}_{k}^{(t)})
\end{align}
where 
\begin{subequations}\label{38}
\begin{align}
    \Delta_{xp_{m,k}^{(t)}\rightarrow\mathbf{p}_{k}^{(t)}}(\mathbf{p}_{k}^{(t)})=&\int_{\vartheta_{m,k,x}^{(t)}}\Delta_{\vartheta_{m,k,x}^{(t)}\rightarrow xp_{m,k}^{(t)}}(\vartheta_{m,k,x}^{(t)})p(\vartheta_{m,k,x}^{(t)}|\mathbf{p}_{k}^{(t)}),\\
    \Delta_{yp_{m,k}^{(t)}\rightarrow\mathbf{p}_{k}^{(t)}}(\mathbf{p}_{k}^{(t)})=&\int_{\vartheta_{m,k,y}^{(t)}}\Delta_{\vartheta_{m,k,y}^{(t)}\rightarrow yp_{m,k}^{(t)}}(\vartheta_{m,k,y}^{(t)})p(\vartheta_{m,k,y}^{(t)}|\mathbf{p}_{k}^{(t)}),\\
    \Delta_{dp_{m,k}^{(t)}\rightarrow\mathbf{p}_{k}^{(t)}}(\mathbf{p}_{k}^{(t)})=&\int_{\varsigma_{m,k}^{(t)}}\Delta_{\varsigma_{m,k}^{(t)}\rightarrow dp_{m,k}^{(t)}}(\varsigma_{m,k}^{(t)})p(\varsigma_{m,k}^{(t)}|\mathbf{p}_{k}^{(t)}).
\end{align}
\end{subequations}
% $\Delta_{\vartheta_{m,k,x}^{(t)}\rightarrow xp_{m,k}^{(t)}}(\vartheta_{m,k,x}^{(t)})p(\vartheta_{m,k,x}^{(t)}|\mathbf{p}_{k}^{(t)})$
% $\Delta_{\vartheta_{m,k,y}^{(t)}\rightarrow yp_{m,k}^{(t)}}(\vartheta_{m,k,y}^{(t)})p(\vartheta_{m,k,y}^{(t)}|\mathbf{p}_{k}^{(t)})$
% $\Delta_{\varsigma_{m,k}^{(t)}\rightarrow dp_{m,k}^{(t)}}(\varsigma_{m,k}^{(t)})p(\varsigma_{m,k}^{(t)}|\mathbf{p}_{k}^{(t)})$
As derived in Appendix \ref{append4}, the product of messages $\Delta_{xp_{m,k}^{(t)}\rightarrow\mathbf{p}_{k}^{(t)}}(\mathbf{p}_{k}^{(t)})$, $\Delta_{yp_{m,k}^{(t)}\rightarrow\mathbf{p}_{k}^{(t)}}(\mathbf{p}_{k}^{(t)})$, and $\Delta_{dp_{m,k}^{(t)}\rightarrow\mathbf{p}_{k}^{(t)}}(\mathbf{p}_{k}^{(t)})$ can be approximated by a Gaussian distribution, i.e.,
\begin{align}\label{40}
    \Delta_{xp_{m,k}^{(t)}\rightarrow\mathbf{p}_{k}^{(t)}}(\mathbf{p}_{k}^{(t)})
    \Delta_{yp_{m,k}^{(t)}\rightarrow\mathbf{p}_{k}^{(t)}}(\mathbf{p}_{k}^{(t)})
    \Delta_{dp_{m,k}^{(t)}\rightarrow\mathbf{p}_{k}^{(t)}}(\mathbf{p}_{k}^{(t)})=\mathcal{N}(\mathbf{p}_{k}^{(t)};\mathbf{m}_{m,k}^{(t)},\mathbf{C}_{m,k}^{(t)}).
\end{align}
$\mathbf{m}_{m,k}^{(t)}$ and $\mathbf{C}_{m,k}^{(t)}$ in \eqref{40} is given by
\begin{subequations}
\begin{align}
    \mathbf{m}_{m,k}^{(t)} &= \mathbf{p}_{\mathrm{R},m}+d_{m,k}^{(t)}\mathbf{E}_{m}\boldsymbol{v}_{m,k}^{(t)},\label{40a}\\
    \mathbf{C}_{m,k}^{(t)} &= \mathbf{E}_{m,k}\text{diag}(\boldsymbol{u}_{m,k}^{(t)})\mathbf{E}_{m,k}^{\mathrm{T}},\label{40b}
\end{align}
\end{subequations}
where $d_{m,k}^{(t)}=\mathrm{c}_{0}(\mu_{\varsigma_{m,k}^{(t)}\rightarrow dp_{m,k}^{(t)}}-\tau_{m})$, $\mathbf{E}_{m}=[\mathbf{e}_{m,x},\mathbf{e}_{m,y},\mathbf{e}_{m,z}]$ with $\mathbf{e}_{m,z}=\mathbf{e}_{m,x}\times\mathbf{e}_{m,y}$, $\boldsymbol{v}_{m,k}^{(t)}=\Big[\mu_{\vartheta_{m,k,x}^{(t)}\rightarrow xp_{m,k}^{(t)}}\!+\phi_{m,x},\mu_{\vartheta_{m,k,y}^{(t)}\rightarrow yp_{m,k}^{(t)}}\!+\phi_{m,y},\Big(\!1-\!\big(\mu_{\vartheta_{m,k,x}^{(t)}\rightarrow xp_{m,k}^{(t)}}\!+\!\phi_{m,x}\big)^{2}\!-\big(\mu_{\vartheta_{m,k,y}^{(t)}\rightarrow yp_{m,k}^{(t)}}\!+\!\phi_{m,y}\big)^{2}\Big)^{\frac{1}{2}}\Big]^{\mathrm{T}}$, $\mathbf{E}_{m}\!=\![\mathbf{e}_{m,x},\mathbf{e}_{m,y},\hat{\mathbf{e}}_{m,k}]$ with $\hat{\mathbf{e}}_{m,k}^{(t)}\!=\!\mathbf{E}_{m}\boldsymbol{v}_{m,k}^{(t)}$, and $\boldsymbol{u}_{m,k}^{(t)}\!=\!\!\left[\frac{(d_{m,k}^{(t)})^{2}}{\kappa_{\vartheta_{m,k,x}^{(t)}\rightarrow xp_{m,k}^{(t)}}},\frac{(d_{m,k}^{(t)})^{2}}{\kappa_{\vartheta_{m,k,y}^{(t)}\rightarrow yp_{m,k}^{(t)}}},\frac{\mathrm{c}_{0}^{2}}{\kappa_{\varsigma_{m,k}^{(t)}\rightarrow dp_{m,k}^{(t)}}}\right]$.
% \begin{subequations}
% \begin{align}
%     % [\boldsymbol{u}_{m,k}^{(t)}]_{1}&=\frac{(d_{m,k}^{(t)})^{2}}{\big(1-\big(\mu_{\vartheta_{m,k,x}^{(t)}\rightarrow xp_{m,k}^{(t)}}+\phi_{m,x}\big)^{2}\big)\kappa_{\vartheta_{m,k,x}^{(t)}\rightarrow xp_{m,k}^{(t)}}},\\
%     % [\boldsymbol{u}_{m,k}^{(t)}]_{2}&=\frac{(d_{m,k}^{(t)})^{2}}{\big(1-\big(\mu_{\vartheta_{m,k,y}^{(t)}\rightarrow yp_{m,k}^{(t)}}+\phi_{m,y}\big)^{2}\big)\kappa_{\vartheta_{m,k,y}^{(t)}\rightarrow yp_{m,k}^{(t)}}},\\
%     % [\boldsymbol{u}_{m,k}^{(t)}]_{2}&=\frac{\mathrm{c}_{0}^{2}}{\kappa_{\varsigma_{m,k}^{(t)}\rightarrow dp_{m,k}^{(t)}}}.
%     [\boldsymbol{u}_{m,k}^{(t)}]_{1}&=\frac{(d_{m,k}^{(t)})^{2}}{\kappa_{\vartheta_{m,k,x}^{(t)}\rightarrow xp_{m,k}^{(t)}}},\\
%     [\boldsymbol{u}_{m,k}^{(t)}]_{2}&=\frac{(d_{m,k}^{(t)})^{2}}{\kappa_{\vartheta_{m,k,y}^{(t)}\rightarrow yp_{m,k}^{(t)}}},\\
%     [\boldsymbol{u}_{m,k}^{(t)}]_{2}&=\frac{\mathrm{c}_{0}^{2}}{\kappa_{\varsigma_{m,k}^{(t)}\rightarrow dp_{m,k}^{(t)}}}.
% \end{align}
% \end{subequations}
\par
Assuming that a Gaussian distribution $p(\mathbf{p}_{k}^{(0)})$ for each $1\leq k\leq K$ is provided in the initial frame, all the messages passing along the Markov chain $\{\mathbf{p}_{k}^{(t)}\}$ are Gaussian messages. Then, $\Delta_{\mathbf{p}_{k}^{(t)}\rightarrow pp_{k}^{(t+1)}}(\mathbf{p}_{k}^{(t)})$ is a Gaussian distribution with mean vector and covariance matrix given by
\begin{subequations}\label{44}
\begin{align}
    \mathbf{m}_{\mathbf{p}_{k}^{(t)}\rightarrow pp_{k}^{(t+1)}}&=\mathbf{C}_{\mathbf{p}_{k}^{(t)}\rightarrow pp_{k}^{(t+1)}}\Big(\sum_{k=1}^{K}(\mathbf{C}_{m,k}^{(t)})^{-1}\mathbf{m}_{m,k}^{(t)}
    +\mathbf{C}_{pp_{k}^{(t)}\rightarrow \mathbf{p}_{k}^{(t)} }^{-1}\mathbf{m}_{pp_{k}^{(t)}\rightarrow \mathbf{p}_{k}^{(t)}}\Big),\\
    \mathbf{C}_{\mathbf{p}_{k}^{(t)}\rightarrow pp_{k}^{(t+1)}}^{-1}&=\sum_{k=1}^{K}(\mathbf{C}_{m,k}^{(t)})^{-1}+\mathbf{C}_{pp_{k}^{(t)}\rightarrow \mathbf{p}_{k}^{(t)} }^{-1}.
\end{align}
\end{subequations}
\subsubsection{Messages from $pp_{k}^{(t+1)}$ to $\mathbf{p}_{k}^{(t+1)}$} For $1\leq k\leq K$,
\begin{align}\label{42}
    \Delta_{pp_{k}^{(t+1)}\rightarrow\mathbf{p}_{k}^{(t+1)}}(\mathbf{p}_{k}^{(t+1)})\propto\int_{\mathbf{p}_{k}^{(t)}}\Delta_{\mathbf{p}_{k}^{(t)}\rightarrow pp_{k}^{(t+1)}}(\mathbf{p}_{k}^{(t)})p(\mathbf{p}_{k}^{(t+1)}|\mathbf{p}_{k}^{(t)}).
\end{align}
Given \eqref{15}, the Gaussian message $\Delta_{pp_{k}^{(t+1)}\rightarrow\mathbf{p}_{k}^{(t+1)}}$ is calculated by
\begin{subequations}\label{47}
\begin{align}
    \mathbf{m}_{pp_{k}^{(t+1)}\rightarrow\mathbf{p}_{k}^{(t+1)}}&=\mathbf{m}_{\mathbf{p}_{k}^{(t)}\rightarrow pp_{k}^{(t+1)}},\\
    \mathbf{C}_{pp_{k}^{(t+1)}\rightarrow\mathbf{p}_{k}^{(t+1)}}&=\mathbf{C}_{\mathbf{p}_{k}^{(t)}\rightarrow pp_{k}^{(t+1)}}+\mathbf{C}_{k}.
\end{align}
\end{subequations}
\subsubsection{Messages from $\zeta_{m,k}^{(t)}$ to $zz_{m,k}^{(t+1)}$} Based on the variational inference in Section \ref{varia}, the message $\Delta_{\zeta_{m,k}^{(t)}\rightarrow zz_{m,k}^{(t+1)}}(\zeta_{m,k}^{(t)})$ is given by $q(\boldsymbol{\zeta}_{m}^{(t)})$, i.e., $\Delta_{\zeta_{m,k}^{(t)}\rightarrow zz_{m,k}^{(t+1)}}(\zeta_{m,k}^{(t)})=\delta(\zeta_{m,k}^{(t)}-\hat{\zeta}_{m,k}^{(t)})$.
\subsubsection{Messages from $zz_{m,k}^{(t+1)}$ to $\zeta_{m,k}^{(t+1)}$} For $1\leq m\leq M$, $1\leq k\leq K$,
\begin{align}\label{42nn}
    \Delta_{zz_{m,k}^{(t+1)}\rightarrow\zeta_{m,k}^{(t+1)}}(\zeta_{m,k}^{(t+1)})&\propto\int_{\zeta_{m,k}^{(t)}}{\Delta_{\zeta_{m,k}^{(t)}\rightarrow zz_{m,k}^{(t+1)}}(\zeta_{m,k}^{(t)})
    p(\zeta_{m,k}^{(t+1)}|\zeta_{m,k}^{(t)})}\notag\\
    &=p(\zeta_{m,k}^{(t+1)}|\hat{\zeta}_{m,k}^{(t)}),
\end{align}
where $p(\zeta_{m,k}^{(t+1)}|\hat{\zeta}_{m,k}^{(t)})$ is provided in \eqref{deth}.
\subsection{Overall Algorithm}
With discussions in the preceding subsections, we summarize our Bayesian multiuser tracking (BMT) algorithm in Algorithm \ref{tracking_algorithm}. BMT is an online tracking algorithm, where the BS estimates the multiuser positions in each frame by executing the message passing and variational inference. Lines 7-10 of Algorithm \ref{tracking_algorithm} correspond to the variational inference. Lines 2-3 and line 15 correspond to the message passing along the Markov chain. As for the initialization of the BMT algorithm in the first pilot frame, we assume that the initial position information of the users is obtained from the Global Positioning System (GPS) with the mean and covariance denoted by $\mathbf{m}_{\mathbf{p}_{k}^{(0)}\rightarrow pp_{k}^{(1)}}$ and
$\mathbf{C}_{\mathbf{p}_{k}^{(0)}\rightarrow pp_{k}^{(1)}}$. The estimation of channel parameters $\{\boldsymbol{\vartheta}_{m,x}^{(t)},\boldsymbol{\vartheta}_{m,y}^{(t)},\boldsymbol{\varsigma}_{m}^{(t)}\}$ are obtained from position estimations $\{\mathbf{m}_{\mathbf{p}_{k}^{(t)}\rightarrow pp_{k}^{(t+1)}}\}$ based on geometric constraint \eqref{11}.\par
% Lines 7-10 in Algorithm \ref{tracking_algorithm} are inspired by the variational inference of line spectra \cite{badiu2017variational,zhu2019grid}. Since the PBF of RISs made up the compression matrix $\boldsymbol{W}_{m}^{(t)}$ as shown in \eqref{9}, the existing algorithms are not applicable. The derivation in lines 7-10 extends the variational inference of line spectra to a more general form. Therefore, the channel parameters $\theta_{m,k,x}^{(t)}$, $\theta_{m,k,y}^{(t)}$, and $\tau_{m,k}^{(t)}$ are well estimated from $\boldsymbol{R}_{m}^{(t)}$ to help with multiuser tracking.\par
The complexity of the BMT algorithm mainly arises from the variational inference, i.e., lines 7-10 in Algorithm \ref{tracking_algorithm}. The complexity of calculating $q({\vartheta}_{m,k,x}^{(t)})$ and $q({\vartheta}_{m,k,x}^{(t)})$ is $\mathcal{O}(K(n_{x}N_{\mathrm{x}}+n_{y}N_{\mathrm{y}}))$ where $n_{x}$ and $n_{y}$ are the iteration numbers of GDM. The complexity of calculating $q(\varsigma_{m,k}^{(t)})$,
$q(\boldsymbol{\zeta}_{m}^{(t)})$, and
$q(\boldsymbol{\varrho}_{m}^{(t)})$ are respectively $\mathcal{O}(KL^2)$, $\mathcal{O}(K^{3}GL)$, and $\mathcal{O}(K^{2}GN_{\mathrm{R}}(N_{\mathrm{x}}+N_{\mathrm{y}}))$. Therefore, the overall complexity of BMT is $\mathcal{O}(n_{1}MK(n_{x}N_{\mathrm{x}}+n_{y}N_{\mathrm{y}}+L^2+K^{2}GL+KGN_{\mathrm{R}}(N_{\mathrm{x}}+N_{\mathrm{y}})))$, where $n_1$ is the number of iterations needed in the variational inference.
\begin{algorithm}[t]
	\caption{BMT Algorithm} 
	\label{tracking_algorithm} 
	%\hspace*{0.02in} {\bf Input:}
	{\bf Input:} Observed signal $\boldsymbol{y}^{(t,g,l)}$ for $1\leq t \leq T$, $1\leq g \leq G$, $1\leq l \leq L$, initial user position estimation $\mathbf{m}_{\mathbf{p}_{k}^{(0)}\rightarrow pp_{k}^{(1)}}$ with its covariance matrix $\mathbf{C}_{\mathbf{p}_{k}^{(0)}\rightarrow pp_{k}^{(1)}}$ for $1\leq k \leq K$.
	\\
	%\hspace*{-0.01in} {\bf Output:} 
	{\bf Output:} Position estimations $\{\mathbf{m}_{\mathbf{p}_{k}^{(t)}\rightarrow pp_{k}^{(t+1)}}\}$, cascaded channel path gain estimations $\{\hat{\varrho}_{m,k}^{(t)}\}$, and blockage estimations $\{\hat{\boldsymbol{\zeta}}_{k}^{(t)}\}$ for $1\leq t \leq T$, $1\leq k \leq K$, $1\leq m \leq M$.\par
	\begin{algorithmic}[1] %1 Indicates that each line displays a number
		\FOR{$t = 1$ to $T$}
		\STATE{For $\forall k$, calculate  $\mathbf{m}_{pp_{k}^{(t)}\rightarrow \mathbf{p}_{k}^{(t)}} $
			and $\mathbf{C}_{pp_{k}^{(t)}\rightarrow \mathbf{p}_{k}^{(t)}}$ by \eqref{47}.}
		\STATE{For $\forall m,k$, calculate $\Delta_{zz_{m,k}^{(t)}\rightarrow\zeta_{m,k}^{(t)}}(\zeta_{m,k}^{(t)})$ by \eqref{42nn}.}
		\FOR{$m=1$ to $M$}
			\STATE{For $\forall g,l$, calculate $ \Delta_{[\boldsymbol{r}_{g,l}^{(t)}]_{m} \rightarrow h_{m}^{(t)}}([\boldsymbol{r}_{g,l}^{(t)}]_{m})$ by \eqref{21}.}
			\STATE{For $\forall k$, calculate $\Delta_{xp_{m,k}^{(t)}\rightarrow \vartheta_{m,k,x}^{(t)} }(\vartheta_{m,k,x}^{(t)})$, $\Delta_{yp_{m,k}^{(t)}\rightarrow \vartheta_{m,k,y}^{(t)} }(\vartheta_{m,k,y}^{(t)})$ and $\Delta_{dp_{m,k}^{(t)}\rightarrow \varsigma_{m,k}^{(t)} }(\varsigma_{m,k}^{(t)})$}.
		\REPEAT
		\STATE{For $\forall k$, calculate $q({\vartheta}_{m,k,x}^{(t)})$ and $q({\vartheta}_{m,k,y}^{(t)})$.}
		\STATE{For $\forall k$, calculate $q(\varsigma_{m,k}^{(t)})$.}
		\STATE{Calculate $q(\boldsymbol{\zeta}_{m}^{(t)})$.}
		\STATE{Calculate $q(\boldsymbol{\varrho}_{m}^{(t)})$.}
		\UNTIL stopping criterion
		\STATE{For $\forall k$, calculate $\Delta_{\vartheta_{m,k,x}^{(t)}\rightarrow xp_{m,k}^{(t)}}(\vartheta_{m,k,x}^{(t)})$,
$\Delta_{\vartheta_{m,k,y}^{(t)}\rightarrow yp_{m,k}^{(t)} }(\vartheta_{m,k,y}^{(t)})$ and $\Delta_{\varsigma_{m,k}^{(t)}\rightarrow dp_{m,k}^{(t)}  }(\varsigma_{m,k}^{(t)})$.}
		\ENDFOR
		\STATE{For $\forall k$, Calculate $\mathbf{m}_{\mathbf{p}_{k}^{(t)}\rightarrow pp_{k}^{(t+1)}}$ and $\mathbf{C}_{\mathbf{p}_{k}^{(t)}\rightarrow pp_{k}^{(t+1)}}$ by \eqref{44}.}
		\ENDFOR
		\STATE \textbf{return} $\{\mathbf{m}_{\mathbf{p}_{k}^{(t)}\rightarrow pp_{k}^{(t+1)}}\}$, $\{\hat{\varrho}_{m,k}^{(t)}\}$, and $\{\hat{\boldsymbol{\zeta}}_{m}^{(t)}\}$ for $1\leq t \leq t_0$, $1\leq i \leq K$, $1\leq m \leq M$.
	\end{algorithmic}
\end{algorithm}
\section{Bayesian Cramer Rao Bound}\label{S4}
For the deterministic parameter estimation problem, a useful lower bound for the MSE of the unbiased estimator is the Cramer Rao bound (CRB), calculated from the inverse of the Fisher information matrix (FIM). For the random parameter estimation problem as considered in this paper, the BCRB proposed in \cite{van2004detection} is desired. In this section, we derive the BCRB for the estimation of the multiuser positions and the channel parameters in a single frame from the BIM. We assume $\zeta_{m,k}^{(t)}=1$ for all $k$, $m$, and $t$ in the derivation. We define two parameter sets involved in the derivation, i.e., $\boldsymbol{\eta}^{(t)}=\big[(\boldsymbol{\eta}_{1}^{(t)})^{\mathrm{T}},...,(\boldsymbol{\eta}_{K}^{(t)})^{\mathrm{T}}\big]^{\mathrm{T}}$ and $\boldsymbol{\gamma}^{(t)}=\big[(\boldsymbol{\gamma}_{1,1}^{(t)})^{\mathrm{T}},(\boldsymbol{\gamma}_{1,2}^{(t)})^{\mathrm{T}},...,(\boldsymbol{\gamma}_{M,K}^{(t)})^{\mathrm{T}}\big]^{\mathrm{T}}$ where\footnote{When the LoS path is blocked, i.e., $\zeta_{m,k}^{(t)}=0$, we remove the correspond parameters $\arg\varrho_{m,k}^{(t)}$ and $\vert\varrho_{m,k}^{(t)}\vert$ from the parameter set $\boldsymbol{\eta}_{k}^{(t)}$ and $\boldsymbol{\gamma}_{m,k}^{(t)}$. The derivation of the BCRB remains the same.} 
\begin{subequations}
\begin{align}
    \boldsymbol{\eta}_{k}^{(t)} &= \big[(\mathbf{p}_{k}^{(t)})^{\mathrm{T}},\arg\varrho_{1,k}^{(t)},\vert\varrho_{1,k}^{(t)}\vert,...,\arg\varrho_{M,k}^{(t)},\vert\varrho_{M,k}^{(t)}\vert\big]^{\mathrm{T}},\\
    \boldsymbol{\gamma}_{m,k}^{(t)} &= \big[\vartheta_{m,k}^{(t)},\vartheta_{m,k}^{(t)},\varsigma_{m,k}^{(t)},\arg\varrho_{m,k}^{(t)},\vert\varrho_{m,k}^{(t)}\vert\big]^{\mathrm{T}}.
\end{align}
\end{subequations}
The BCRB for the parameter set $\boldsymbol{\eta}^{(t)}$ in each frame $t$ is calculated from the BIM denoted by $\boldsymbol{\Lambda}_{\mathrm{B}}^{(t)}$. It is shown in \cite{tichavsky1998posterior} that the BIM for discrete-time filtering obeys the recursion
\begin{align}
    \boldsymbol{\Lambda}_{\mathrm{B}}^{(t)} =\boldsymbol{\Lambda}^{(t)} + \mathbf{G}_{22}^{(t)} + \mathbf{G}_{21}^{(t)}(\boldsymbol{\Lambda}_{\mathrm{B}}^{(t-1)} + \mathbf{G}_{11}^{(t)})^{-1}\mathbf{G}_{12}^{(t)},
\end{align}
where $\boldsymbol{\Lambda}^{(t)}$ is the FIM of parameter set $\boldsymbol{\eta}^{(t)}$ and the other items are defined as
\begin{align}\label{54}
    \mathbf{G}_{ij}^{(t)} = \mathbb{E}\left[-\frac{\partial^2\log p(\boldsymbol{\eta}^{(t)}|\boldsymbol{\eta}^{(t-1)})}{\partial\boldsymbol{\eta}^{(t+i-2)}\partial(\boldsymbol{\eta}^{(t+j-2)})^{\mathrm{T}}}\right],\ \mathrm{for}\ i,j\in\{1,2\},
\end{align}
where the expectation is taken over $p(\boldsymbol{\eta}^{(t)},\boldsymbol{\eta}^{(t-1)})$. Given the conditional probability \eqref{15} and the conditional independence between $\varrho_{m,k}^{(t)}$ and $\varrho_{m,k}^{(t-1)}$, the nonzero submatrices in \eqref{54} are
\begin{align}
    \mathbb{E}\left[-\frac{\partial^2\log p(\boldsymbol{\eta}^{(t)}|\boldsymbol{\eta}^{(t-1)})}{\partial\mathbf{p}_{k}^{(t+i-2)}\partial(\mathbf{p}_{k}^{(t+j-2)})^{\mathrm{T}}}\right]=
    (-1)^{i+j}\mathbf{C}_{k}^{-1},\ \mathrm{for}\ i,j\in\{1,2\}.
\end{align}
Given the complex-Gaussian-noise-disturbed received signal in \eqref{10}, the FIM of parameter set $\boldsymbol{\eta}^{(t)}$ is calculated by \cite{kay1993fundamentals}
\begin{align}\label{49}
    \boldsymbol{\Lambda}^{(t)}&=\frac{2}{\nu}\sum_{g=1}^{G}{\sum_{l=1}^{L}{\Re \left\{ \left( \frac{\partial \boldsymbol{m}^{(t,g,l)}}{\partial \boldsymbol{\eta}^{(t)}} \right) ^{\mathrm{H}}\left( \frac{\partial \boldsymbol{m}^{(t,g,l)}}{\partial \boldsymbol{\eta}^{(t)}} \right) \right\}}},\\
    &\overset{\left( a \right)}{=}\frac{2}{\nu}(\boldsymbol{T}^{(t)})^{\mathrm{T}}\sum_{g=1}^{G}{\sum_{l=1}^{L}{\Re \left\{ 
    \left(\frac{\partial \boldsymbol{r}_{g,l}^{(t)}}{\partial \boldsymbol{\gamma}^{(t)}}\right)^{\mathrm{H}}\mathbf{A}_{\mathrm{B}}^{\mathrm{H}}(\boldsymbol{\theta})
    \mathbf{A}_{\mathrm{B}}(\boldsymbol{\theta})\left(\frac{\partial \boldsymbol{r}_{g,l}^{(t)}}{\partial \boldsymbol{\gamma}^{(t)}}\right)\right\}}}\boldsymbol{T}^{(t)},
\end{align}
where $\overset{\left( a \right)}{=}$ utilizes the the
chain rule, $\boldsymbol{m}^{(t,g,l)}=\mathbf{A}_{\mathrm{B}}(\boldsymbol{\theta})\boldsymbol{r}_{g,l}^{(t)}$ is the noise-free version of the received signal $\boldsymbol{y}^{(t,g,l)}$,
and $\boldsymbol{T}^{(t)}=\frac{\partial \boldsymbol{\gamma}^{(t)}}{\partial\boldsymbol{\eta}^{(t)}}$ with the nonzero entries being calculated based on \eqref{11} given by
\begin{subequations}\label{51}
\begin{align}
    \frac{\partial\vartheta_{m,k,x}^{(t)}}{\partial\mathbf{p}_{k}^{(t)}}&=\frac{\mathbf{e}_{m,x}-(\mathbf{e}_{m,k}^{(t)})^{\mathrm{T}}\mathrm{e}_{m,x}\mathbf{e}_{m,k}^{(t)}}{\|\mathbf{p}_{k}^{(t)}-\mathbf{p}_{\mathrm{R},m}\|^{2}},\\
    \frac{\partial\vartheta_{m,k,y}^{(t)}}{\partial\mathbf{p}_{k}^{(t)}}&=\frac{\mathbf{e}_{m,y}-(\mathbf{e}_{m,k}^{(t)})^{\mathrm{T}}\mathrm{e}_{m,y}\mathbf{e}_{m,k}^{(t)}}{\|\mathbf{p}_{k}^{(t)}-\mathbf{p}_{\mathrm{R},m}\|^{2}},\\
    \frac{\partial\varsigma_{m,k}^{(t)}}{\partial\mathbf{p}_{k}^{(t)}}&=\frac{\mathbf{e}_{m,k}^{(t)}}{\mathrm{c}_{0}},\\
    \frac{\partial\arg\varrho_{m,k}^{(t)}}{\partial\arg\varrho_{m,k}^{(t)}}&=1,\ 
    \frac{\partial\vert\varrho_{m,k}^{(t)}\vert}{\partial\vert\varrho_{m,k}^{(t)}\vert}=1,
\end{align}
\end{subequations}
where $\mathbf{e}_{m,k}^{(t)}=\frac{\mathbf{p}_{k}^{(t)}-\mathbf{p}_{\mathrm{R},m}}{\|\mathbf{p}_{k}^{(t)}-\mathbf{p}_{\mathrm{R},m}\|}$ is the unit vector pointing from the $k$-th RIS to the $m$-th user. The nonzero entries of  $\frac{\partial \boldsymbol{r}_{g,l}^{(t)}}{\partial \boldsymbol{\gamma}^{(t)}}$ are calculated based on \eqref{8} given by
\begin{subequations}\label{52}
\begin{align}
    % \frac{\partial [\boldsymbol{r}_{g,l}^{(t)}]_{m}}{\partial \boldsymbol{\gamma}_{m,k}^{(t)}}=(\boldsymbol{\omega}_{k}^{(g)})^{\mathrm{T}}\mathbf{B}(\vartheta_{m,k,x}^{(t)},\vartheta_{m,k,y}^{(t)})\mathrm{diag}(\boldsymbol{c}_{m,k}^{(l)}),\\
    \frac{\partial[\boldsymbol{r}_{g,l}^{(t)}]_{m}}{\partial\vartheta_{m,k,x}^{(t)}}&=x_{k}^{(t,l)}\varrho_{m,k}^{(t)} [\mathbf{a}_{L}(\varsigma_{m,k}^{(t)})]_{l}(\boldsymbol{\omega}_{m}^{(t,g)})^{\mathrm{T}}
    \frac{\partial\mathbf{a}_{\mathrm{x}}(\vartheta_{m,k,x}^{(t)})}{\partial\vartheta_{m,k,x}^{(t)}}\otimes\mathbf{a}_{\mathrm{y}}(\vartheta_{m,k,y}^{(t)}),\\
    \frac{\partial[\boldsymbol{r}_{g,l}^{(t)}]_{m}}{\partial\vartheta_{m,k,y}^{(t)}}&=x_{k}^{(t,l)}\varrho_{m,k}^{(t)} [\mathbf{a}_{L}(\varsigma_{m,k}^{(t)})]_{l}(\boldsymbol{\omega}_{m}^{(t,g)})^{\mathrm{T}}
    \mathbf{a}_{\mathrm{x}}(\vartheta_{m,k,x}^{(t)})\otimes\frac{\partial\mathbf{a}_{\mathrm{y}}(\vartheta_{m,k,y}^{(t)})}{\partial\vartheta_{m,k,y}^{(t)}},\\
    \frac{\partial[\boldsymbol{r}_{g,l}^{(t)}]_{m}}{\partial\varsigma_{m,k}^{(t)}}&=x_{k}^{(t,l)}\varrho_{m,k}^{(t)}\left[\frac{\partial\mathbf{a}_{L}(\varsigma_{m,k}^{(t)})}{\partial\varsigma_{m,k}^{(t)}}\right]_{l}(\boldsymbol{\omega}_{m}^{(t,g)})^{\mathrm{T}}
    \mathbf{a}_{\mathrm{x}}(\vartheta_{m,k,x}^{(t)})\otimes\mathbf{a}_{\mathrm{y}}(\vartheta_{m,k,y}^{(t)}),\\
    \frac{\partial[\boldsymbol{r}_{g,l}^{(t)}]_{m}}{\partial\arg\varrho_{m,k}^{(t)}}&=jx_{k}^{(t,l)}\varrho_{m,k}^{(t)} [\mathbf{a}_{L}(\varsigma_{m,k}^{(t)})]_{l}(\boldsymbol{\omega}_{m}^{(t,g)})^{\mathrm{T}}
    \mathbf{a}_{\mathrm{x}}(\vartheta_{m,k,x}^{(t)})\otimes\mathbf{a}_{\mathrm{y}}(\vartheta_{m,k,y}^{(t)}),\\
    \frac{\partial[\boldsymbol{r}_{g,l}^{(t)}]_{m}}{\partial\vert\varrho_{m,k}^{(t)}\vert}&=x_{k}^{(t,l)}
    \frac{\varrho_{m,k}^{(t)}}{\vert\varrho_{m,k}^{(t)}\vert}[\mathbf{a}_{L}(\varsigma_{m,k}^{(t)})]_{l}(\boldsymbol{\omega}_{m}^{(t,g)})^{\mathrm{T}}
    \mathbf{a}_{\mathrm{x}}(\vartheta_{m,k,x}^{(t)})\otimes\mathbf{a}_{\mathrm{y}}(\vartheta_{m,k,y}^{(t)}).
\end{align}
\end{subequations}
% where $\dot{\mathbf{a}}_{\mathrm{x}}(\vartheta_{m,k,x}^{(t)})$, $\dot{\mathbf{a}}_{\mathrm{y}}(\vartheta_{m,k,y}^{(t)})$, and $\dot{\mathbf{a}}_{L}(\varsigma_{m,k}^{(t)})$ denote $\frac{\partial\mathbf{a}_{\mathrm{x}}(\vartheta_{m,k,x}^{(t)})}{\partial\vartheta_{m,k,x}^{(t)}}$, $\frac{\partial\mathbf{a}_{\mathrm{y}}(\vartheta_{m,k,y}^{(t)})}{\partial\vartheta_{m,k,y}^{(t)}}$, and $\frac{\partial\mathbf{a}_{L}(\varsigma_{m,k}^{(t)})}{\partial\varsigma_{m,k}^{(t)}}$. 

The BCRB of the parameter estimation $\hat{\boldsymbol{\eta}}^{(t)}$ is given by
\begin{align}
    \mathbb{E}[(\hat{\boldsymbol{\eta}}^{(t)}-\boldsymbol{\eta}^{(t)})
    (\hat{\boldsymbol{\eta}}^{(t)}-\boldsymbol{\eta}^{(t)})]\succeq(\boldsymbol{\Lambda}_{\mathrm{B}}^{(t)})^{-1}.
\end{align}
As for the BCRB of channel parameters $\vartheta_{m,k}^{(t)}$, $\vartheta_{m,k}^{(t)}$, and $\varsigma_{m,k}^{(t)}$, we apply the parameter transformation and obtain the BCRB of parameter estimation $\hat{\boldsymbol{\gamma}}^{(t)}$ given by
\begin{align}
     \mathbb{E}[(\hat{\boldsymbol{\gamma}}^{(t)}-\boldsymbol{\gamma}^{(t)})
    (\hat{\boldsymbol{\gamma}}^{(t)}-\boldsymbol{\gamma}^{(t)})]\succeq\boldsymbol{T}^{(t)}(\boldsymbol{\Lambda}_{\mathrm{B}}^{(t)})^{-1}(\boldsymbol{T}^{(t)})^{\mathrm{T}}.
\end{align}
The BCRB acts as a benchmark for our BMT algorithm. In the next section, we design the PBF of multiple RISs based on the derived BCRB.
\section{Passive Beamforming Design}\label{S5}
% Due to the large number of elements in the RIS and the limited communication resource, the OFDM symbols number $G$ is smaller than $N_{\mathrm{R}}$.
To enable high-precision multiuser tracking, the PBF vectors need to be appropriately designed.
In \eqref{9}, the matrix $\boldsymbol{W}_{m}^{(t)}\in\mathbb{C}^{N_{\mathrm{R}}\times G}$ collects the PBF vector of the $m$-th RIS over $G$ OFDM symbols regarded as a measurement matrix.  Since the BCRB provides a metric for evaluating the multiuser localization performance in each frame, we minimize the BCRB of the user positions $\{\mathbf{p}_{k}^{(t)}\}$ by optimizing the PBF of the RISs. The optimization problem in the $t$-th frame is formulated as
\begin{mini*}|s|
	{\{\mathbf{W}_{m}^{(t)}\}_{m=1}^{M}}
	{\sum_{j\in\mathcal{R}}{\left[\left( \boldsymbol{\Lambda}_{\mathrm{B}}^{(t)}\left(
% 	\{\mathbf{p_{m}^{(t)}}\}_{m=1}^{M},\{\boldsymbol{\varrho}_{m}^{(t)}\}_{k=1}^{K}
\{\mathbf{W}_{m}^{(t)}\}_{m=1}^{M}
	;\boldsymbol{\eta}^{(t)}\right) \right) ^{-1}\right]_{j,j}}}
	{}
	{\text{(P1)}:}
	\addConstraint{|[\mathbf{W}_{m}^{(t)}]_{n,g} |=1,\ \mathrm{for}\ 1\leq m\leq M,\ 1\leq n\leq N_{\mathrm{R}},\ 1\leq g\leq G,}
\end{mini*}
where $\mathcal{R}$ indexes the position variables $\{\mathbf{p}_{k}^{(t)}\}$ in the parameter set $\boldsymbol{\eta}^{(t)}$ and $|[\mathbf{W}_{m}^{(t)}]_{n,g} |=1$ is the unit modulus constraint of the RIS phase shift. Note that the objective function in (P1) is parameterized by $\boldsymbol{\eta}^{(t)}$ which is not available in practice. We utilize the estimations in the previous frame and reformulate the objective function in (P1) as
% obtain the approximated $\boldsymbol{\eta}^{(t)}$ based on $\Delta_{pp_{k}^{(t)}\rightarrow\mathbf{p}_{k}^{(t)} }(\mathbf{p}_{k}^{(t)})$ and $q(\boldsymbol{\varrho}_{m}^{(t-1)}|\boldsymbol{\zeta}_{m}^{(t-1)},\boldsymbol{R}_{m}^{(t-1)})$. 
% We define $\bar{\boldsymbol{\eta}}^{(t)}=\big[(\bar{\boldsymbol{\eta}}_{1}^{(t)})^{\mathrm{T}},...,(\bar{\boldsymbol{\eta}}_{M}^{(t)})^{\mathrm{T}}\big]^{\mathrm{T}}$ with $\bar{\boldsymbol{\eta}}_{m}^{(t)} = \big[\mathbf{m}_{pp_{k}^{(t)}\rightarrow \mathbf{p}_{k}^{(t)}} ^{\mathrm{T}},\arg\hat{\varrho}_{1,m}^{(t-1)},\vert\hat{\varrho}_{1,m}^{(t-1)}\vert,...,\arg\hat{\varrho}_{m,k}^{(t-1)},\vert\hat{\varrho}_{m,k}^{(t-1)}\vert\big]^{\mathrm{T}}$,
% The objective function in (P1) is reformulated by
\begin{align}\label{59}
    \mathbb{E}\left[\sum_{j\in\mathcal{R}}{\left[\left( \boldsymbol{\Lambda}_{\mathrm{B}}^{(t)}\left(
\{\mathbf{W}_{m}^{(t)}\}_{m=1}^{M}
	;{\boldsymbol{\eta}}^{(t)}\right) \right) ^{-1}\right]_{j,j}}\right],
\end{align}
where the expectation is taken over $\Delta_{pp_{k}^{(t)}\rightarrow\mathbf{p}_{k}^{(t)} }(\mathbf{p}_{k}^{(t)})$ with $\boldsymbol{\varrho}_{m}^{(t)}$ fixing by the mean of surrogate pdf $q(\boldsymbol{\varrho}_{m}^{(t-1)})$. In the approximated posterior distribution $q(\boldsymbol{\varrho}_{m}^{(t-1)})$ given in \eqref{35}, the estimation of cascaded channel path gain $\varrho_{m,k}^{(t-1)}$ is zero when the LoS path is blocked, i.e., $q(\varrho_{m,k}^{(t-1)}|\zeta_{m,k}^{(t-1)}=0)=\delta(\varrho_{m,k}^{(t-1)})$. Without special design, the beamforming design in (P1) 
neglects these blocked paths even if they are no longer blocked in subsequent frames which results in the BMT algorithm not being able to exploit all the geometric constraints.
Following the common free space path loss model, we substitute these zero estimations of $\delta(\varrho_{m,k}^{(t-1)})$ by 
\begin{align}
    q(\varrho_{m,k}^{(t-1)}|\zeta_{m,k}^{(t-1)}=0)=\delta\left(\varrho_{m,k}^{(t-1)}-\frac{\lambda^{2}\mathrm{e}^{-\jmath\frac{2\pi}{\lambda}
    \left(\left\|\mathbf{p}_{\mathrm{R},m}-\mathbf{m}_{pp_{k}^{(t)}\rightarrow\mathbf{p}_{k}^{(t)}}\right\|_2+\left\|\mathbf{p}_{\mathrm{B}}-\mathbf{p}_{\mathrm{R},m}\right\|_2\right)}}{16\pi^2
    \left\|\mathbf{p}_{\mathrm{R},m}-\mathbf{m}_{pp_{k}^{(t)}\rightarrow\mathbf{p}_{k}^{(t)}}\right\|_{2}\left\|\mathbf{p}_{\mathrm{B}}-\mathbf{p}_{\mathrm{R},m}\right\|_2}\right).
\end{align}\par
% in the calculation of expectation in \eqref{59}.
It is difficult to explicitly carry out
the expectation in \eqref{59}. We use the sample average to approximate the expectation calculation. In addition, we 
directly optimize the phase shifts of reflecting elements for RISs and obtain the optimization problem as 
\begin{mini*}|s|
	{\{\varphi_{m,n}^{(t,g)}\}}
	{\frac{1}{N_\mathrm{s}}\sum_{i=1}^{N_\mathrm{s}}\sum_{j\in\mathcal{R}}{\left[\left( \boldsymbol{\Lambda}_{\mathrm{B}}^{(t)}\left(
% 	\{\mathbf{p_{m}^{(t)}}\}_{m=1}^{M},\{\boldsymbol{\varrho}_{m}^{(t)}\}_{k=1}^{K}
\{\varphi_{m,n}^{(t,g)}\}
	;\boldsymbol{\eta}^{(t)}\right) \right) ^{-1}\right]_{j,j}},}
	{}
	{\text{(P2)}:}
\end{mini*}
where $\varphi_{m,n}^{(t,g)}=\arg[\mathbf{W}_{m}^{(t)}]_{n,g}$, $\mathbf{p}_{k}^{(t)}$ in parameter set $\boldsymbol{\eta}^{(t)}$ is sampled from $\Delta_{pp_{k}^{(t)}\rightarrow\mathbf{p}_{k}^{(t)} }(\mathbf{p}_{k}^{(t)})$ for each $k$, and $N_{\mathrm{s}}$ is the number of samples. Due to the non-convexity of the objective function, the global optimal of (P2) is difficult to be found. We use the GDM to obtain a local minimum of (P2). In each iterative search of GDM, we use the Armijo rule to determine the step size. In the first frame, the gradient descent search starts from a random phase shift setting $\{\varphi_{m,n}^{(0,g)}\}$. In any frame $t$, the gradient descent search starts from the optimized phase shifts setting in the preceding frame $t-1$.
\begin{table}[t]
	\centering
	\small
	\caption{System Parameters}
	\begin{tabular}{p{\columnwidth/5}p{\columnwidth*4/6}}
		\hline
		\textbf{Parameter}&\textbf{Value}\\
		\hline
		 $\lambda$&$10.7$ mm\\
		 $G$&$15$\\
% 		 \{10,15,25,35,55,75,100\}
		 $L$&$40$\\
		 $f_s$&$250\ \mathrm{kHz}$\\
		$N_\mathrm{B}$&32\\
		$(N_{\mathrm{x}}, N_{\mathrm{y}})$&$(10,10)$\\
% 		\{(7,7),(10,10),(13,13),(15,15),(17,17),(20,20)\}
% 		$(p_{l,m,k}^{(t)},p_{d,m,k}^{(t)})$&$1$\\
		$\mathbf{p}_\mathrm{B}$&$(-20\ \mathrm{m},0\ \mathrm{m},0\ \mathrm{m})$\\
		$\mathbf{e}_\mathrm{B}$&{$(0,1,0)$}\\
		$\mathbf{p}_{\mathrm{R},i}$&{$(0\ \mathrm{m},20\ \mathrm{m},10\ \mathrm{m}),(0\ \mathrm{m},-20\ \mathrm{m},10\ \mathrm{m})$}\\
		%$\mathbf{e}_\mathrm{R,i}$&(1, 0, 0)\\
		$\mathbf{e}_{\mathrm{R},i,x}$&{$(1,0,0)$, $(-1,0,0)$}\\
		$\mathbf{e}_{\mathrm{R},i,y}$&{$(0,0,1)$, $(0,0,1)$}\\
		\hline
	\end{tabular}
	\label{table2}
\end{table}
\section{Numerical Results}\label{S6}
In this section, we conduct numerical experiments to assess the performance of the BMT algorithm proposed in Section \ref{S3} with different system settings, compared with the BCRB derived in Section \ref{S4}. We execute the BMT algorithm under the PBF designed in Section \ref{S5} as well as other state-of-art PBF designs.
\subsection{Settings of Numerical Experiment}
In the numerical experiments, the main parameter settings of the multi-RIS aided MIMO-OFDM systems are listed in Table \ref{table2}, unless otherwise specified. There are $K=3$ users in the system and the trajectory of each user is generated based on the conditional probability $p(\mathbf{p}_{k}^{(t)}|\mathbf{p}_{k}^{(t-1)})=\mathcal{N}(\mathbf{p}_{k}^{(t)};\mathbf{p}_{k}^{(t-1)},\mathbf{C}_{p})$ with $\mathbf{C}_{p}=\mathrm{diag}([0.03,0.03,0.03]^{\mathrm{T}})$. The initial location of users $\{\mathbf{p}_{k}^{(0)}\}_{k=1}^{3}$ are set to $(-5\ \mathrm{m},0\ \mathrm{m},3.5\ \mathrm{m})$, $(10\ \mathrm{m},10\ \mathrm{m},1\ \mathrm{m})$, and $(10\ \mathrm{m},-10\ \mathrm{m},1\ \mathrm{m})$. Each trajectory contains $T=300$ discrete user positions. We generate the
binary variables $\zeta_{m,k}^{(t)}$ based on the birth and death process in \eqref{deth} with $p_{l,m,k}^{(t)}=0.9$ and $p_{d,m,k}^{(t)}=0.05$, given the initialization $\zeta_{m,k}^{(0)}=1$ for all $m$, $k$ and $t$. Given the position of the BS, the RISs, and the users, we generate the complex channel path gains $\rho_{m,k}^{(t)}$ and $\rho_{m}$ by following the free space path loss model and generate $\varrho_{m,k}^{(t)}$ by
\begin{align}
    \varrho_{m,k}^{(t)}=\zeta_{m,k}^{(t)}\frac{\lambda^{2}\mathrm{e}^{-\jmath\frac{2\pi}{\lambda}
    \left(\left\|\mathbf{p}_{\mathrm{R},m}-\mathbf{p}_{k}^{(t)}\right\|_2+\left\|\mathbf{p}_{\mathrm{B}}-\mathbf{p}_{\mathrm{R},m}\right\|_2\right)}}{16\pi^2
    \left\|\mathbf{p}_{\mathrm{R},m}-\mathbf{p}_{k}^{(t)}\right\|_{2}\left\|\mathbf{p}_{\mathrm{B}}-\mathbf{p}_{\mathrm{R},m}\right\|_2}.
\end{align}
The pilot signals of different users are orthogonally set by $\boldsymbol{x}_{k}=\mathbf{a}_{L}(\frac{k}{f_s})$.
We evaluate the BMT algorithm performance by the RMSE of user localization and the success rate of $\zeta_{m,k}^{(t)}$ estimation averaging from $20$ independently generated trajectories.
% \begin{figure}[t]
% \centering
%     \subfloat[]{\centering
%         \includegraphics[width=0.48\linewidth]{p1_3.pdf}
%         \label{fig:4a}
%     }
%     \subfloat[]{\centering
%         \includegraphics[width=0.48\linewidth]{p1_4.pdf}
%         \label{fig:4b}
%     }
%     \caption{The BMT algorithm performance v.s. SNR for $M=2$ and $M=3$. (a) MSE of the multiuser position $\mathbf{p}_{k}^{(t)}$ estimation; (b) accuracy of blockage ${\zeta}_{m,k}^{(t)}$ estimation}
%     \label{fig:4}
% \end{figure}
\subsection{Results and Discussions}
% \begin{figure}[htbp]
% \centering
% \begin{minipage}[t]{0.48\textwidth}
% \centering
% \includegraphics[width=1\textwidth]{p1_1.pdf}
% \caption{Different $\rho$ settings, $\mathbb{E}\left[\|\mathbf{e}_{{\rm c},k}^{(t)}\|_{2}^{2}\right]$ versus optimization round, Monte Carlo 50 times }
% \label{fig:rho_one}
% \end{minipage}
% \begin{minipage}[t]{0.48\textwidth}
% \centering
% \includegraphics[width=1\textwidth]{p1_2.pdf}
% \caption{Different $\bar{\rho}$ settings, learning accuracy versus communication round, concentrated, Monte Carlo 50 times }
% \label{fig:corr_acc}
% \end{minipage}
% \end{figure}
% \begin{figure}[htbp]
% \centering
% \begin{minipage}[t]{0.47\textwidth}
%     \centering
%     \includegraphics[width=1\textwidth]{p1_1.pdf}
%     \caption{Different $\boldsymbol{\rho}_{k}^{(t)}$ settings, test accuray versus communication round, MNIST, Monte Carlo 100 times }
%     \label{fig:corr_acc_one}
% \end{minipage}
% \qquad
% \begin{minipage}[t]{0.47\textwidth}
% \centering
%     \includegraphics[width=1\textwidth]{p1_2.pdf}
%     \caption{Different $\boldsymbol{\rho}_{k}^{(t)}$ settings, test accuracy versus communication round, Fashion-MNIST, Monte Carlo 100 times }
%     \label{fig:corr_acc_two}
% \end{minipage}
% \end{figure}
\begin{figure}[t]
\centering
    \subfloat[]{\centering
        \includegraphics[width=0.48\linewidth]{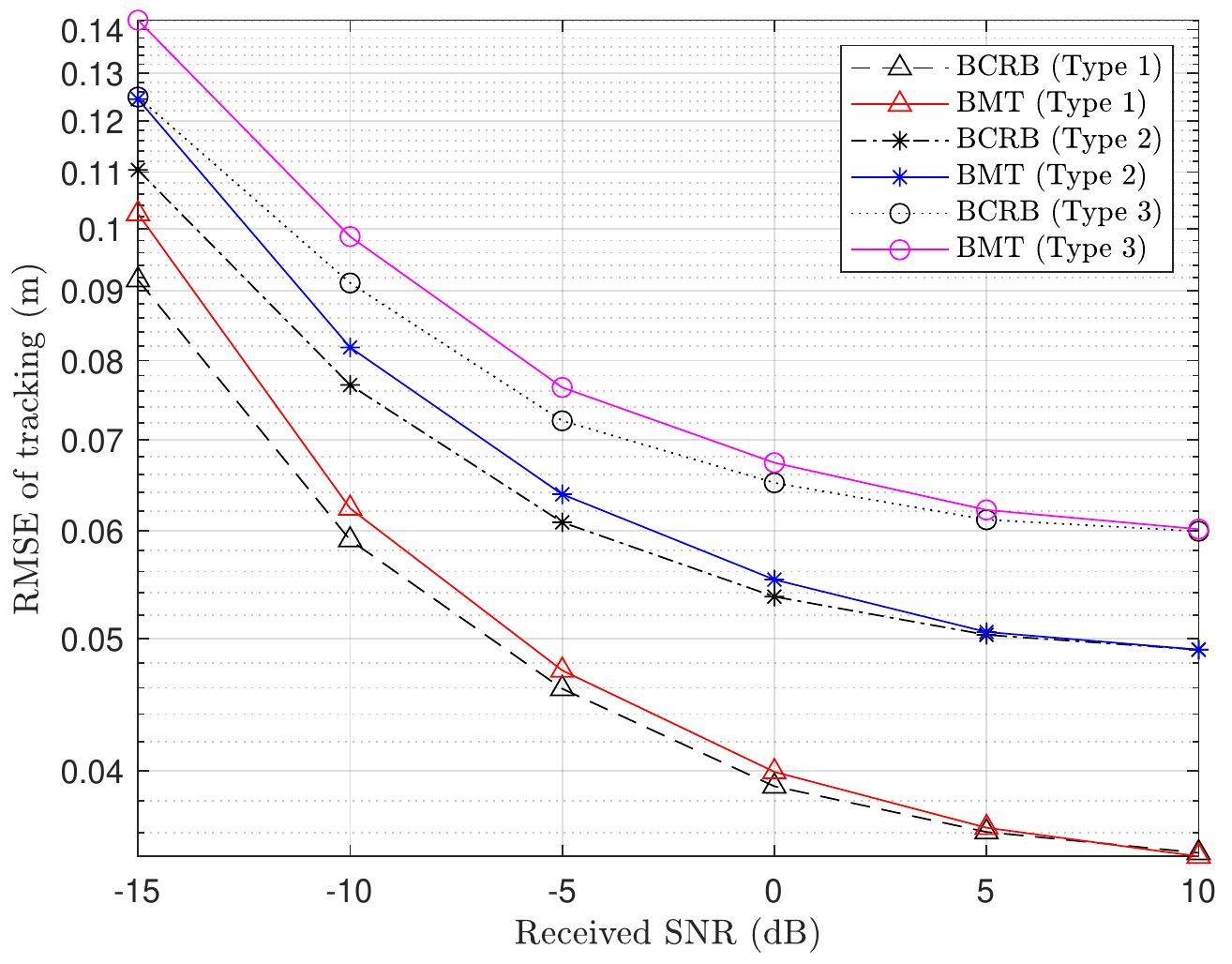}
        \label{fig:5a}
    }
    \subfloat[]{\centering
        \includegraphics[width=0.48\linewidth]{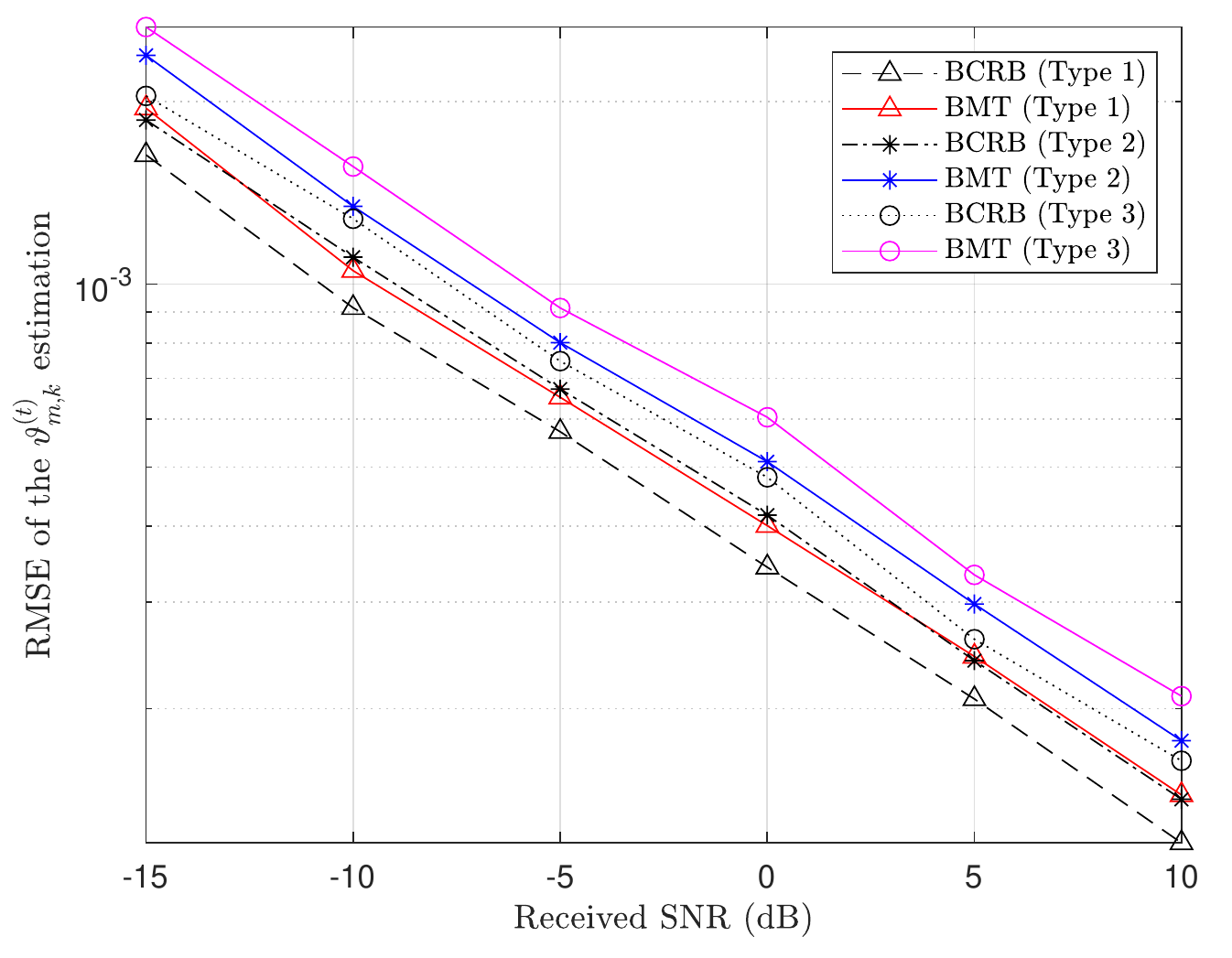}
        \label{fig:5b}
    }\\
    \subfloat[]{\centering
        \includegraphics[width=0.48\linewidth]{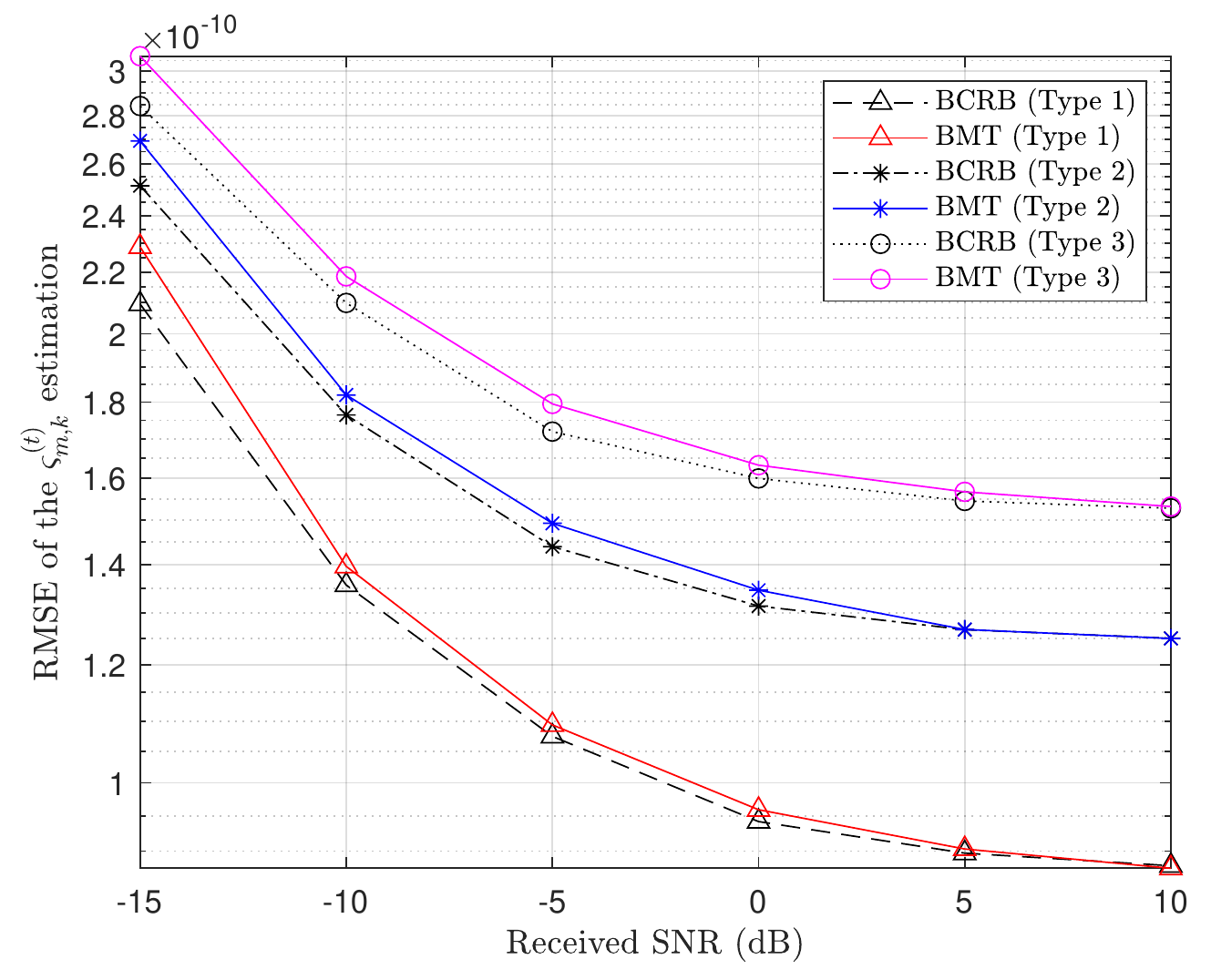}
        \label{fig:5c}
    }
    \subfloat[]{\centering
        \includegraphics[width=0.48\linewidth]{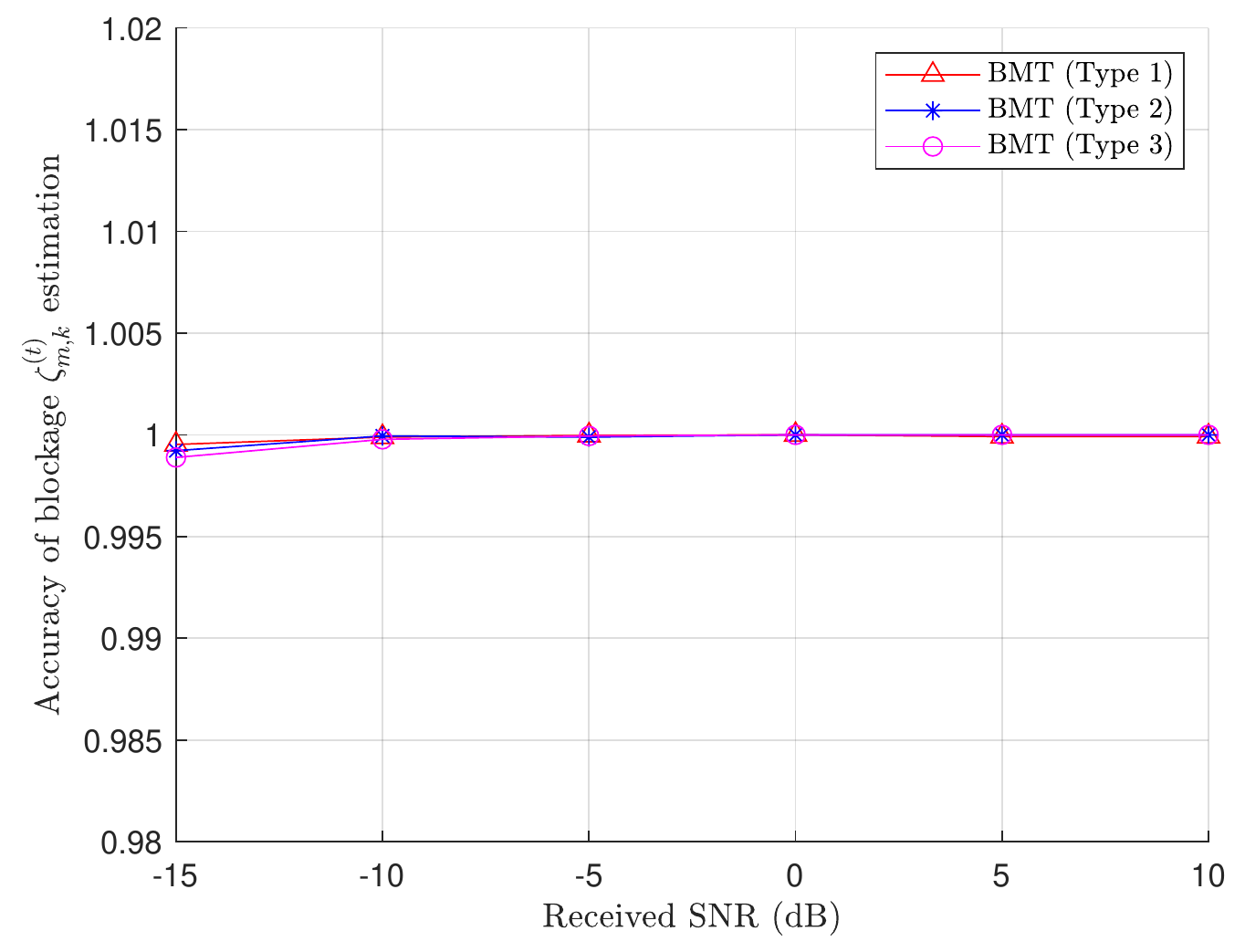}
        \label{fig:5d}
    }
    \caption{The BMT algorithm performance v.s. SNR for different user mobility. (a) RMSE of the multiuser position $\mathbf{p}_{k}^{(t)}$ estimations; (b) RMSE of the $\vartheta_{m,k}^{(t)}$ estimations; (c) RMSE of the $\varsigma_{m,k}^{(t)}$ estimations; (d) accuracy of blockage ${\zeta}_{m,k}^{(t)}$ estimations}
    \label{fig:5}
\end{figure}
\subsubsection{Effect of User Mobility} We first investigate the impact of user mobility on the performance of the BMT algorithm. In the trajectories generation, user mobility is characterized by the covariance matrix $\mathbf{C}_{p}$. Given $\mathbf{C}_{p}=\mathrm{diag}([0.03,0.03,0.03]^{\mathrm{T}})$, we have $\mathbb{E}\left[{\left\|\mathbf{p}_{k}^{(t+1)}-\mathbf{p}_{k}^{(t)}\right\|}_2^{2}\right] = 0.09$. We generate multiuser trajectories with three different $\mathbf{C}_{p}$ settings and evaluate the tracking performance. Concretely, $\mathbf{C}_{p}$ is set by $\mathbf{C}_{p}=\mathrm{diag}([0.01,0.01,0.01]^{\mathrm{T}})$, $\mathbf{C}_{p}=\mathrm{diag}([0.03,0.03,0.03]^{\mathrm{T}})$, and $\mathbf{C}_{p}=\mathrm{diag}([0.05,0.05,0.05]^{\mathrm{T}})$ and are respectively denoted by type 1, type 2 and type 3 in Fig. \ref{fig:5}. As shown in Fig \ref{fig:5a}, the user tracking performance of slow-moving users significantly outperforms fast-moving users while the tracking performances of all three scenarios are close to their respective BCRBs. When executing the BMT algorithm,
the stronger the user mobility, the weaker the temporal correlation of the user position, the less information is carried between adjacent frames, and the poorer the algorithm performance. In Fig. \ref{fig:5b}, the accuracy of blockage ${\zeta}_{m,k}^{(t)}$ estimation in the considered SNR range is close to 1. In the following experiments, we no longer evaluate the performance of ${\zeta}_{m,k}^{(t)}$ estimation since the results are very similar.
\begin{figure}[t]
    \subfloat[]{
        \includegraphics[width=0.47\linewidth]{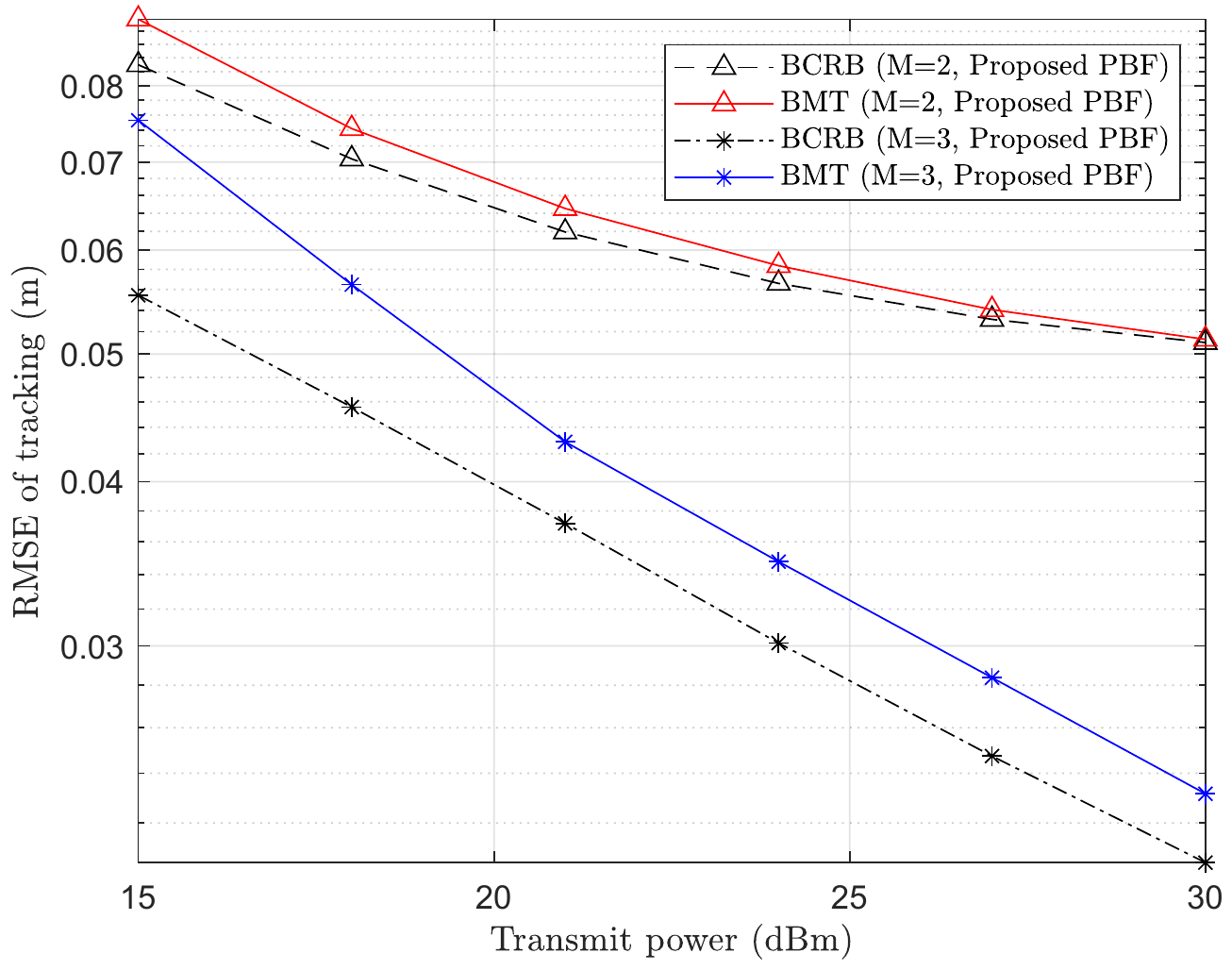}
        \label{fig:3a}
    }
    \subfloat[]{
        \includegraphics[width=0.47\linewidth,keepaspectratio]{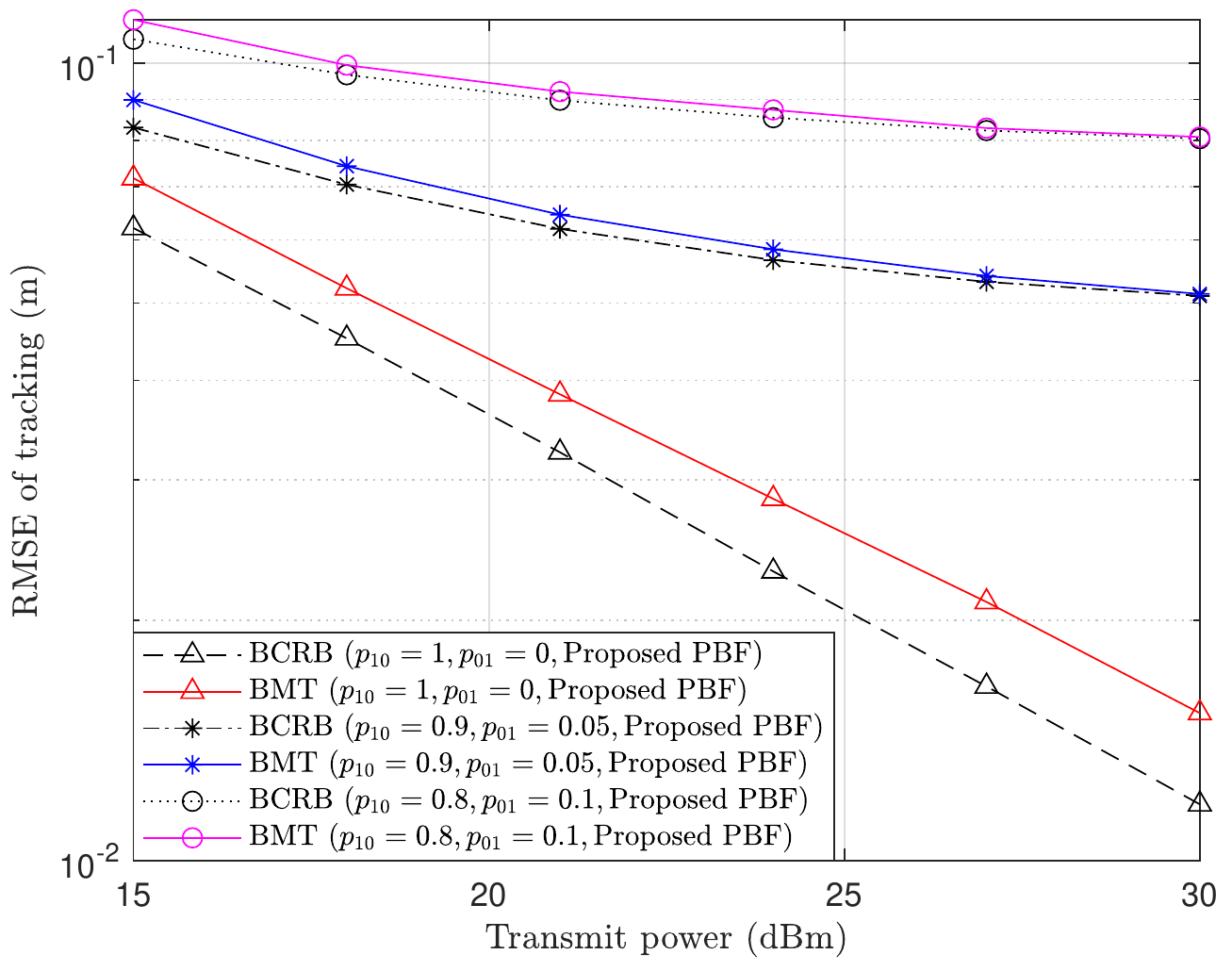}
        \label{fig:3b}
    }
    \caption{The BMT algorithm performance v.s. SNR (a) MSE of the multiuser position $\mathbf{p}_{k}^{(t)}$ estimation for $M=2$ and $M=3$; (b) MSE of the multiuser position $\mathbf{p}_{k}^{(t)}$ estimation with different $p_{l,m,k}^{(t)}$ and $p_{d,m,k}^{(t)}$ settings.}
    \label{fig:3}
\end{figure}
\subsubsection{Effect of the RIS Number and the Blockage of LoS Path} To study the effect of the number of RISs, we add a third RIS located in $(10\ \mathrm{m},20\ \mathrm{m},5\ \mathrm{m})$ with $\mathbf{e}_{\mathrm{R},i,x}=(1,0,0)$ and $\mathbf{e}_{\mathrm{R},i,y}=(0,0,1)$ in the MIMO-OFDM system. Fixing the power of interference-plus-noise by $\nu=-125\ \mathrm{dBm}$, we vary the transmit power and perform the BMT algorithm respectively aided by $M=2$ RISs and $M=3$ RISs. The results are shown in Fig. \ref{fig:3a}. For both the MSE of multiuser tracking and the BCRB, the additional RIS leads to better tracking performance. To study the effect of the blockage of the LoS path, we generate $\zeta_{m,k}^{(t)}$ based on the birth and death process with different $p_{l,m,k}^{(t)}$ and $p_{d,m,k}^{(t)}$ and evaluate the tracking performance. As shown in Fig. \ref{fig:3b}, we 
achieve the best tracking performance when all the LoS path exists (i.e., $p_{l,m,k}^{(t)}=1$, $p_{d,m,k}^{(t)}=0$ for 
all $m$, $k$, and $t$). As more LoS paths are blocked, the number of available geometric constraints in \eqref{11} reduces, which results in progressive deterioration of the tracking performance. 
\begin{figure}[t]
    \centering
    \includegraphics[width=0.5\columnwidth]{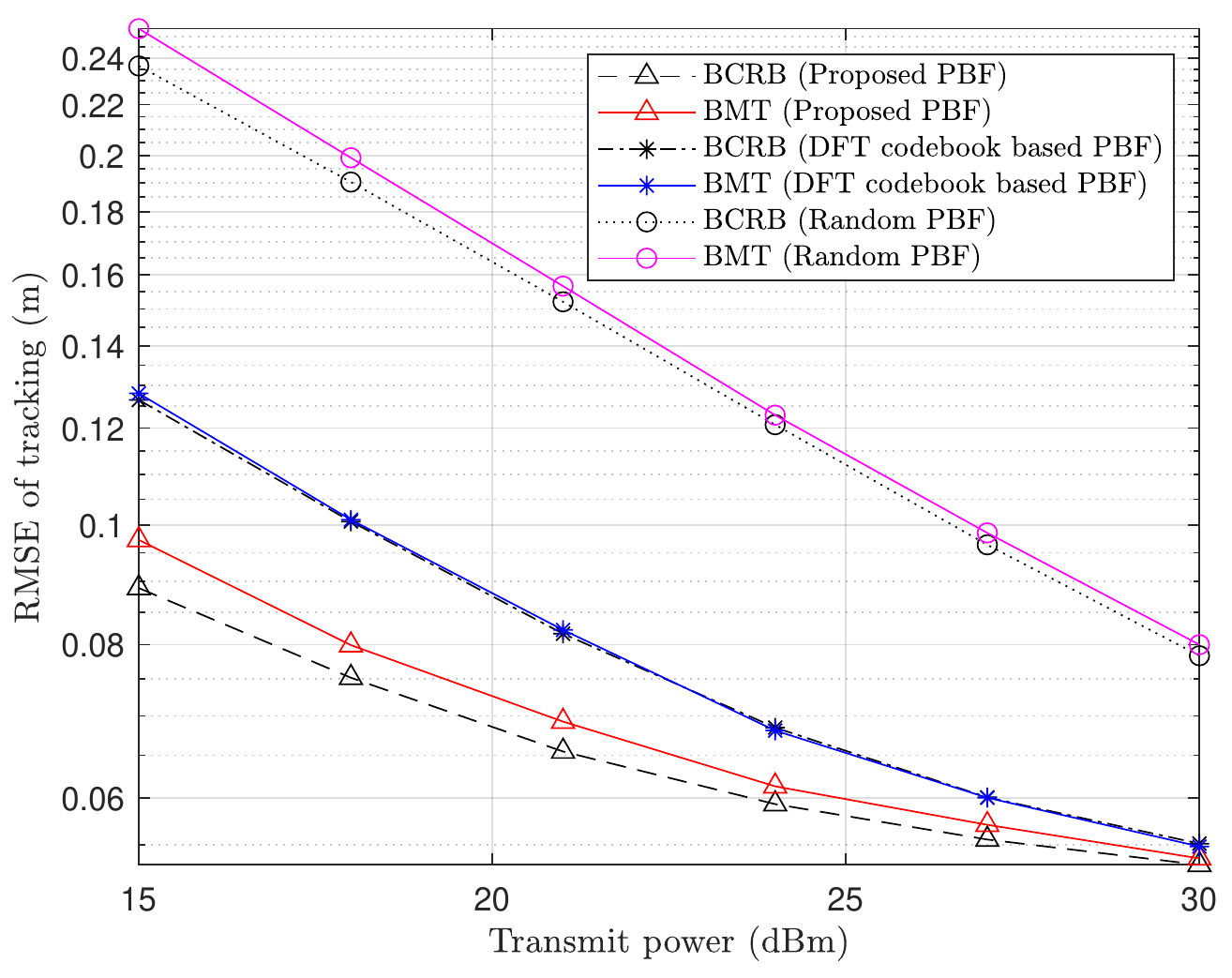}
    \caption{The BMT algorithm performance v.s. SNR with exploiting different passive beamforming methods. The OFDM overhead is fixed by $G=12$ for comparing PBF schemes.}
    \label{fig:6}
\end{figure}
\subsubsection{Comparison of Passive Beamforming Methods} We consider two PBF designs as benchmark schemes, namely, the random PBF design and the discrete Fourier transform (DFT) codebook-based PBF design \cite{noh2017multi}. For the random PBF design, the RIS phase shift settings are independently drawn from the uniform distribution $U(0,2\pi)$. For the $m$-th RIS and $k$-th user, the DFT codebook is given by
\begin{align}\label{60}
    \boldsymbol{W}_{m,k}^{(t)}=[\boldsymbol{D}^{N_{\mathrm{x}}}]_{1:N_{\mathrm{x}},(h_{m,k,\mathrm{x}}^{(t)}-\frac{H-1}{2}):(h_{m,k,\mathrm{x}}^{(t)}+\frac{H-1}{2})}\otimes[\boldsymbol{D}^{N_{\mathrm{y}}}]_{1:N_{\mathrm{y}},(h_{m,k,\mathrm{y}}^{(t)}-\frac{H-1}{2}):(h_{m,k,\mathrm{x}}^{(t)}+\frac{H-1}{2})},
\end{align}
where $\boldsymbol{D}^{N}\in\mathbb{C}^{N\times N}$ denotes a DFT matrix, $h_{m,k,\mathrm{x}}^{(t)}\overset{\mathrm{def}}{=}\arg\min_{h}\|[\boldsymbol{D}^{N_{\mathrm{x}}}]_{1:N_{\mathrm{x}},h}-\hat{\mathbf{a}}_{\mathrm{x}}(\vartheta_{m,k,x}^{(t-1)})\|_{2}$, and $h_{m,k,\mathrm{y}}^{(t)}\overset{\mathrm{def}}{=}\arg\min_{h}\|[\boldsymbol{D}^{N_{\mathrm{y}}}]_{1:N_{\mathrm{y}},h}-\hat{\mathbf{a}}_{\mathrm{y}}(\vartheta_{m,k,y}^{(t-1)})\|_{2}$. Based on \eqref{60}, the DFT codebook for the $m$-th RIS is $\boldsymbol{W}_{m}^{(t)}=[\boldsymbol{W}_{m,1}^{(t)},...,\boldsymbol{W}_{m,K}^{(t)}]\in\mathbb{C}^{N_{\mathrm{R}}\times KH^{2}}$. As shown in Fig. \ref{fig:6}, the BCRB-based PBF design significantly outperforms the random PBF design and the DFT codebook-based PBF design in multiuser tracking performance.
\begin{figure}[t]
    \centering
    \includegraphics[width=0.5\columnwidth]{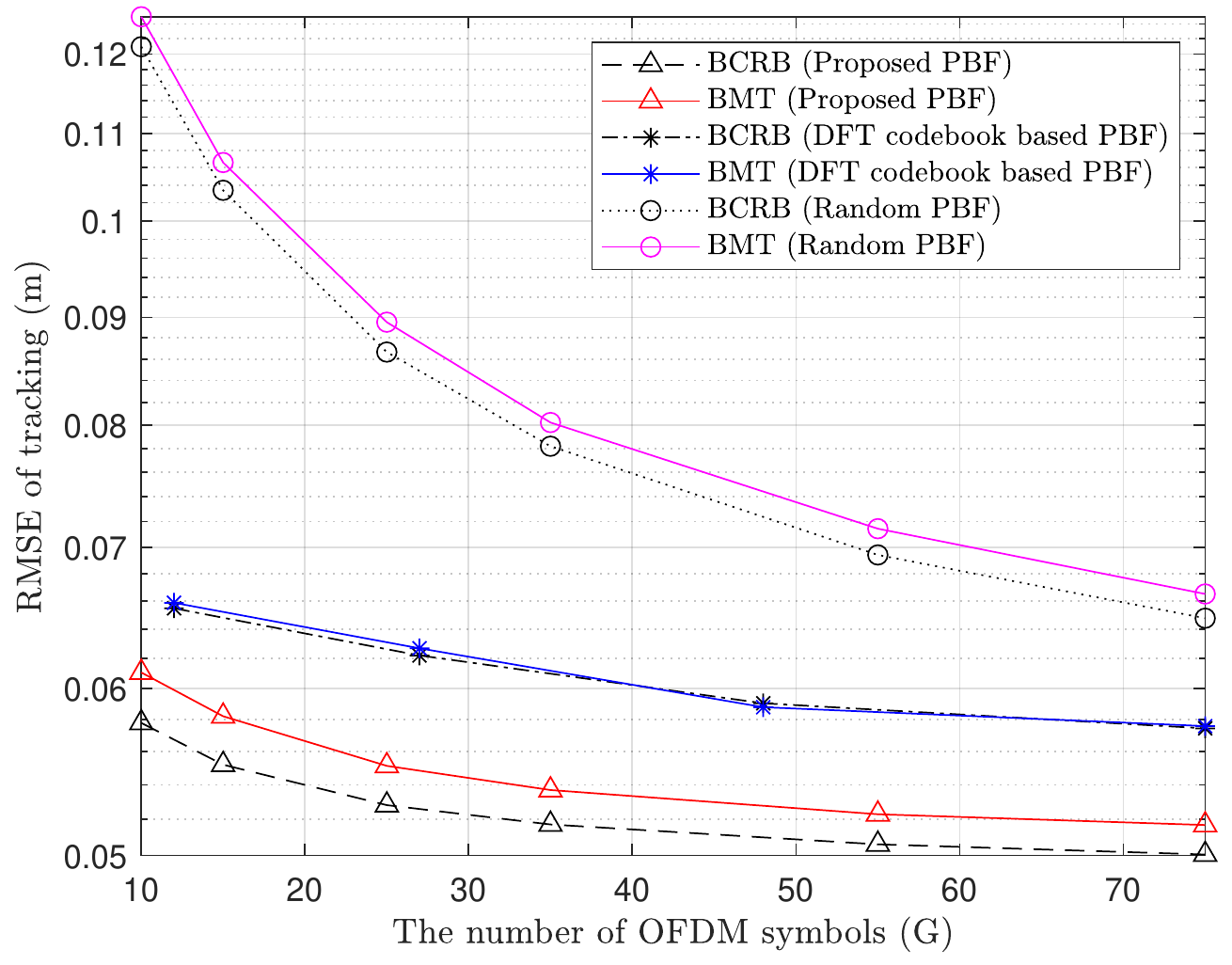}
    \caption{The BMT algorithm performance v.s. the number of OFDM symbols $G$ by exploiting different passive beamforming methods.}
    \label{fig:4}
\end{figure}
\subsubsection{Effect of the Number of OFDM Symbols}
In this subsection, we set the transmit power by $25$ dBm and vary the OFDM symbols overhead $G=[12,27,48,75]$ for the DFT codebook PBF design and $G=[10,15,25,35,55,75]$ for the random PBF design and proposed BCRB-based PBF design. The results are shown in Fig. \ref{fig:4}, where the RMSE of the multiuser position estimation decreases as the OFDM symbols overhead increases for all three PBF designs. For proposed BCRB-based PBF and DFT codebook-based PBF, with the RIS elements set by $N_\mathrm{R}=100$ ($10\times10$), the tracking performance is nearly saturated when the OFDM symbols overhead $G$ exceeds 40.
\begin{figure}[t]
\centering
    \includegraphics[width=0.5\columnwidth]{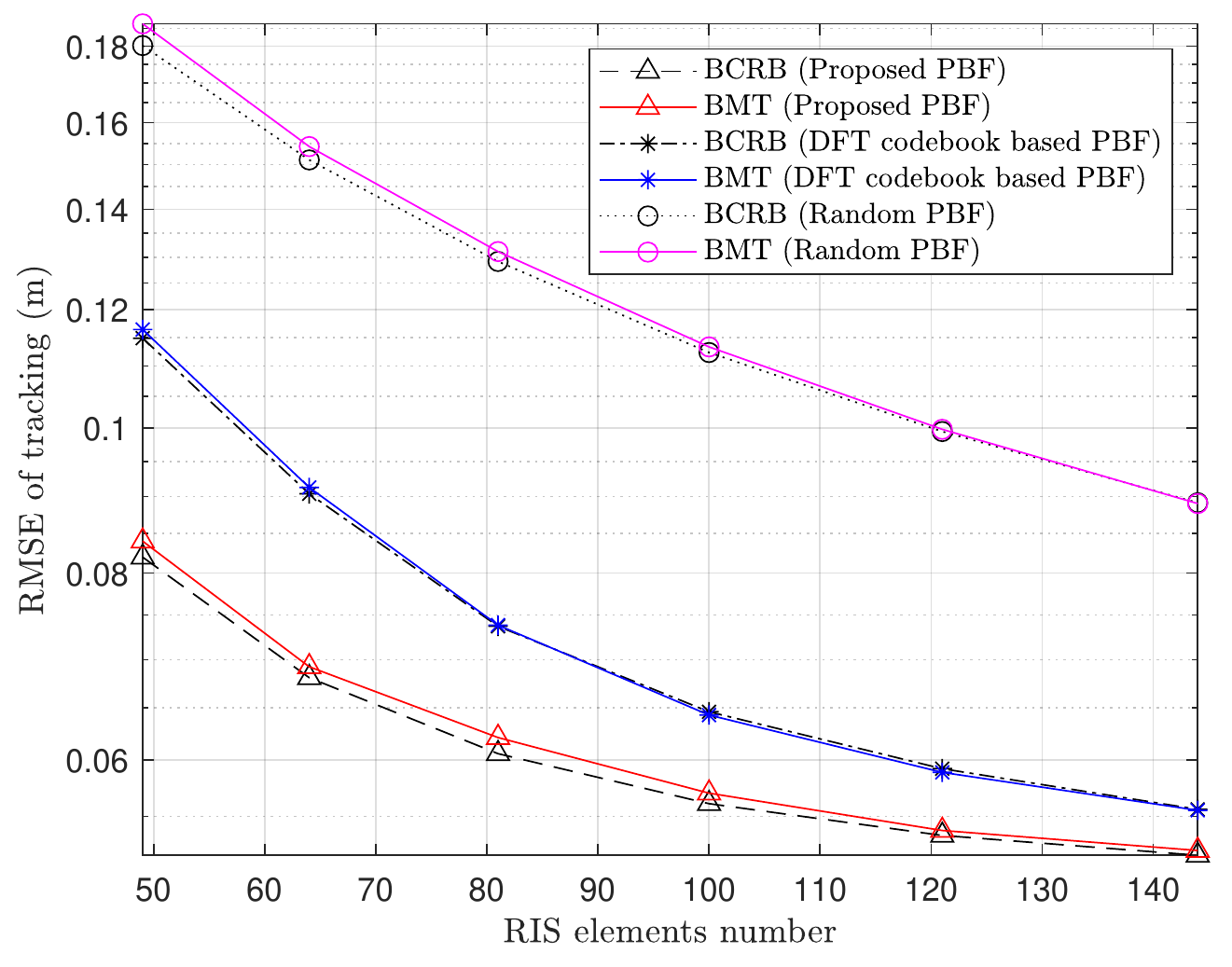}
    \caption{The BMT algorithm performance v.s. RIS elements number $N_{\mathrm{R}}$ with exploiting different passive beamforming methods. The received noise power is set by $-125\ \mathrm{dBm}$. The OFDM overhead is fixed by $G=12$ for comparing PBF schemes.}
    \label{fig:7}
\end{figure}
\subsubsection{Effect of the Number of RIS Elements} As shown in Fig. \ref{fig:7}, we study the impact of the RIS element number on the multiuser tracking performance. The result shows that the increment of the RIS elements number can significantly improve the tracking performance. More RIS elements number lead to a higher receive SNR since the passive beamforming gain is proportional to the RIS element number. Moreover, with equipping more reflecting elements, the RIS-aided tracking system has a higher angle domain resolution thus leading to more precise multiuser tracking. While the proposed PBF 
significantly outperforms benchmark schemes when the number of RIS elements is small, the tracking performance gap between DFT codebook-based PBF and proposed PBF reduces when $N_{\mathrm{R}}$ becomes large.
\section{Conclusion}\label{S7}
In this paper, we studied the multiuser tracking problem in the multi-RIS-aided MIMO-OFDM system. We first built a probability model for multiuser positions and channel parameters by utilizing the geometric relationship of the LoS paths and the temporal correlation of user positions. The blockage of the LoS paths between multiple RIS and users was also considered in our model. Then, we proposed an online Bayesian multiuser tracking algorithm based on the message passing principle where the Gaussian message approximation and variational inference are exploited to calculate the messages. We developed the BCRB to analyze the performance limits of
the tracking problem which further guides the PBF design of the multiple RISs. Numerical simulation results showed that the proposed BMT algorithm can achieve centimeter-level multiuser tracking accuracy and the proposed BCRB-based PBF design
outperforms the benchmark PBF schemes in improving the multiuser tracking accuracy. 
\appendices
% \section{}\label{append1}
% Calculation of $ \Delta_{[\boldsymbol{r}_{g,l}^{(t)}]_{m} \rightarrow h_{m,g,l}^{(t)}}([\boldsymbol{r}_{g,l}^{(t)}]_{m})$
\section{Derivation of \eqref{29}}\label{append2}
The optimal form of $q({\vartheta}_{k,x})$ maximizing $\mathcal{L}$ is calculated by \cite{bishop2006pattern}
\begin{align}\label{28}
    \ln{q({\vartheta}_{k,x})}=\mathbb{E}_{\backslash {\vartheta}_{k,x}}[\ln{\Delta(\boldsymbol{\vartheta}_{x},\boldsymbol{\vartheta}_{y},\boldsymbol{\varsigma},\boldsymbol{\zeta},\boldsymbol{\varrho})}]+\text{const},
\end{align}
where $\mathbb{E}_{\backslash {\vartheta}_{k,x}}[\cdot]$ denote the expectation over $\frac{q(\boldsymbol{\vartheta}_{x},\boldsymbol{\vartheta}_{y},\boldsymbol{\varsigma},\boldsymbol{\zeta},\boldsymbol{\varrho})}{q({\vartheta}_{k,x})}$.
We further obtain
\begin{align}
    \ln&{q({\vartheta}_{k,x})}
    \overset{\left( a \right)}{=}\ln{\Delta_{xp_{k}\rightarrow \vartheta_{k,x} }(\vartheta_{k,x})} + \mathbb{E}_{\backslash {\vartheta}_{k,x}}\left[\int_{\boldsymbol{R}}\prod_{g=1}^{G}\prod_{l=1}^{L} p({r}_{g,l}|\boldsymbol{\vartheta}_{x},\boldsymbol{\vartheta}_{y},\boldsymbol{\varsigma}_{k},\boldsymbol{\varrho})\Delta_{{r}_{g,l} \rightarrow h}({r}_{g,l})\right] + \text{const}\notag\\
    \overset{\left( b \right)}{=}&\ln{\Delta_{xp_{k}\rightarrow \vartheta_{k,x} }(\vartheta_{k,x})}
    - \mathbb{E}_{\backslash {\vartheta}_{k,x}}\left[\frac{N_{\mathrm{B}}}{\nu}\left\|\boldsymbol{R}-\boldsymbol{W}^{\mathrm{T}}\sum_{j=1}^{K} 
    \varrho_{j}\mathbf{a}_{\mathrm{R}}(\vartheta_{j,x},\vartheta_{j,y})\boldsymbol{x}_{j}^{\mathrm{T}}\odot\mathbf{a}_{L}^{\mathrm{T}}(\varsigma_{j})
    \right\|_{\mathrm{F}}^{2}\right]
    + \text{const}
    \notag\\
    =&\ln{\Delta_{xp_{k}\rightarrow \vartheta_{k,x} }(\vartheta_{k,x})}
    + \frac{2N_{\mathrm{B}}}{\nu}\mathbb{E}_{\backslash {\vartheta}_{k,x}}\left[\Re\left\{\boldsymbol{x}_{k}^{\mathrm{T}}\odot\mathbf{a}_{L}^{\mathrm{T}}(\varsigma_{k})\boldsymbol{R}^{\mathrm{H}}\boldsymbol{W}^{\mathrm{T}}
    \varrho_{k}\mathbf{a}_{\mathrm{R}}(\vartheta_{k,x},\vartheta_{k,y})\right\}\right]\notag\\
    &-\frac{2N_{\mathrm{B}}}{\nu}
    \mathbb{E}_{\backslash {\vartheta}_{k,x}}\left[\Re\left\{\sum_{j\neq k}
    \varrho_{j}^{*}\varrho_{k}\boldsymbol{x}_{k}^{\mathrm{T}}\odot\mathbf{a}_{L}^{\mathrm{T}}(\varsigma_{k})\boldsymbol{x}_{j}^{*}\odot\mathbf{a}_{L}^{*}(\varsigma_{j})\mathbf{a}_{\mathrm{R}}^{\mathrm{H}}(\vartheta_{j,x},\vartheta_{j,y})\boldsymbol{W}^{*}\boldsymbol{W}^{\mathrm{T}}
    \mathbf{a}_{\mathrm{R}}(\vartheta_{k,x},\vartheta_{k,y})\right\}\right]
    \notag\\
    &-\frac{N_{\mathrm{B}}}{\nu}
    \mathbb{E}_{\backslash {\vartheta}_{k,x}}\!\left[
    \varrho_{k}^{*}\varrho_{k}\boldsymbol{x}_{k}^{\mathrm{T}}\odot\mathbf{a}_{L}^{\mathrm{T}}(\varsigma_{k})\boldsymbol{x}_{k}^{*}\odot\mathbf{a}_{L}^{*}(\varsigma_{k})\mathbf{a}_{\mathrm{R}}^{\mathrm{H}}(\vartheta_{k,x},\vartheta_{k,y})\boldsymbol{W}^{*}\boldsymbol{W}^{\mathrm{T}}
    \mathbf{a}_{\mathrm{R}}(\vartheta_{k,x},\vartheta_{k,y})\right]\!+\! \text{const},
\end{align}
where $\overset{\left( a \right)}{=}$ and $\overset{\left( b \right)}{=}$ exploit \eqref{22nn} and \eqref{21} respectively.
Calculating the expectation, we obtain \eqref{29}.
\section{Calculation of the Expectation Involved in \eqref{29} and \eqref{36}}\label{append3}
We construct the matrix $\boldsymbol{\Omega}_{g}\in\mathbb{C}^{N_\mathrm{y}\times N_{\mathrm{x}}}$ corresponding to the $g$-th column of $\boldsymbol{W}$ satisfying $\left[\boldsymbol{\Omega}_{g}\right]_{i,j}=[\boldsymbol{\omega}_{g}]_{(j-1)N_{\mathrm{y}}+i}$. We have
\begin{align}
\big\|\boldsymbol{W}^{\mathrm{T}}\mathbf{a}_{\mathrm{R}}(\vartheta_{k,x},\vartheta_{k,y})\big\|_{2}^{2}
    &=\sum\nolimits_{g=1}^{G}\mathbf{a}_{\mathrm{R}}^{\mathrm{H}}(\vartheta_{k,x},\vartheta_{k,y})\boldsymbol{\omega}_{g}^{*}\boldsymbol{\omega}_{g}^{\mathrm{T}}\mathbf{a}_{\mathrm{R}}(\vartheta_{k,x},\vartheta_{k,y})\notag\allowdisplaybreaks\\
    &\overset{\left( a \right)}{=}\mathbf{a}_{\mathrm{x}}^{\mathrm{H}}(\vartheta_{k,x})\sum\nolimits_{g=1}^{G}\left(\boldsymbol{\Omega}_{g}^{\mathrm{H}}\mathbf{a}_{\mathrm{y}}^{*}(\vartheta_{k,y})\mathbf{a}_{\mathrm{y}}^{\mathrm{T}}(\vartheta_{k,y})\boldsymbol{\Omega}_{g}\right)\mathbf{a}_{\mathrm{x}}(\vartheta_{k,x}),
\end{align}
where $\overset{\left( a \right)}{=}$ utilizes $\boldsymbol{\omega}_{g}^{\mathrm{T}}\mathbf{a}_{\mathrm{R}}(\vartheta_{k,x},\vartheta_{k,y})=\mathbf{a}_{\mathrm{y}}^{\mathrm{T}}(\vartheta_{k,y})\boldsymbol{\Omega}_{g}\mathbf{a}_{\mathrm{x}}(\vartheta_{k,x})$. We calculate
\begin{align}
    \mathbb{E}_{q(\vartheta_{k,y}|\boldsymbol{R})}\left[\big\|\boldsymbol{W}^{\mathrm{T}}\mathbf{a}_{\mathrm{R}}(\vartheta_{k,x},\vartheta_{k,y})\big\|_{2}^{2}\right]\!=\mathbf{a}_{\mathrm{x}}^{\mathrm{H}}(\vartheta_{k,x})\sum_{g=1}^{G}\left(\boldsymbol{\Omega}_{g}^{\mathrm{H}}\mathbb{E}_{q(\vartheta_{k,y}|\boldsymbol{R})}\left[\mathbf{a}_{\mathrm{y}}^{*}(\vartheta_{k,y})\mathbf{a}_{\mathrm{y}}^{\mathrm{T}}(\vartheta_{k,y})\right]\boldsymbol{\Omega}_{g}\right)\mathbf{a}_{\mathrm{x}}(\vartheta_{k,x}),
\end{align}
where $\mathbb{E}_{q(\vartheta_{k,y}|\boldsymbol{R})}\left[\mathbf{a}_{\mathrm{y}}^{*}(\vartheta_{k,y})\mathbf{a}_{\mathrm{y}}^{\mathrm{T}}(\vartheta_{k,y})\right]$ is calculated by $\text{Toep}(\hat{\mathbf{a}}_{\mathrm{y}}^{*}(\vartheta_{k,y}),\hat{\mathbf{a}}_{\mathrm{y}}^{\mathrm{T}}(\vartheta_{k,y}))$ with $\text{Toep}(\boldsymbol{u},\boldsymbol{v})$ denotes the Toeplitz matrix constructed by setting $\boldsymbol{u}$ as the first column and setting $\boldsymbol{v}$ as the first row. A more compact form is obtained as
\begin{align}
        \mathbb{E}_{q(\vartheta_{i,y}|\boldsymbol{R})}\big[\left\|\boldsymbol{W}^{\mathrm{T}}\mathbf{a}_{\mathrm{R}}(\vartheta_{i,x},\vartheta_{i,y})\right\|_{2}^{2}\big]=2\Re\{\boldsymbol{\beta}_{x,k}^{\mathrm{H}}{\mathbf{a}}_{\mathrm{x}}(\vartheta_{k,x})\},
\end{align}
where
the $n$-th term of $\boldsymbol{\beta}_{x,k}$ is $[\boldsymbol{\beta}_{x,k}]_{n} = \sum_{(i,j)\in\mathcal{C}}\left[\sum_{g=1}^{G}\left(\boldsymbol{\Omega}_{g}^{\mathrm{H}}\text{Toep}(\hat{\mathbf{a}}_{\mathrm{y}}^{*}(\vartheta_{k,y}),\hat{\mathbf{a}}_{\mathrm{y}}^{\mathrm{T}}(\vartheta_{k,y}))\boldsymbol{\Omega}_{g}\right)\right]_{i,j}$ with $\mathcal{C}=\{(i,j)|1\leq i,j\leq N_{\mathrm{x}},i-j+1=n\}$. We further
obtain 
\begin{align}
    \mathbb{E}_{q(\vartheta_{i,x}|\boldsymbol{R})q(\vartheta_{i,y}|\boldsymbol{R})}\big[\left\|\boldsymbol{W}^{\mathrm{T}}\mathbf{a}_{\mathrm{R}}(\vartheta_{i,x},\vartheta_{i,y})\right\|_{2}^{2}\big]=2\Re\{\boldsymbol{\beta}_{x,k}^{\mathrm{H}}\hat{\mathbf{a}}_{\mathrm{x}}(\vartheta_{k,x})\}.
\end{align}
\section{Derivation of \eqref{40}}\label{append4}
Substituting \eqref{12} and \eqref{366} in \eqref{38}, we have
\begin{align}\label{64}
    \ln&\left(\Delta_{xp_{m,k}^{(t)}\rightarrow\mathbf{p}_{k}^{(t)}}(\mathbf{p}_{k}^{(t)})
    \Delta_{yp_{m,k}^{(t)}\rightarrow\mathbf{p}_{k}^{(t)}}(\mathbf{p}_{k}^{(t)})
    \Delta_{dp_{m,k}^{(t)}\rightarrow\mathbf{p}_{k}^{(t)}}(\mathbf{p}_{k}^{(t)})\right)\notag\\
    &=\kappa_{\vartheta_{m,k,x}^{(t)}\rightarrow xp_{m,k}^{(t)}}\cos\left((\mathbf{e}_{m,k}^{(t)})^{\mathrm{T}}\mathbf{e}_{m,x}-\phi_{m,x}-\mu_{\vartheta_{m,k,x}^{(t)}\rightarrow xp_{m,k}^{(t)}}\right)\notag\\
    &+\kappa_{\vartheta_{m,k,y}^{(t)}\rightarrow yp_{m,k}^{(t)} }\cos\left((\mathbf{e}_{m,k}^{(t)})^{\mathrm{T}}\mathbf{e}_{m,y}-\phi_{m,y}-\mu_{\vartheta_{m,k,y}^{(t)}\rightarrow yp_{m,k}^{(t)}}\right)\notag\\
    &+\kappa_{\varsigma_{m,k}^{(t)}\rightarrow dp_{m,k}^{(t)}}\cos\left(\tau_{m}+ \frac{\left\|\mathbf{p}_{k}^{(t)}-\mathbf{p}_{\mathrm{R},m}\right\|_{2}}{\mathrm{c}_{0}}-\mu_{\varsigma_{m,k}^{(t)}\rightarrow dp_{m,k}^{(t)}}\right),
\end{align}
where $\mathbf{e}_{m,k}^{(t)}=\frac{\mathbf{p}_{k}^{(t)}-\mathbf{p}_{\mathrm{R},m}}{\|\mathbf{p}_{k}^{(t)}-\mathbf{p}_{\mathrm{R},m}\|}$. The mean of Gaussian distribution in \eqref{40} is given by the maximum of \eqref{64}, the covariance matrix is given by the Hessian matrix at the maximum. Letting each cosine term equal $1$, the maximum is obtained as expressed in \eqref{40a}. We approximate the cosine term in \eqref{64} by $\cos(\theta)=1-\frac{\theta^{2}}{2}$ and substitute $\|\mathbf{p}_{k}^{(t)}-\mathbf{p}_{\mathrm{R},m}\|$ by $d_{m,k}^{(t)}$. Keeping the quadratic term of $(\mathbf{p}_{k}^{(t)}-\mathbf{p}_{\mathrm{R},m})$, we obtain
\begin{align}
    \ln&\left(\Delta_{xp_{m,k}^{(t)}\rightarrow\mathbf{p}_{k}^{(t)}}(\mathbf{p}_{k}^{(t)})
    \Delta_{yp_{m,k}^{(t)}\rightarrow\mathbf{p}_{k}^{(t)}}(\mathbf{p}_{k}^{(t)})
    \Delta_{dp_{m,k}^{(t)}\rightarrow\mathbf{p}_{k}^{(t)}}(\mathbf{p}_{k}^{(t)})\right)\notag\\
    =&-(\mathbf{p}_{k}^{(t)}-\mathbf{p}_{\mathrm{R},m})^{\mathrm{T}}
    \left(\frac{\kappa_{\vartheta_{m,k,x}^{(t)}\rightarrow xp_{m,k}^{(t)}}\mathbf{e}_{m,x}\mathbf{e}_{m,x}^{\mathrm{T}}}{2(d_{m,k}^{(t)})^{2}}
    +\frac{\kappa_{\vartheta_{m,k,y}^{(t)}\rightarrow yp_{m,k}^{(t)}}\mathbf{e}_{m,y}\mathbf{e}_{m,y}^{\mathrm{T}}}{2(d_{m,k}^{(t)})^{2}}\right.\notag\\
    &\left.+\frac{\kappa_{\varsigma_{m,k}^{(t)}\rightarrow dp_{m,k}^{(t)}}(\mathbf{p}_{k}^{(t)}-\mathbf{p}_{\mathrm{R},m})(\mathbf{p}_{k}^{(t)}-\mathbf{p}_{\mathrm{R},m})^{\mathrm{T}}}{2\mathrm{c}_{0}^{2}(d_{m,k}^{(t)})^{2}}\right)
    (\mathbf{p}_{k}^{(t)}-\mathbf{p}_{\mathrm{R},m}) + \text{else}.
\end{align}
Calculating the Hessian matrix at the maximum further yields \eqref{40b}.
\bibliographystyle{IEEEtran}
\bibliography{mybib}
\end{document}